\documentclass{article}

\usepackage{arxiv}

\usepackage[utf8]{inputenc} 
\usepackage[T1]{fontenc}    
\usepackage{hyperref}       
\usepackage{url}            
\usepackage{booktabs}       
\usepackage{amsfonts}       
\usepackage{nicefrac}       
\usepackage{microtype}      
\usepackage{lipsum}		
\usepackage{graphicx}
\usepackage{natbib}
\usepackage{doi}

\usepackage[]{acro}
\usepackage{amsmath,amsthm,amssymb}
\usepackage[capitalise]{cleveref}
\usepackage{subcaption}

\DeclareAcronym{bnn}{
	short = BNN ,
	short-plural = s ,
	long = Bayesian neural network ,
	long-plural = s
}

\DeclareAcronym{nisq}{
	short = NISQ ,
	long = noisy intermediate-scale quantum
}

\DeclareAcronym{ipe}{
	short = IPE ,
	long = inner product estimation
}

\DeclareAcronym{qram}{
	short = QRAM ,
	long = quantum random access memory
}

\DeclareAcronym{mcmc}{
	short = MCMC ,
	long = Markov Chain Monte Carlo
}

\DeclareAcronym{sgld}{
	short = SGLD ,
	long = stochastic gradient Langevin dynamics
}

\DeclareAcronym{hmc}{
	short = HMC ,
	long = Hamiltonian Monte Carlo
}

\DeclareAcronym{nuts}{
	short = NUTS ,
	long = No-U-Turn Sampler
}

\DeclareAcronym{jvp}{
	short = JVP ,
	short-plural = s ,
	long = Jacobian-vector product, 
	long-plural = s
}

\newcommand{\params}{\theta}
\newcommand{\data}{\mathcal{D}}
\newcommand{\given}{\, \vert \,}
\newcommand{\diff}{\mathrm{d}}
\newcommand{\bigO}{\mathcal{O}}

\def\ket#1{\mathinner{|{#1}\rangle}}
\newcommand{\braket}[2]{\langle #1|#2\rangle}
\newcommand{\norm}[1]{\left\lVert#1\right\rVert}

\title{Quantum Bayesian Neural Networks}

\date{}	

\author{ Noah Berner\\
	Department of Information Technology\\and Electrical Engineering\\
	ETH Z\"{u}rich\\
	Z\"{u}rich, Switzerland \\
	\texttt{noberner@student.ethz.ch} \\
	\And
	Vincent Fortuin\\
	Department of Computer Science\\
	ETH Z\"{u}rich\\
	Z\"{u}rich, Switzerland \\
	\texttt{fortuin@inf.ethz.ch} \\
	\And
	Jonas Landman\\
	IRIF\\
	Universit\'{e} Paris Diderot\\
	Paris, France \\
	\texttt{landman@irif.fr} \\
}

\renewcommand{\headeright}{}
\renewcommand{\undertitle}{}

\renewcommand{\shorttitle}{Quantum Bayesian Neural Networks}

\hypersetup{
pdftitle={Quantum Bayesian Neural Networks},
pdfauthor={Noah~Berner, Vincent Fortuin, Jonas Landman},
pdfkeywords={Machine Learning, Quantum Computing, Quantum Machine Learning, Bayesian Neural Networks},
}

\begin{document}
\maketitle

\begin{abstract}
	Quantum machine learning promises great speedups over classical algorithms, but it often requires repeated computations to achieve a desired level of accuracy for its point estimates.
    Bayesian learning focuses more on sampling from posterior distributions than on point estimation, thus it might be more forgiving in the face of additional quantum noise.
    We propose a quantum algorithm for Bayesian neural network inference, drawing on recent advances in quantum deep learning, and simulate its empirical performance on several tasks.
    We find that already for small numbers of qubits, our algorithm approximates the true posterior well, while it does not require any repeated computations and thus fully realizes the quantum speedups.
\end{abstract}


\section{Introduction}
\label{introduction}

Quantum machine learning generally comes in two flavors: Variational quantum circuits that mimic the training of neural networks \citep{cerezo2020variational}, which run on \ac{nisq} devices \citep{NISQpreskill}, and quantum algorithms aimed at replacing classical training and prediction algorithms for neural networks \citep{allcock2020quantum}, which run on (future) error-corrected quantum computers.
The latter often use a quantum algorithm for estimating the inner product calculations that occur when training and evaluating neural networks. This \ac{ipe} can evaluate inner products with lower asymptotic complexity than classical algorithms but does so at lower accuracy. When evaluating standard neural networks, this lowered accuracy in the inner product calculation has to be corrected by running the quantum algorithm multiple times to get a better estimate.

However, for \acp{bnn}, the goal is not to get the best point estimate of the parameters $\params^* = \mathrm{arg~max}_\params \; p(\data \given \params)$, as it would be in maximum-likelihood learning.
Instead, one wishes to obtain $K$ samples from the Bayesian posterior over the parameters $\params$ given the data $\data$, that is, $\params_i \sim p(\params \given \data) \propto p(\params) \, p(\data \given \params)$.
These samples can then be used to approximate the posterior predictive for unseen data $\data^*$, that is, $p(\data^* \given \data) = \int p(\data^* \given \params) \, p(\params \given \data) \, \diff \params \approx \frac{1}{K} \sum_{i=1}^K p(\data^* \given \params_i)$.

Since these samples are noisy by stipulation, they might allow for a larger margin of error in the quantum computations.
Ideally, under zero-mean quantum noise with sufficiently small variance, one might achieve meaningful results with just running one single quantum computation per sample, thus realizing the maximum possible quantum speedup.
In this work, we investigate this idea empirically and demonstrate a proof of concept for \ac{bnn} inference on quantum computers.
We find that already for decently small numbers of qubits, our quantum algorithm approximates the true posterior well with just one computation per sample.

\section{Quantum Bayesian Neural Networks}

This paper focuses on reducing the asymptotic runtime of the inference and prediction in \acp{bnn} using quantum algorithms. The algorithms described in this paper are derived from the quantum deep learning algorithms introduced by \citet{allcock2020quantum} for feedforward neural networks. We will describe the alterations made for \acp{bnn} and their consequences in this section. 

\subsection{Quantum Inner Product Estimation}
\label{sec:ipe_without_me}

The main quantum speedups are gained in our algorithm by replacing the classical inner product $v_i^\top v_j$ of two vectors in $\mathbb{R}^{d}$ by its quantum estimate.
To this end, we use the modified \ac{ipe} algorithm, based on the work by \citet{kerenidis2018qmeans}.
It achieves an asymptotic runtime of 

\begin{equation}
    \tilde{\bigO} \left( T\frac{\|v_i\| \|v_j\|}{\epsilon} \right), 
    \label{eq:ipe_runtime}
\end{equation}

where $T$ is the time to load the input vectors into a superposition quantum state, which becomes $T = \bigO(\text{polylog}(d))$ if we assume \ac{qram} or an equivalent quantum memory model \citep{kerenidis_recommendation_system}. $\epsilon$ is an error bound on the inner product estimate, which in turn depends on the number of qubits $n$ used in the quantum phase estimation subroutine. $\tilde{\bigO}$ hides polylogarithmic factors.

Our modification for the usage in \acp{bnn} dispenses of median evaluation from the \ac{ipe} algorithm. Instead of evaluating the inner product multiple times, we use a single estimate. Thus, the asymptotic runtime is reduced by a factor of $\log(1/\Delta)$, where $\Delta$ ensured a specific probability to attain the error $\epsilon$ in the inner product. Consequently, our \ac{ipe} algorithm only has a constant probability of estimating the inner product within an error $\epsilon$.

\subsubsection{Quantum Noise in the Inner Product Estimation}
\label{sec:noise}

\begin{figure}
\vskip 0.2in
    \centering
    \includegraphics[width=0.6\linewidth]{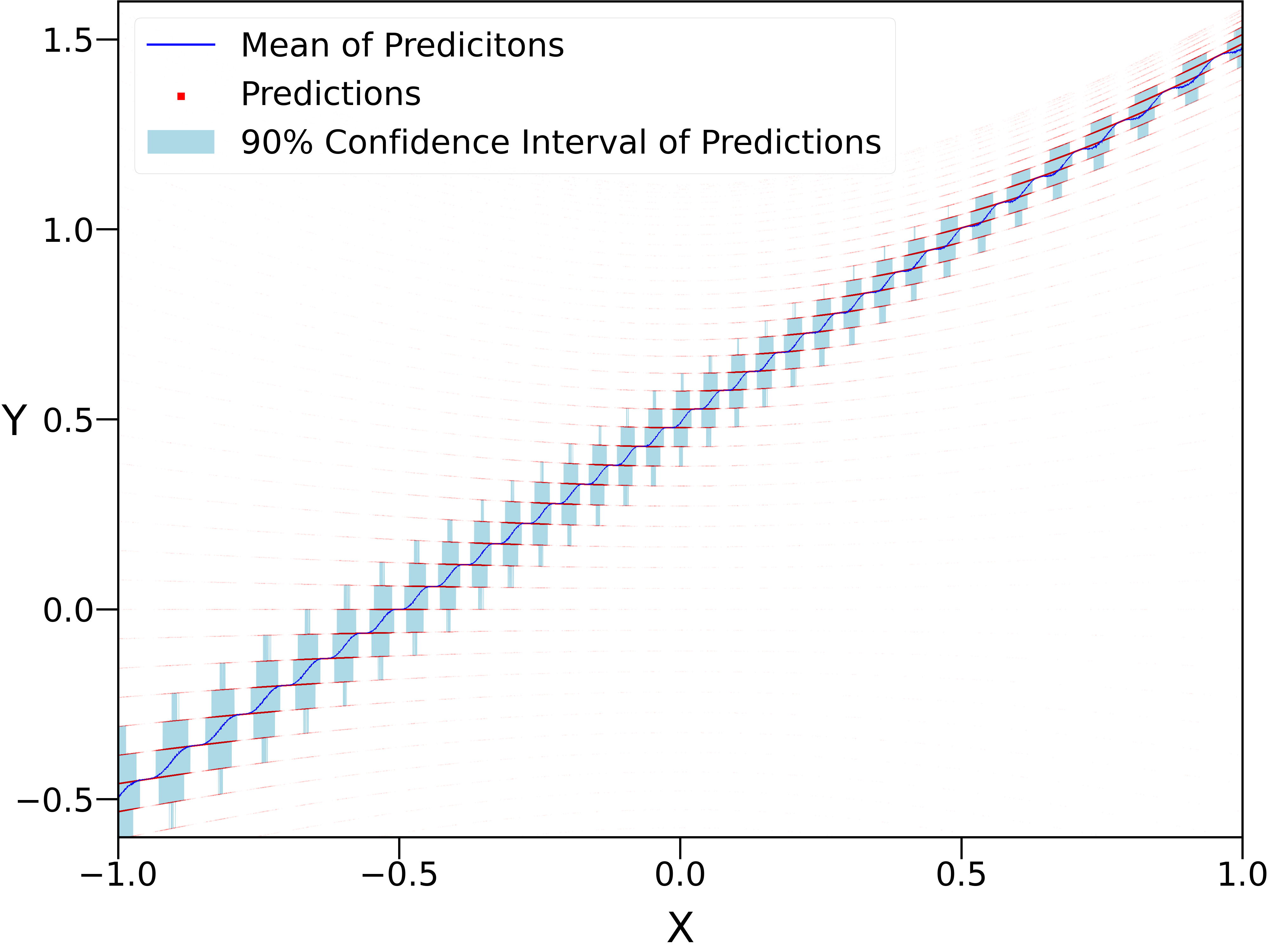}
    \caption{Noise pattern of the \ac{ipe} quantum algorithm. The simulation estimates the inner product $y$ between $(x, 1.0)$ and $(1.0, 0.5)$ using $n=7$ qubits. The noise is non-isotropic and periodic.}
    \label{fig:ipe_noise}
\vskip -0.2in
\end{figure}

The periodicity and non-isotropy of the noise seen in \cref{fig:ipe_noise} stems mostly from the phase estimation subroutine (see \cref{sec:phase_estimation}). The representation of the inner products in the $n$ available qubits means there are only $2^n$ possible values available for the inner product estimate. Also, the \ac{ipe} algorithm uses probabilistic subroutines to estimate the inner product. Thus, the best estimate in the $2^n$ possible values is not attained with certainty. For a more thorough treatment of the noise characteristics of the \ac{ipe} algorithm, see \cref{sec:ipe_error_analysis}.

\begin{figure}[t]
\vskip 0.2in
\begin{center}
    \begin{subfigure}[b]{0.24\linewidth}
        \includegraphics[width=\textwidth]{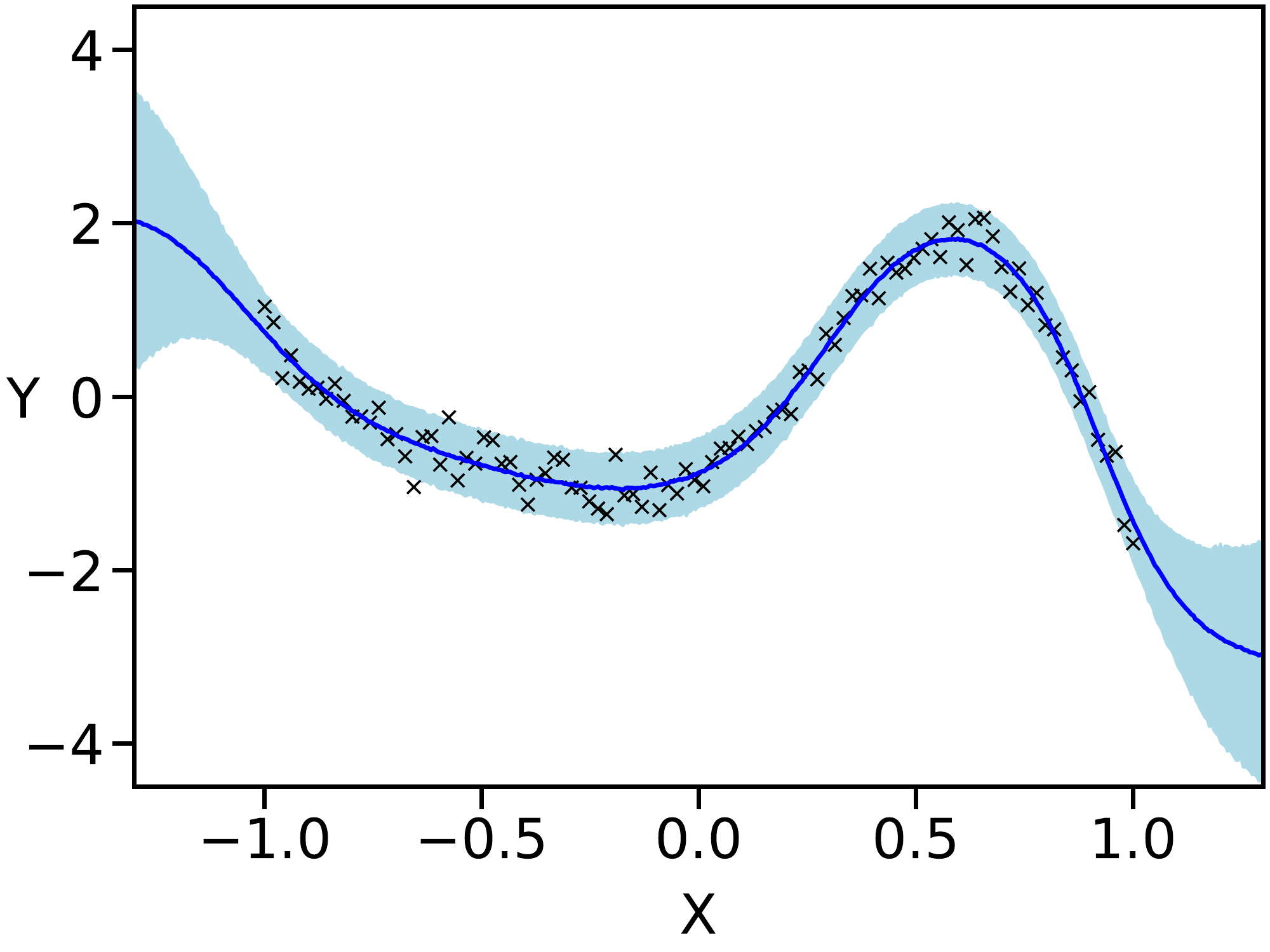}
        \caption{CICP (Reference)}
        \label{fig:lr_cicp_0}
    \end{subfigure}
    \begin{subfigure}[b]{0.24\linewidth}
        \includegraphics[width=\textwidth]{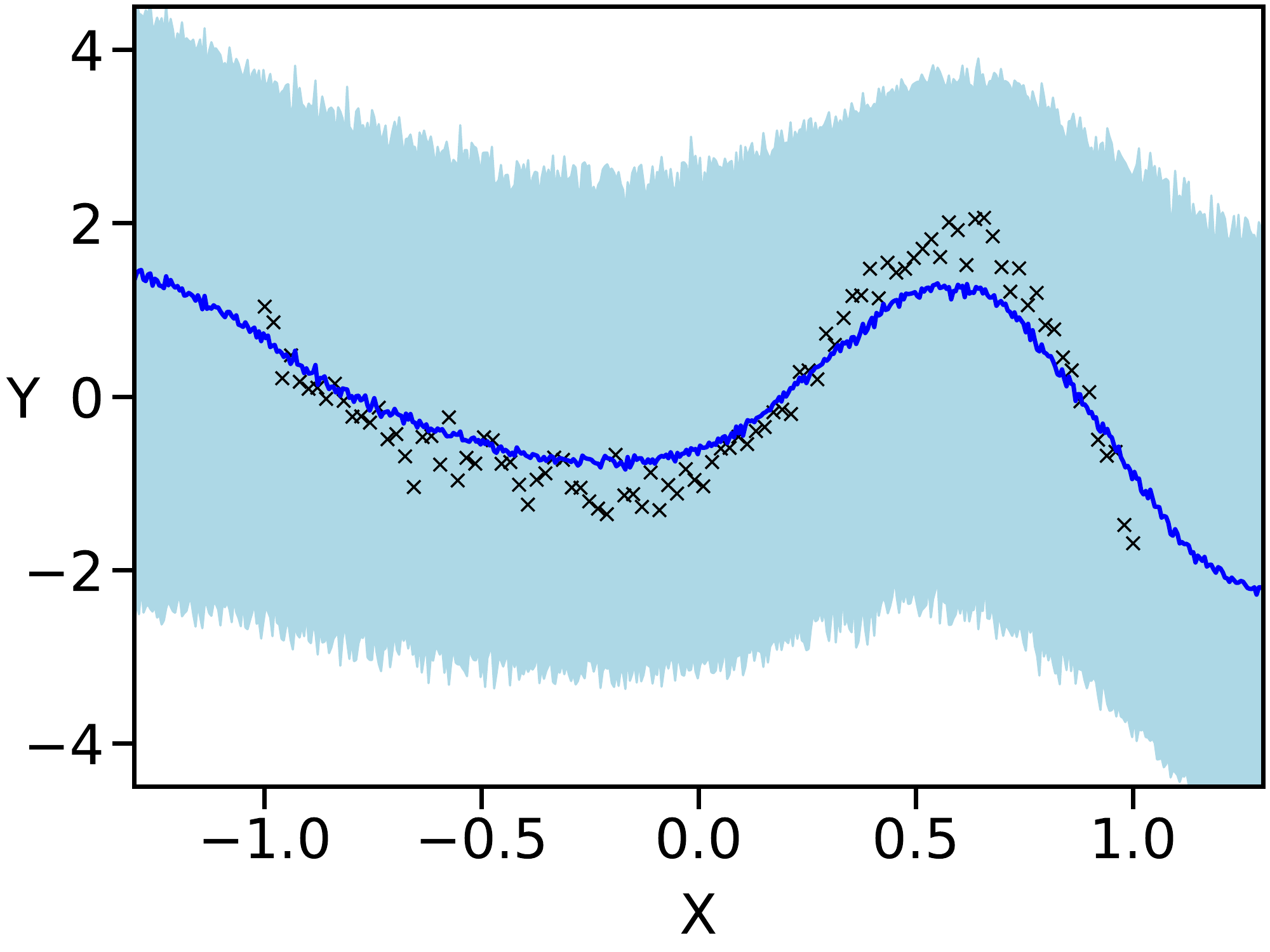}
        \caption{CIQP, $n=5$}
        \label{fig:lr_ciqp_5_0}
    \end{subfigure}
    \begin{subfigure}[b]{0.24\linewidth}
        \includegraphics[width=\textwidth]{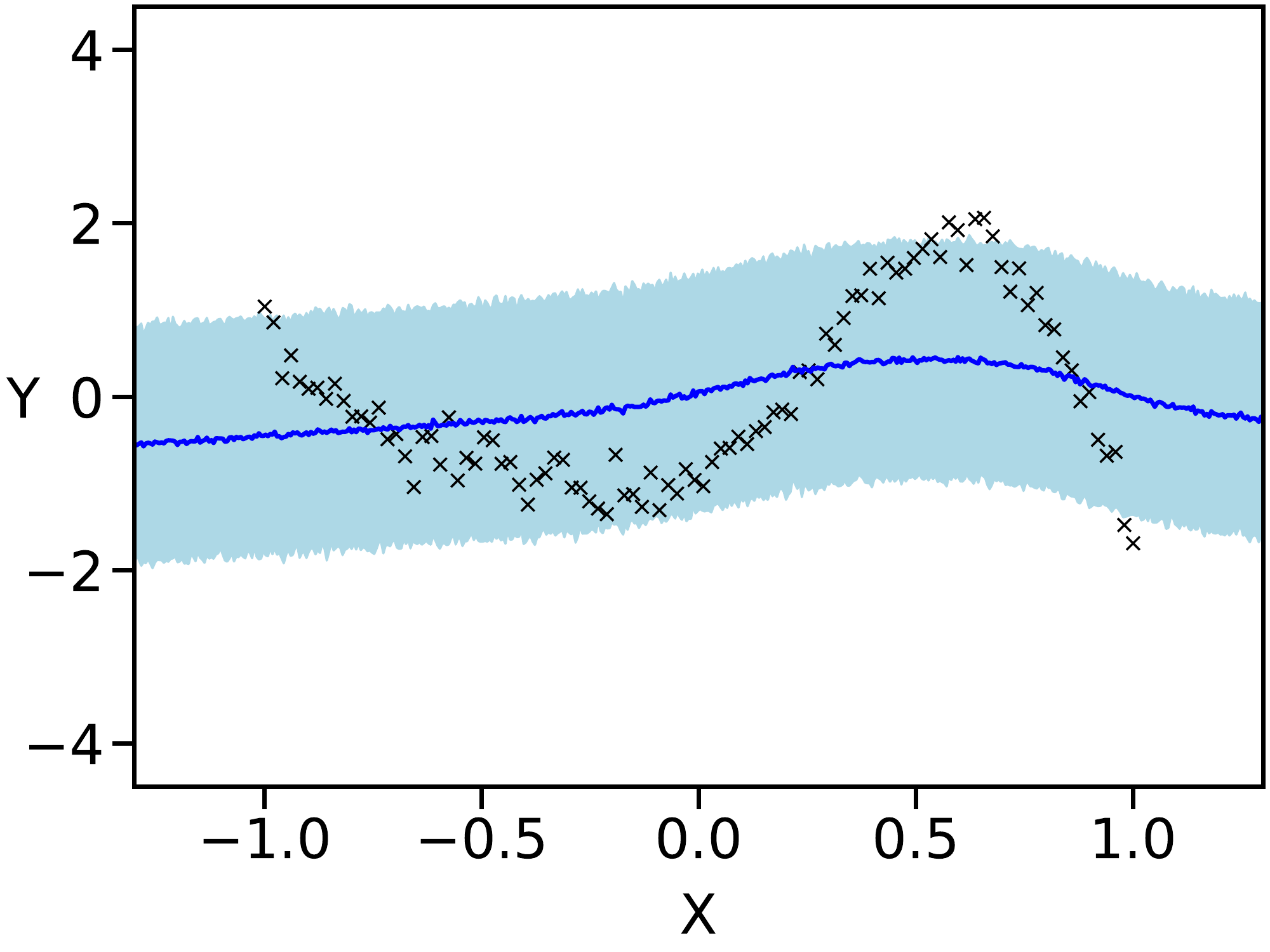}
        \caption{QICP, $n=5$}
        \label{fig:lr_qicp_5_0}
    \end{subfigure}
    \begin{subfigure}[b]{0.24\linewidth}
        \includegraphics[width=\textwidth]{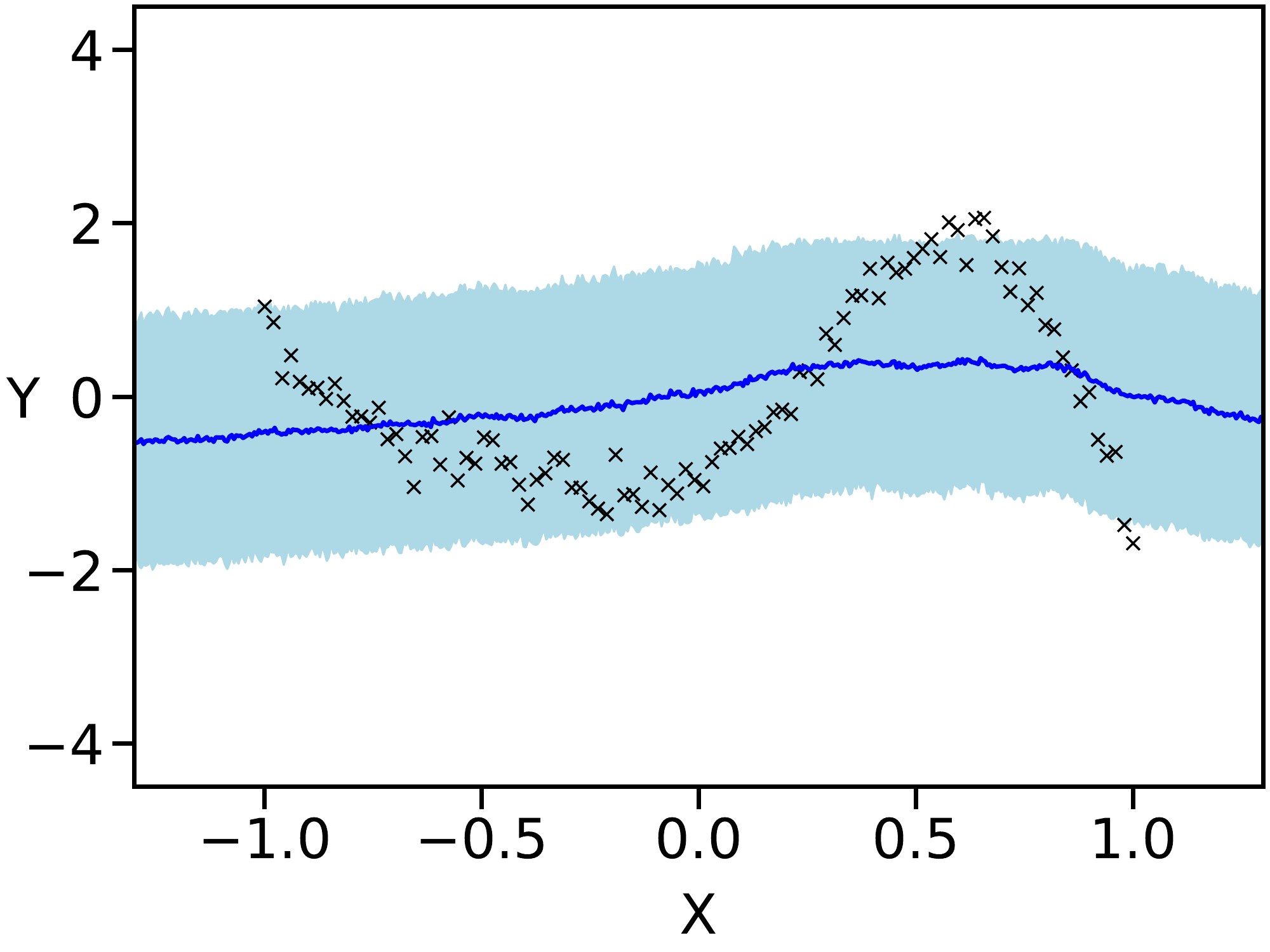}
        \caption{QIQP, $n=5$}
        \label{fig:lr_qiqp_5_0}
    \end{subfigure}
    \vskip 1.0em
    \begin{subfigure}[b]{0.24\linewidth}
        \includegraphics[width=\textwidth]{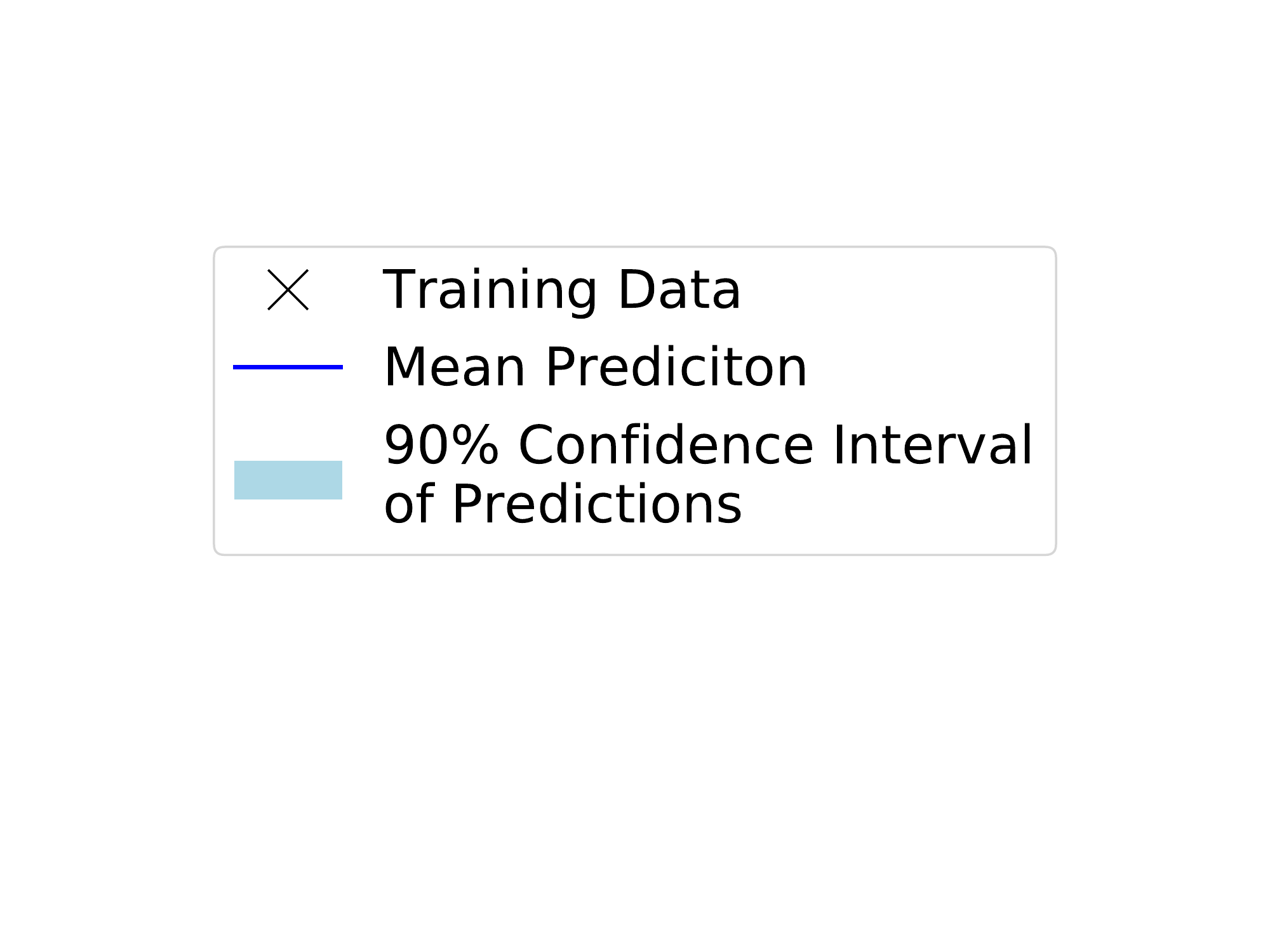}
        \caption{Legend}
    \end{subfigure}
    \begin{subfigure}[b]{0.24\linewidth}
        \includegraphics[width=\textwidth]{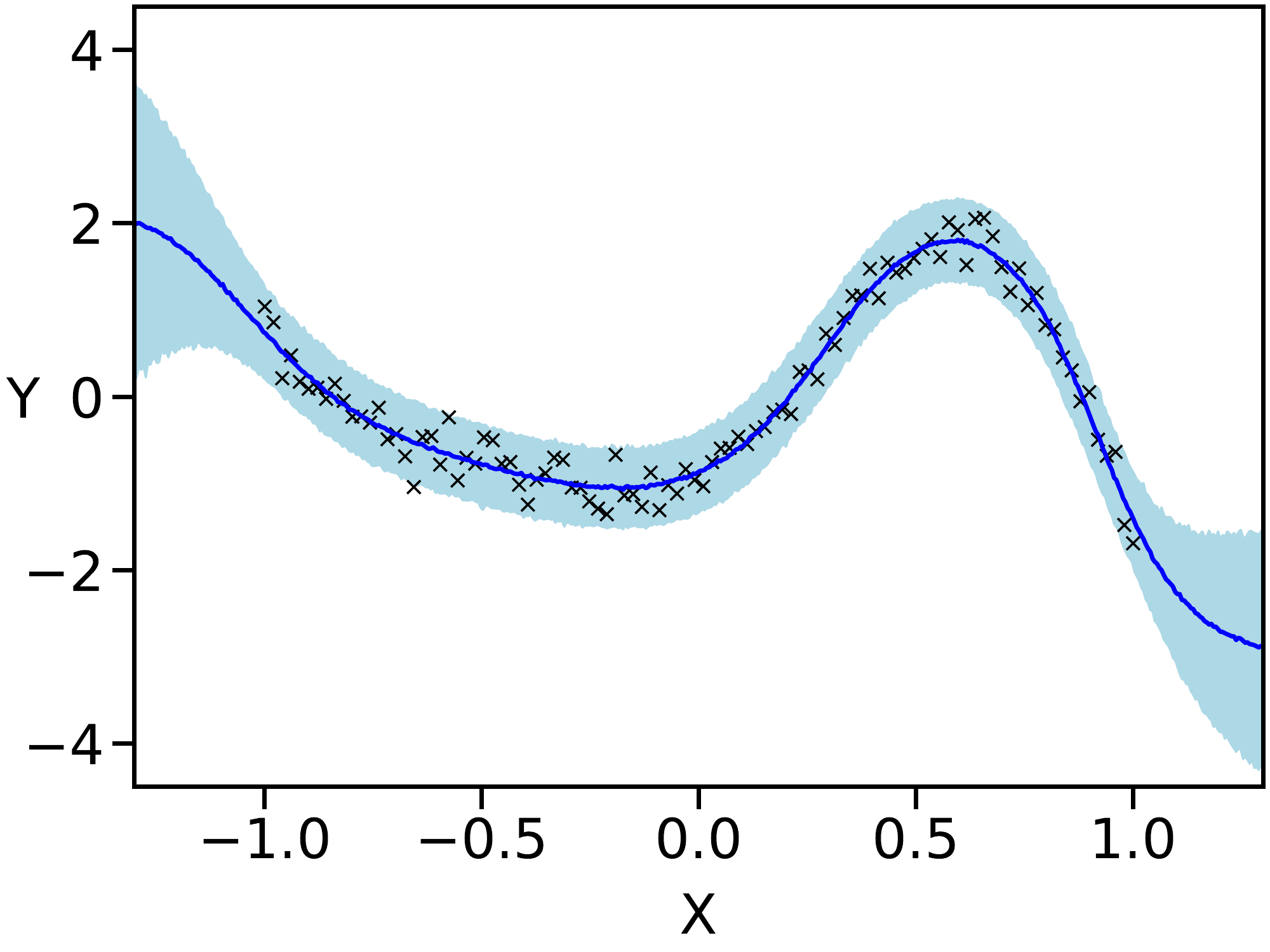}
        \caption{CIQP, $n=10$}
        \label{fig:lr_ciqp_10_0}
    \end{subfigure}
    \begin{subfigure}[b]{0.24\linewidth}
        \includegraphics[width=\textwidth]{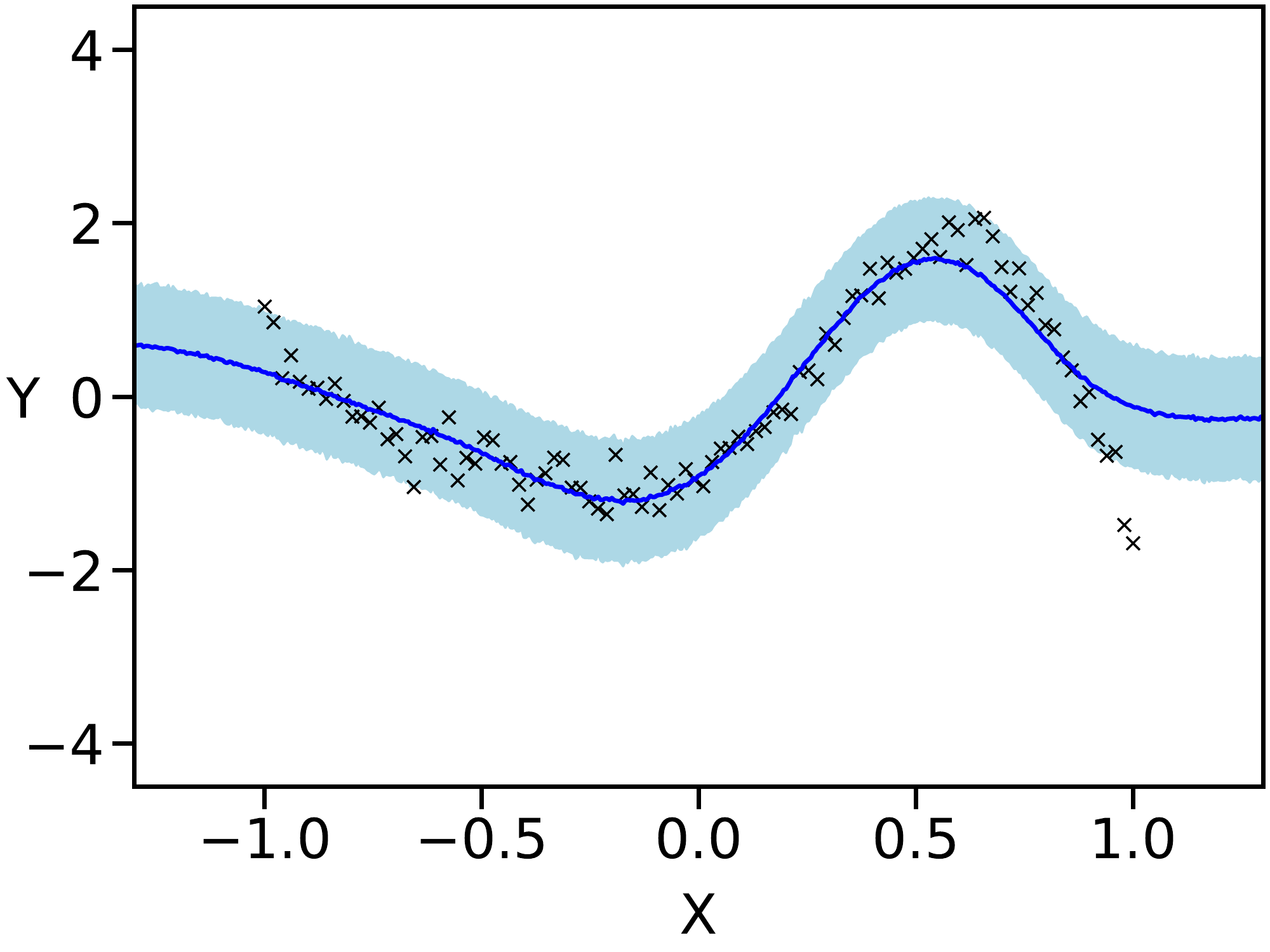}
        \caption{QICP, $n=10$}
        \label{fig:lr_qicp_10_0}
    \end{subfigure}
    \begin{subfigure}[b]{0.24\linewidth}
        \includegraphics[width=\textwidth]{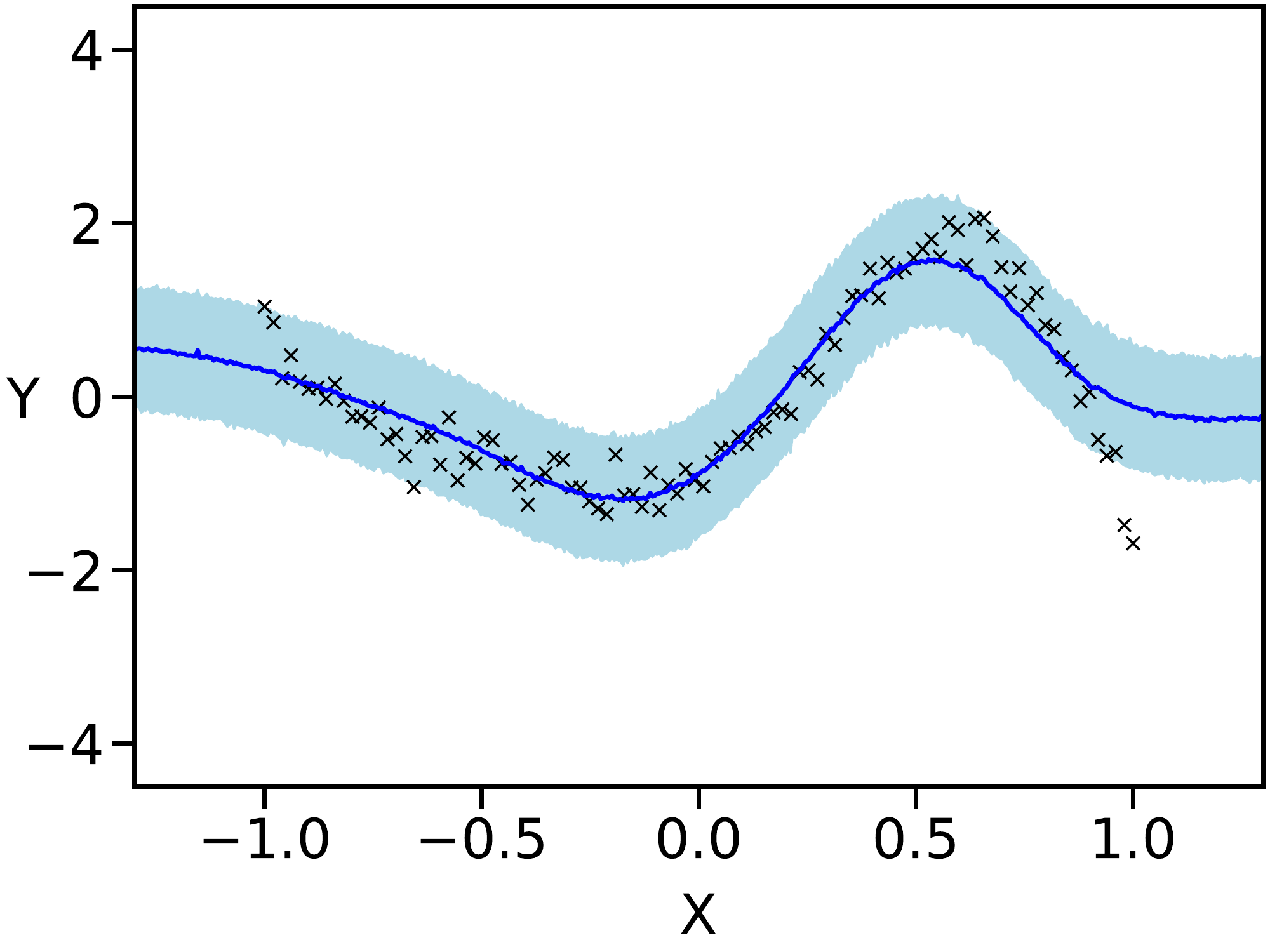}
        \caption{QIQP, $n=10$}
        \label{fig:lr_qiqp_10_0}
    \end{subfigure}
    
\caption{Linear Regression with \ac{bnn}: \emph{C} and \emph{Q} stand for \emph{Classical} and \emph{Quantum} respectively. \emph{I} and \emph{P} stand for \emph{Inference} and \emph{Prediction}. The Figure shows the expected increase in accuracy for higher qubit numbers $n$.}
\label{fig:lr}
\end{center}
\vskip -0.2in
\end{figure}

\subsection{Quantum Inference Algorithm for Bayesian Neural Networks}
\label{sec:quantum_inference}

\Ac{mcmc} methods, such as \ac{sgld}, \ac{hmc}, and \ac{nuts}, are guided by gradients and thus they require backpropagation through the \ac{bnn}. Current software relies on computing these gradients using automatic differentiation and \acp{jvp}. The \ac{jvp} of an inner product contains two inner products (see \cref{sec:jvp}). These can be replaced with our \ac{ipe} algorithm to compute estimates of the true gradient.

Our quantum inference algorithm differs only in, for the asymptotic runtime, negligible parts from the quantum training algorithm described by \citet{allcock2020quantum}. For a single backpropagation, $\bigO(\Omega)$ inner products have to be calculated, where $\Omega$ is the number of neurons in the network. If we draw $K$ samples from our posterior and our training dataset $\data$ has a cardinality of $N = |\data|$, we need to calculate $K N$ backpropagations. Thus our quantum inference algorithm for \acp{bnn} has an asymptotic runtime of 

\begin{equation}
    \tilde{\bigO}\left( (K N)^{1.5} \Omega \frac{1}{\epsilon} R \right),
    \label{eq:quantum_inference_runtime}
\end{equation}

where $R$ is a variable defined in \cref{eq:R} and can be expected to be reasonably small for practical problems. The error component $\frac{1}{\epsilon}$ of the \ac{ipe} depends on the number of qubits $n$ used in the phase estimation subroutine. For a small qubit number $n$, $\frac{1}{\epsilon}$ is also small. The additional factor of $\sqrt{K N}$ is a computational overhead of storing the weight matrices implicitly (see \cref{sec:low_rank}).

A classical inference would incur a runtime of $\bigO(K N P)$, where $P$ is the number of weights inside the neural network and for a fully-connected \ac{bnn} is proportional to $\Omega^2$. The quantum algorithm has an advantage over the classical algorithm if $\sqrt{K N} \ll \Omega$, that is, for large networks.  


\subsubsection{Low-Rank Initialization and Implicit Storage of Weight Matrices}
\label{sec:low_rank}

After a backpropagation through the network, the updated weight matrices need to be stored. \ac{qram} allows for fast load times into a quantum state, but storing is linear in the input size. Thus, if we were to store the weight matrices explicitly, we would incur a runtime of $\bigO(\Omega^2)$ per backpropagation. This would negate the speedup seen in \cref{sec:quantum_inference}.  \citet{allcock2020quantum} propose to solve this problem via low-rank initialization and implicit storage of the weight matrices.

As a caveat, it should be noted that the low-rank initialization of the weight matrices needs to be compatible with the initialization using the prior $p(\params)$.
Moreover, the algorithm for implicit storage of the weight matrices is dependent on the inference algorithm and its feasibility needs to be evaluated on a case-by-case basis.

For our simulations, we operate in a full-rank prior regime. However, we also present simulation results that show that if we move to a low-rank prior regime, the results are still viable (see Appendix~\ref{sec:add_results}).
We do assume that the sampler allows for implicit storage of the weight matrices. It will however be an exciting direction for future work to study how these requirements could be relaxed.

\subsection{Quantum Prediction Algorithm for Bayesian Neural Networks}

The prediction algorithm presented in this paper follows the same outline as the evaluation algorithm in \citet{allcock2020quantum}, but with the modified \ac{ipe} introduced in \cref{sec:ipe_without_me}. To get all predictions, the modified evaluation algorithm is executed $K M$ times, where $K$ is the number of weight samples drawn from the posterior and $M$ is the cardinality of the prediction dataset.
The quantum prediction algorithm for a \ac{bnn} has an asymptotic runtime of

\begin{equation}
    \tilde{\bigO} \left(K^{1.5} \sqrt{N}  M  \Omega \frac{1}{\epsilon} R_e \right), 
    \label{eq:quantum_prediction_runtime}
\end{equation}

Here, the additional factor of $\sqrt{K N}$ is a consequence of storing the weight matrices implicitly during the quantum inference algorithm. If we were to use classical inference, this factor would disappear. Again, $R_e$ is a variable defined in \cref{eq:R_e} and can be expected to be reasonably small for practical problems. The rest of the runtime analysis is analogous to the one in \cref{sec:quantum_inference}.

A classical algorithm for evaluating a \ac{bnn} with the same inputs will have an asymptotic complexity of $\tilde{\bigO}(K M P)$. Similarly to the case of training, the quantum algorithm thus provides a speedup over the classical algorithm if $\sqrt{K M} \ll \Omega$, that is, for large networks. If classical inference is used, the speedup occurs unconditionally.

\begin{figure}[t]
\vskip 0.2in
\begin{center}
    \begin{subfigure}[b]{0.24\linewidth}
        \includegraphics[width=\textwidth]{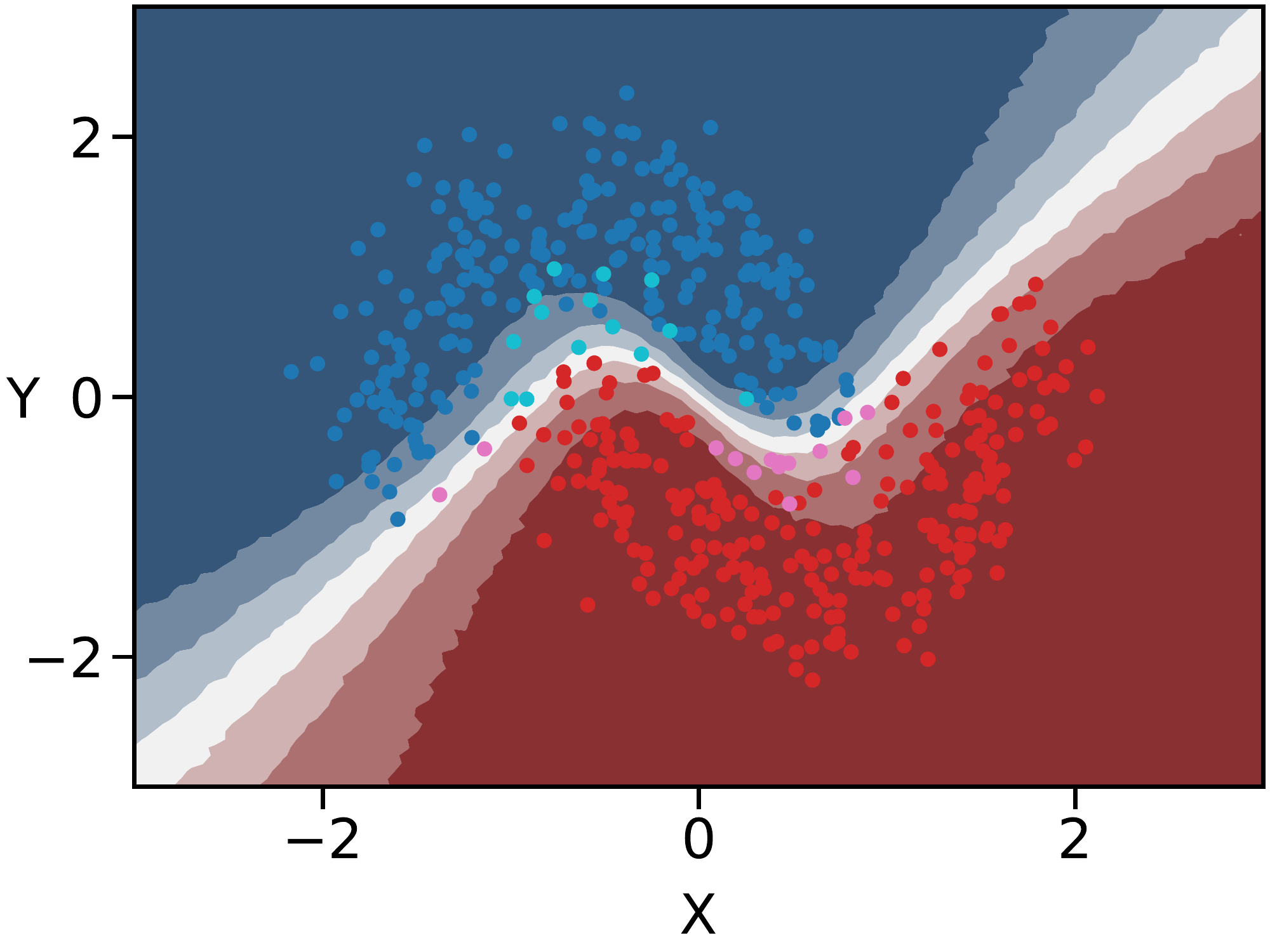}
        \caption{CICP (Reference)}
        \label{fig:bc_mean_cicp_0}
    \end{subfigure}
    \begin{subfigure}[b]{0.24\linewidth}
        \includegraphics[width=\textwidth]{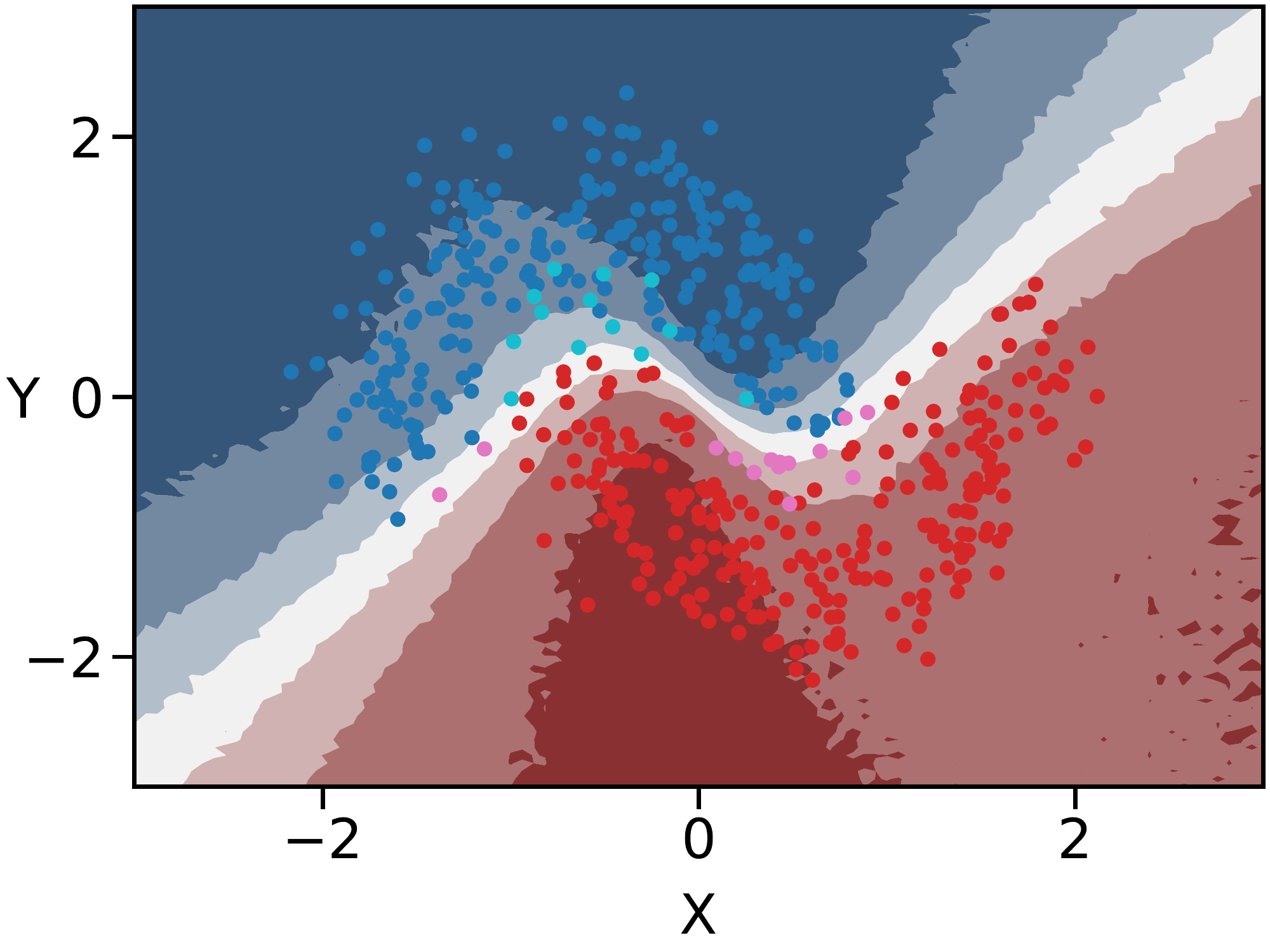}
        \caption{CIQP, $n=5$}
        \label{fig:bc_mean_ciqp_5_0}
    \end{subfigure}
    \begin{subfigure}[b]{0.24\linewidth}
        \includegraphics[width=\textwidth]{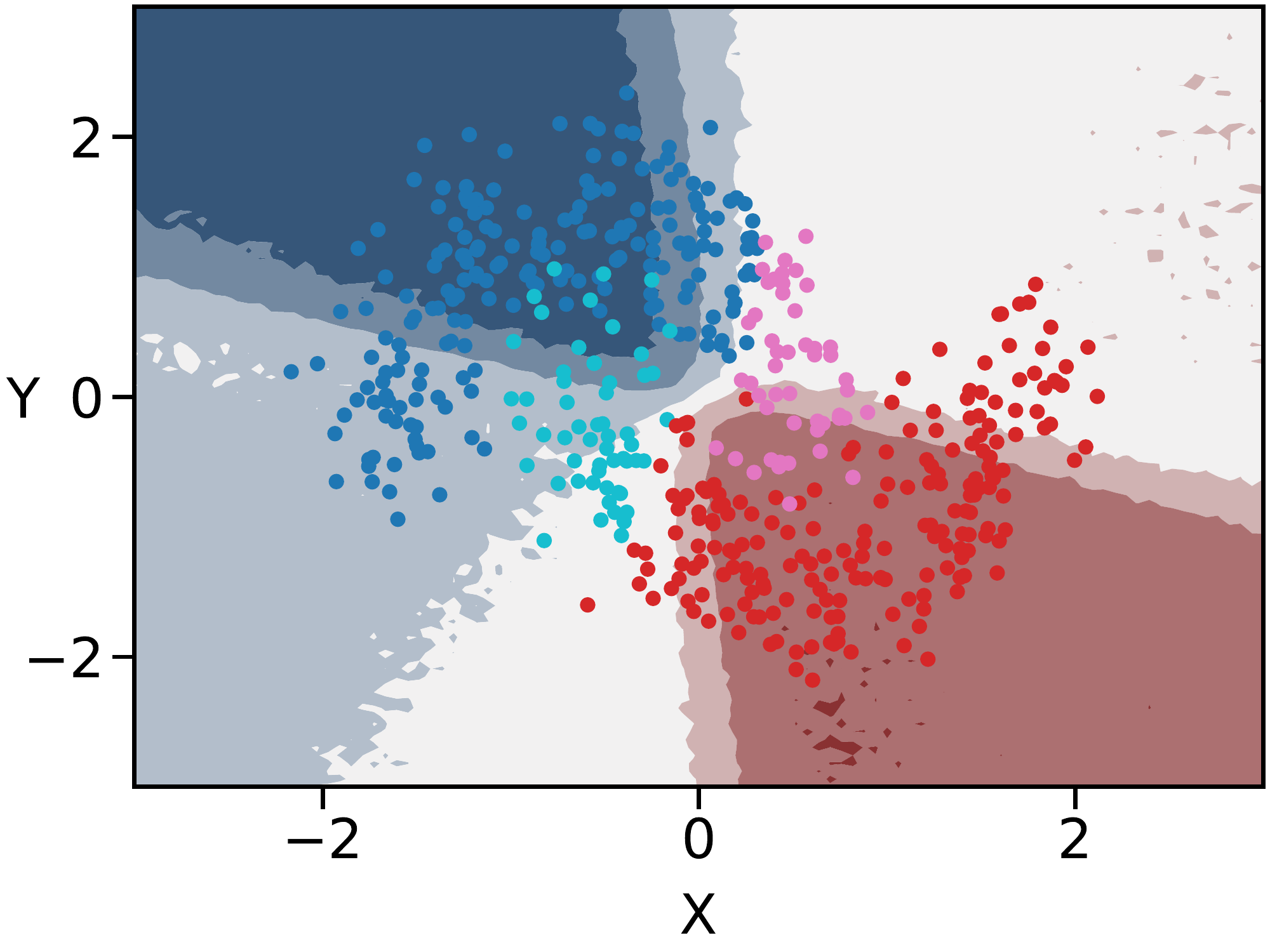}
        \caption{QICP, $n=5$}
        \label{fig:bc_mean_qicp_5_0}
    \end{subfigure}
    \begin{subfigure}[b]{0.24\linewidth}
        \includegraphics[width=\textwidth]{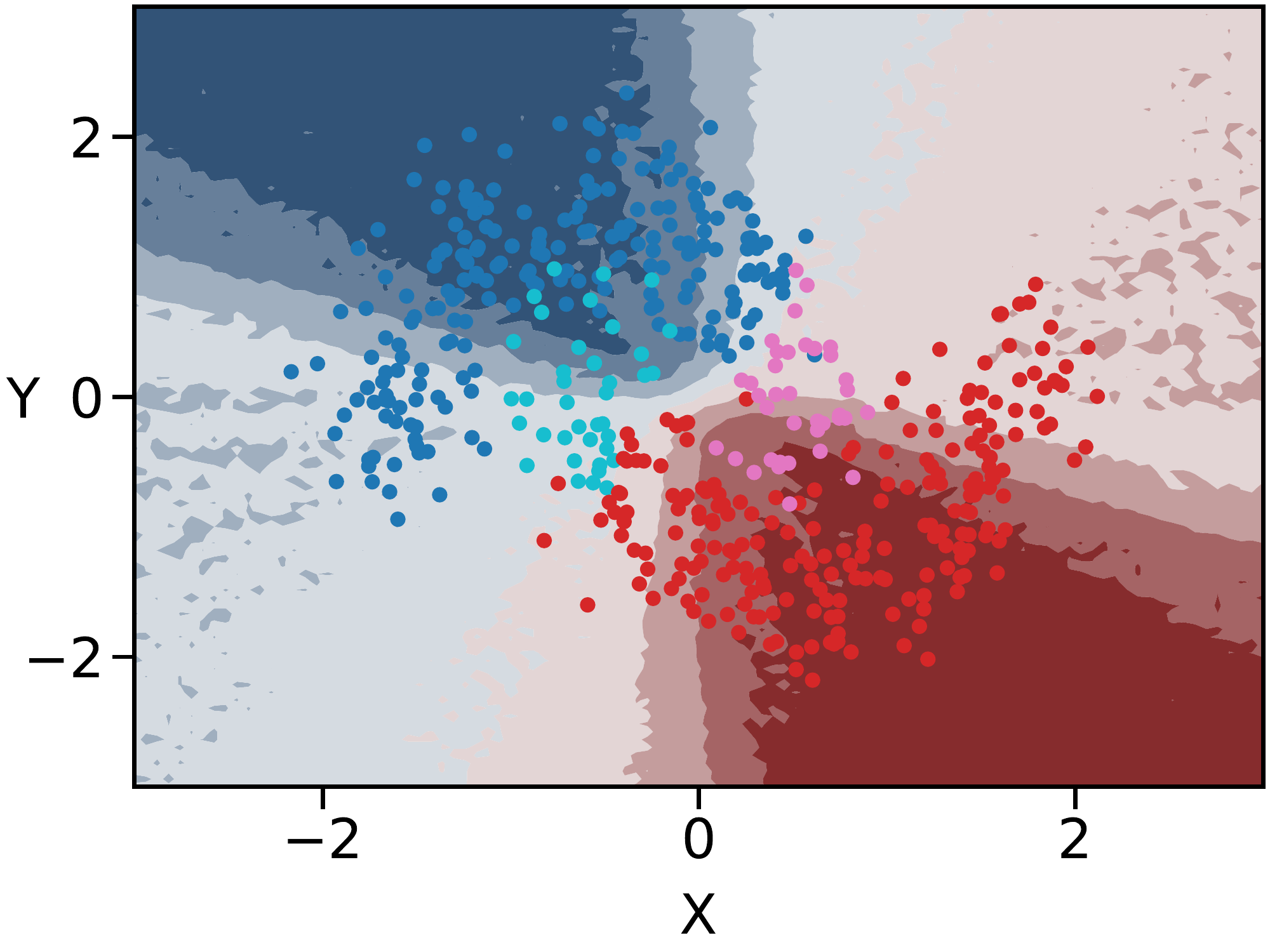}
        \caption{QIQP, $n=5$}
        \label{fig:bc_mean_qiqp_5_0}
    \end{subfigure}
    \vskip 1.0em
    \begin{subfigure}[b]{0.24\linewidth}
        \includegraphics[width=\textwidth]{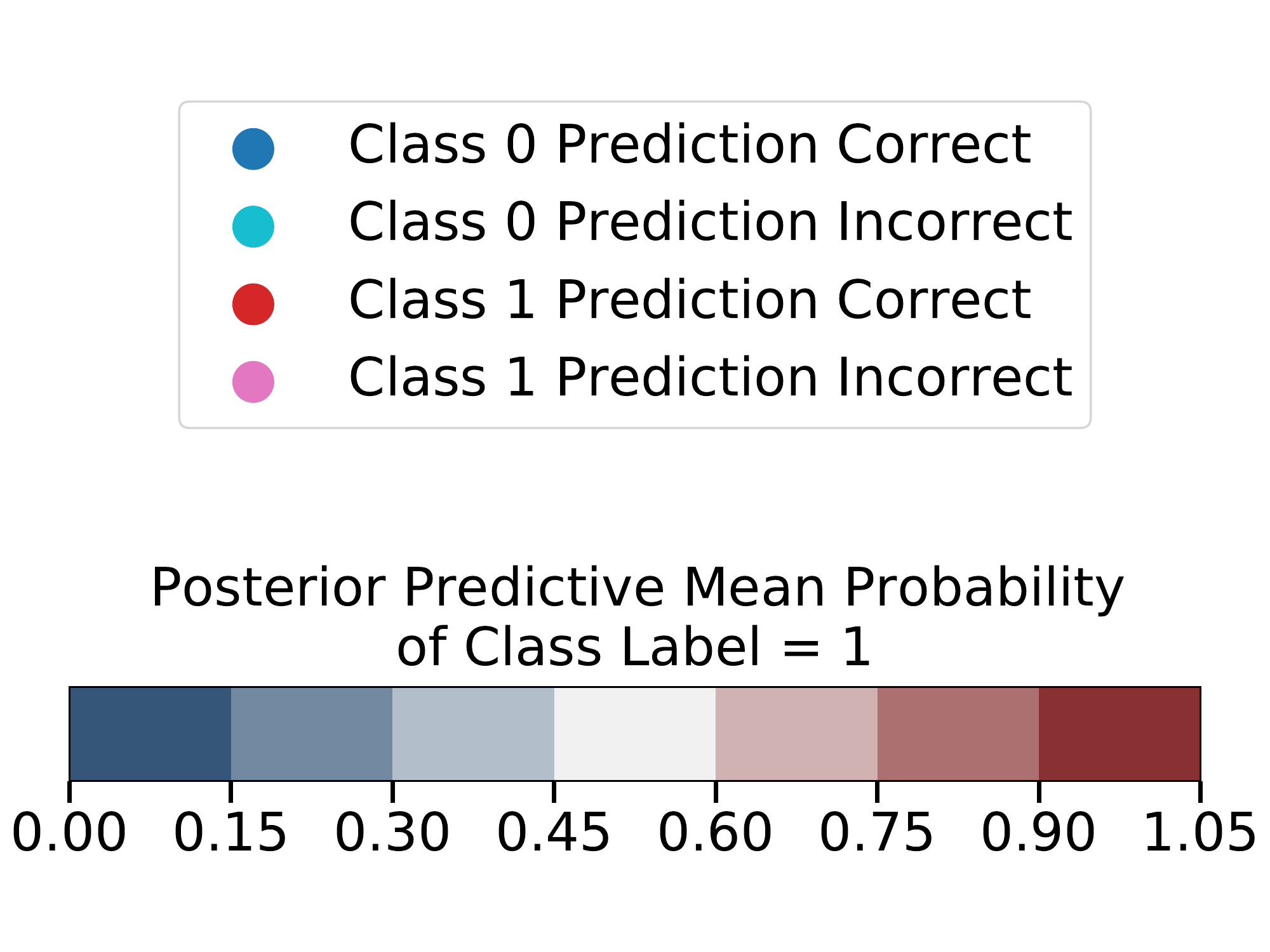}
        \caption{Legend}
    \end{subfigure}
    \begin{subfigure}[b]{0.24\linewidth}
        \includegraphics[width=\textwidth]{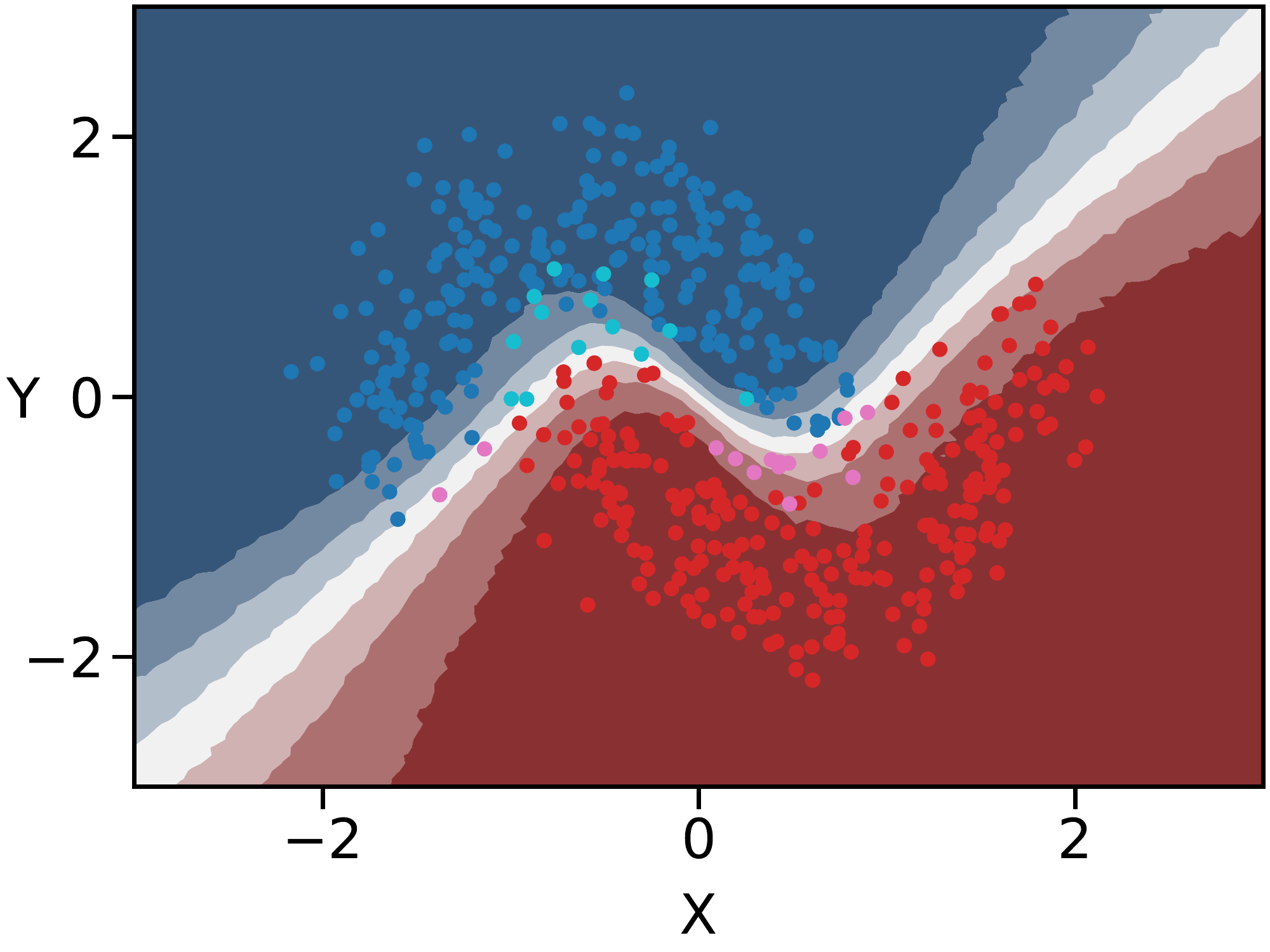}
        \caption{CIQP, $n=10$}
        \label{fig:bc_mean_ciqp_10_0}
    \end{subfigure}
    \begin{subfigure}[b]{0.24\linewidth}
        \includegraphics[width=\textwidth]{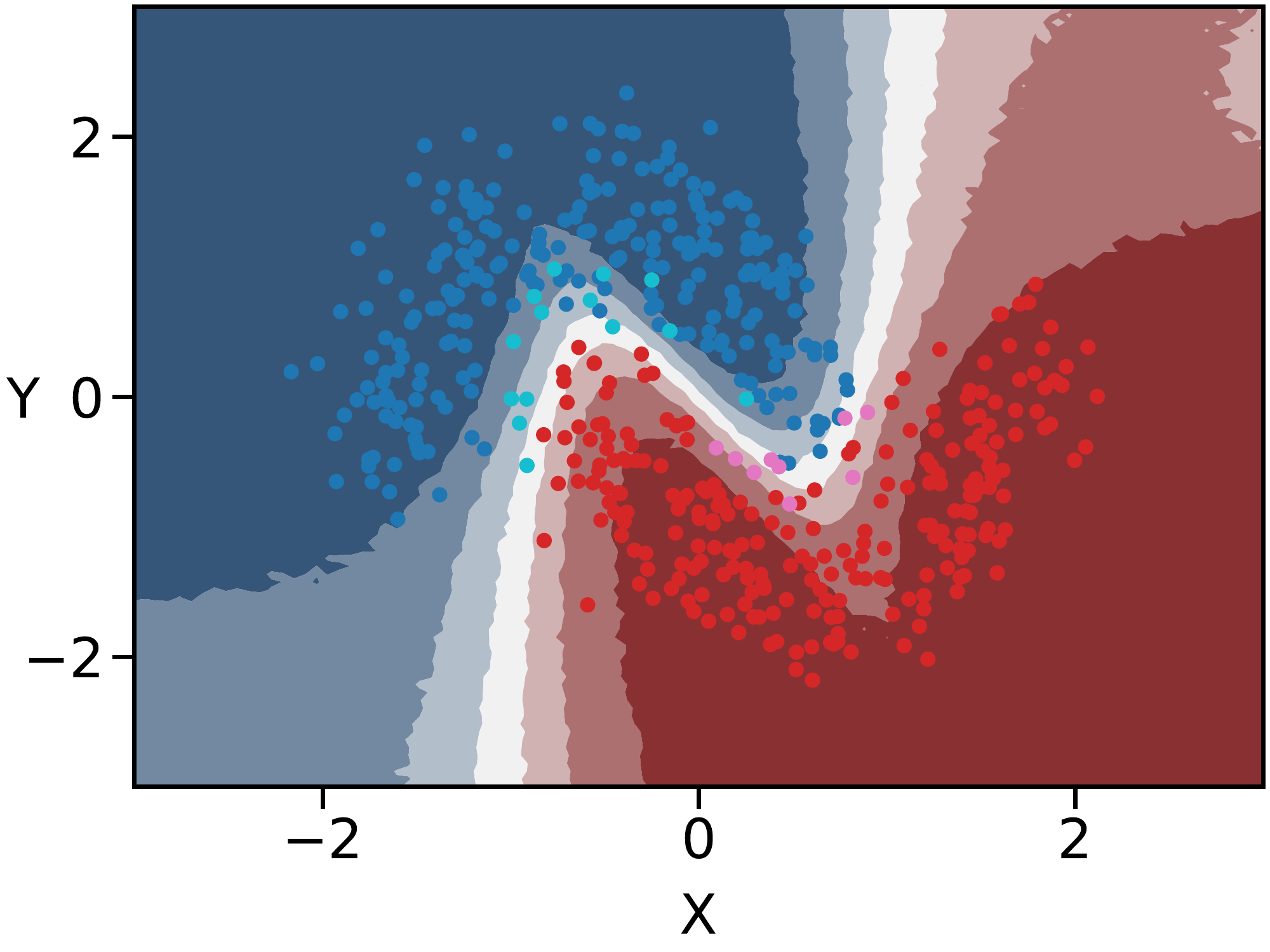}
        \caption{QICP, $n=10$}
        \label{fig:bc_mean_qicp_10_0}
    \end{subfigure}
    \begin{subfigure}[b]{0.24\linewidth}
        \includegraphics[width=\textwidth]{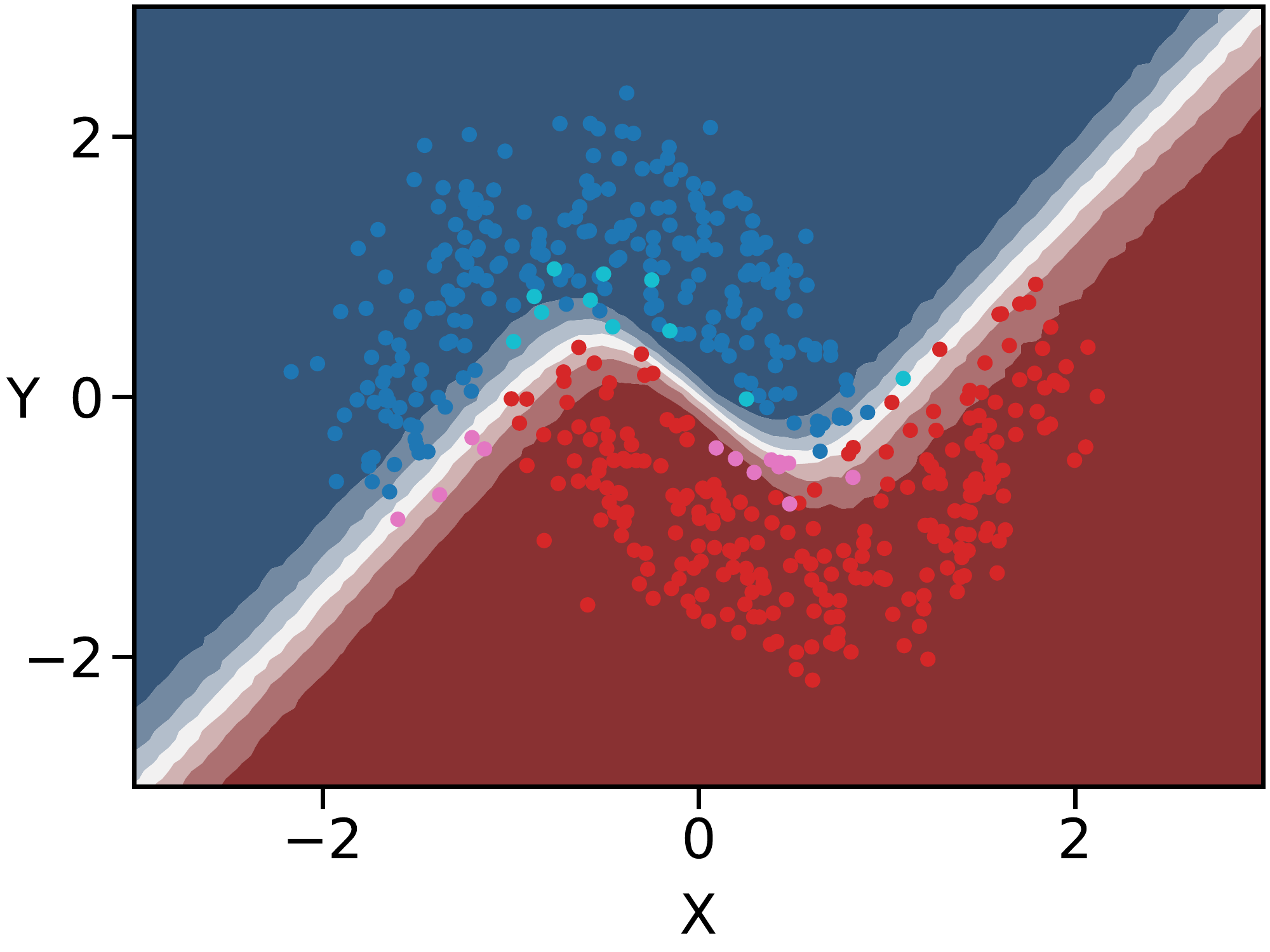}
        \caption{QIQP, $n=10$}
        \label{fig:bc_mean_qiqp_10_0}
    \end{subfigure}
    
\caption{Binary Classification with \ac{bnn}: \emph{C} and \emph{Q} stand for \emph{Classical} and \emph{Quantum} respectively. \emph{I} and \emph{P} stand for \emph{Inference} and \emph{Prediction}. The Figure shows the expected increase in accuracy for higher qubit numbers $n$.}
\label{fig:bc_mean}
\end{center}
\vskip -0.2in
\end{figure}

\section{Results}

We provide results for a linear regression task and a binary classification task. The \ac{bnn} we use in both tasks has two hidden layers with five neurons each. The results are obtained using a simulation of the \ac{ipe} algorithm on a classical computer. We vary the number of qubits ($n$) used in the phase estimation algorithm for the \ac{ipe} procedure. We expect a higher accuracy on the inner product estimate for a larger number of qubits.

For both tasks, we compare the fully classical algorithm (i.e., classical inference and classical prediction, \emph{CICP}), classical inference with quantum prediction (\emph{CIQP}), quantum inference with classical prediction (\emph{QICP}), and quantum inference with quantum prediction (\emph{QIQP}).
While the QIQP setting promises the largest speedups, the other settings can also be interesting in certain applications.
For instance, CIQP could be used when a predictive model is only trained once (offline), but then used for prediction repeatedly in real-time (online).

\begin{figure}[t]
\vskip 0.2in
\begin{center}
    \begin{subfigure}[b]{0.33\linewidth}
        \includegraphics[width=\textwidth]{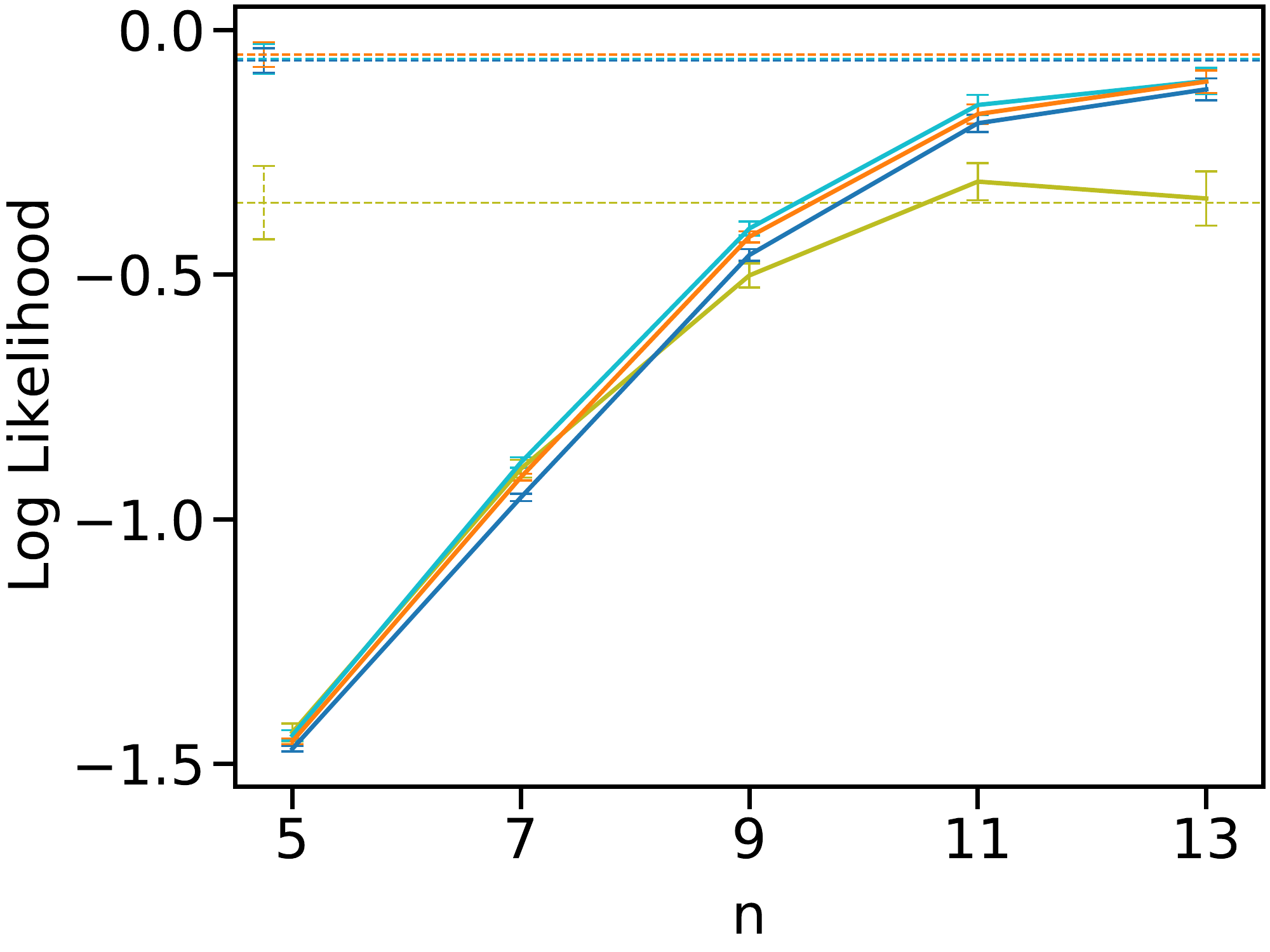}
        \caption{Boston}
        \label{fig:uci_boston}
    \end{subfigure}
    \begin{subfigure}[b]{0.33\linewidth}
        \includegraphics[width=\textwidth]{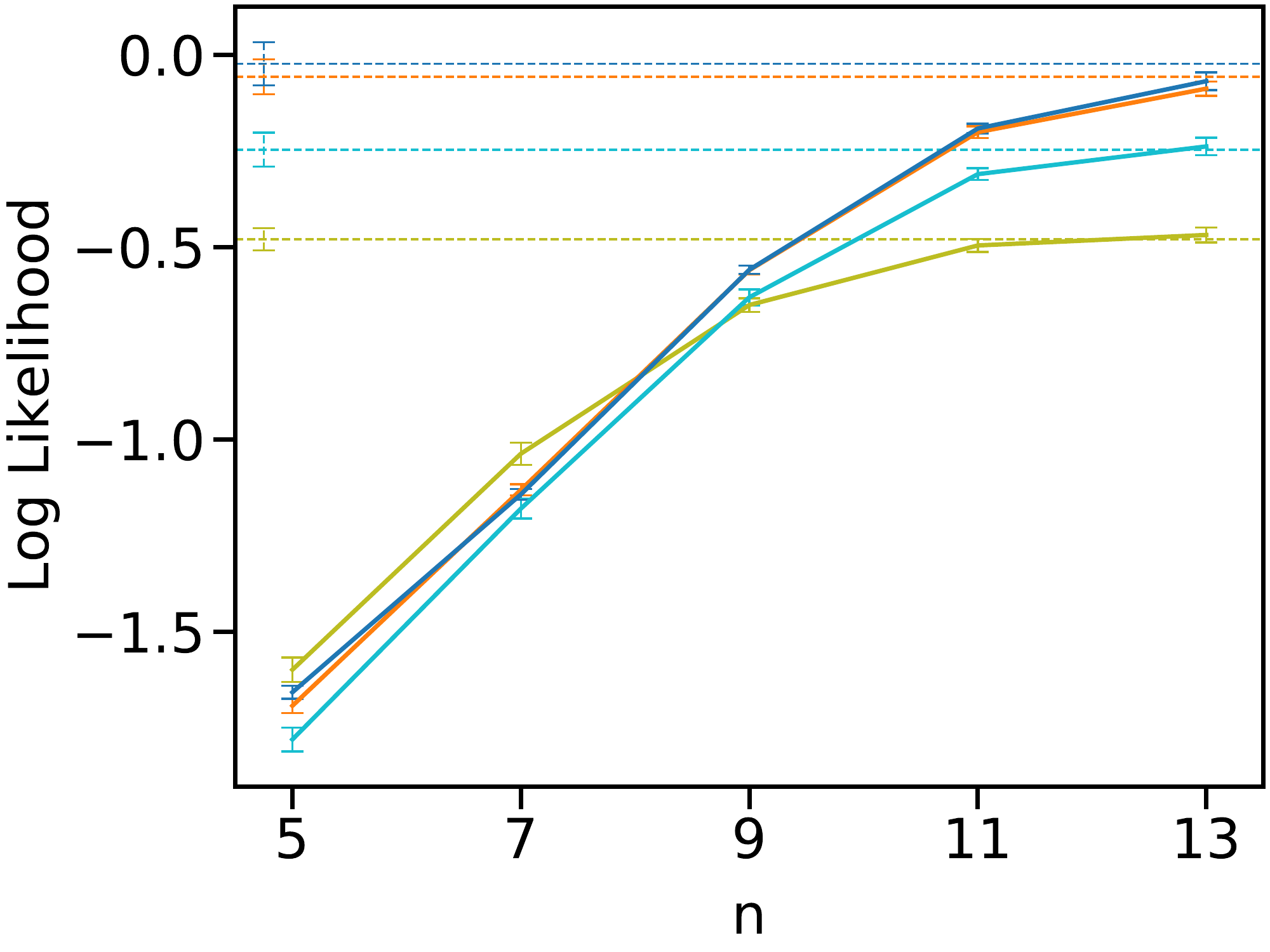}
        \caption{Concrete}
        \label{fig:uci_concrete}
    \end{subfigure}
    \begin{subfigure}[b]{0.33\linewidth}
        \includegraphics[width=\textwidth]{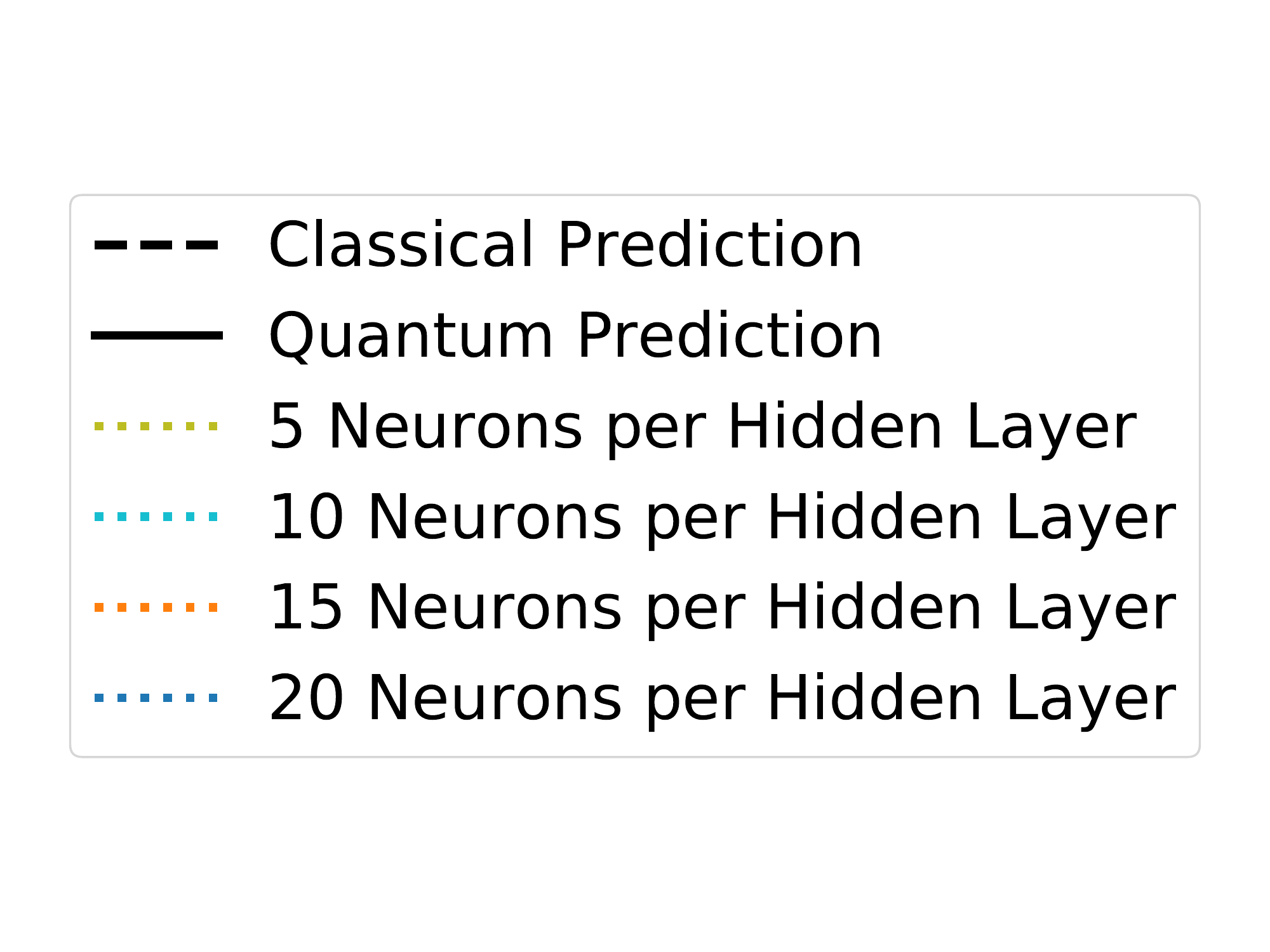}
        \caption{Legend}
        \label{fig:uci_legend}
    \end{subfigure}
    \vskip 1.0em
    \begin{subfigure}[b]{0.33\linewidth}
        \includegraphics[width=\textwidth]{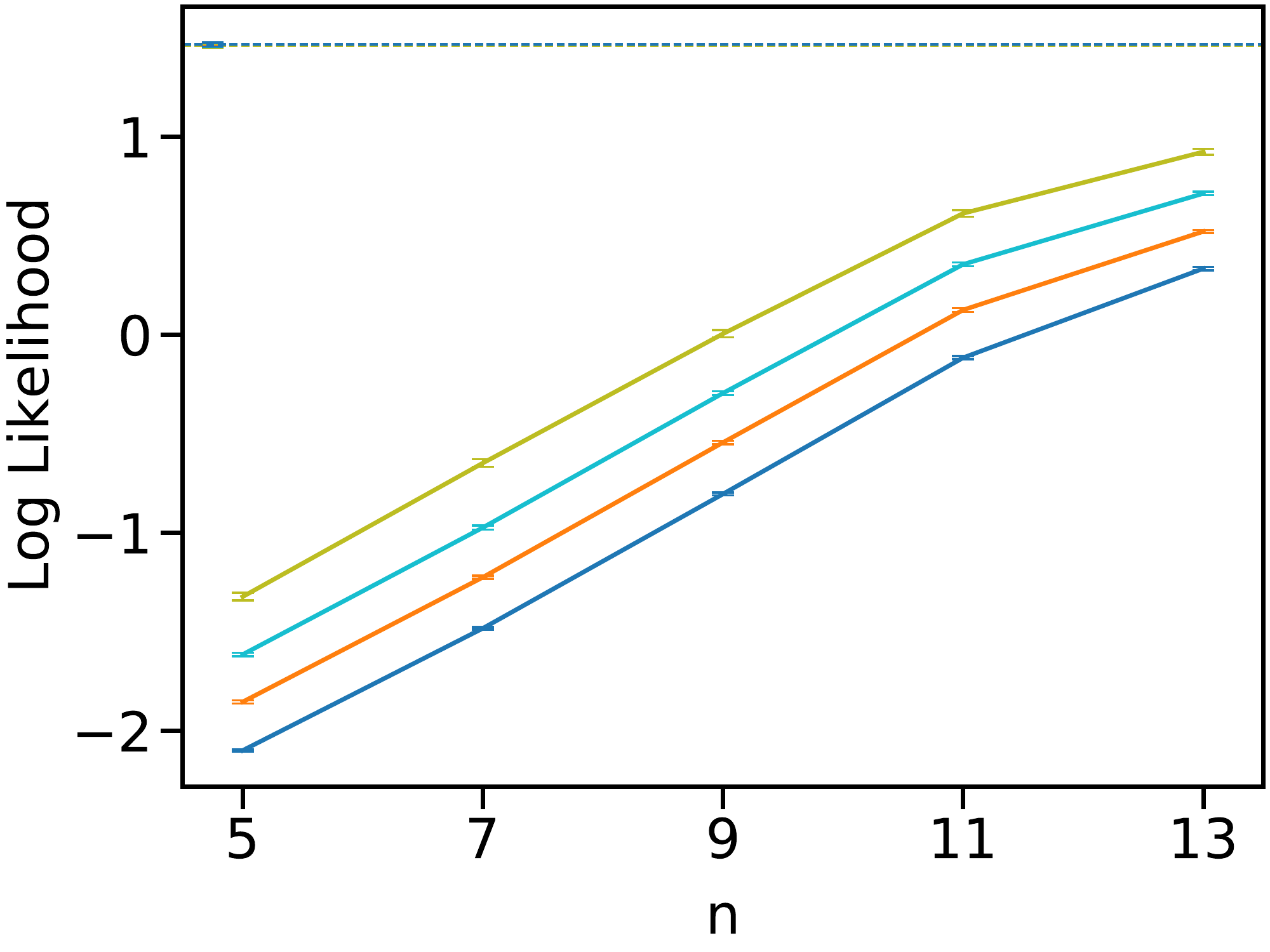}
        \caption{Energy}
        \label{fig:uci_energy}
    \end{subfigure}
    \begin{subfigure}[b]{0.33\linewidth}
        \includegraphics[width=\textwidth]{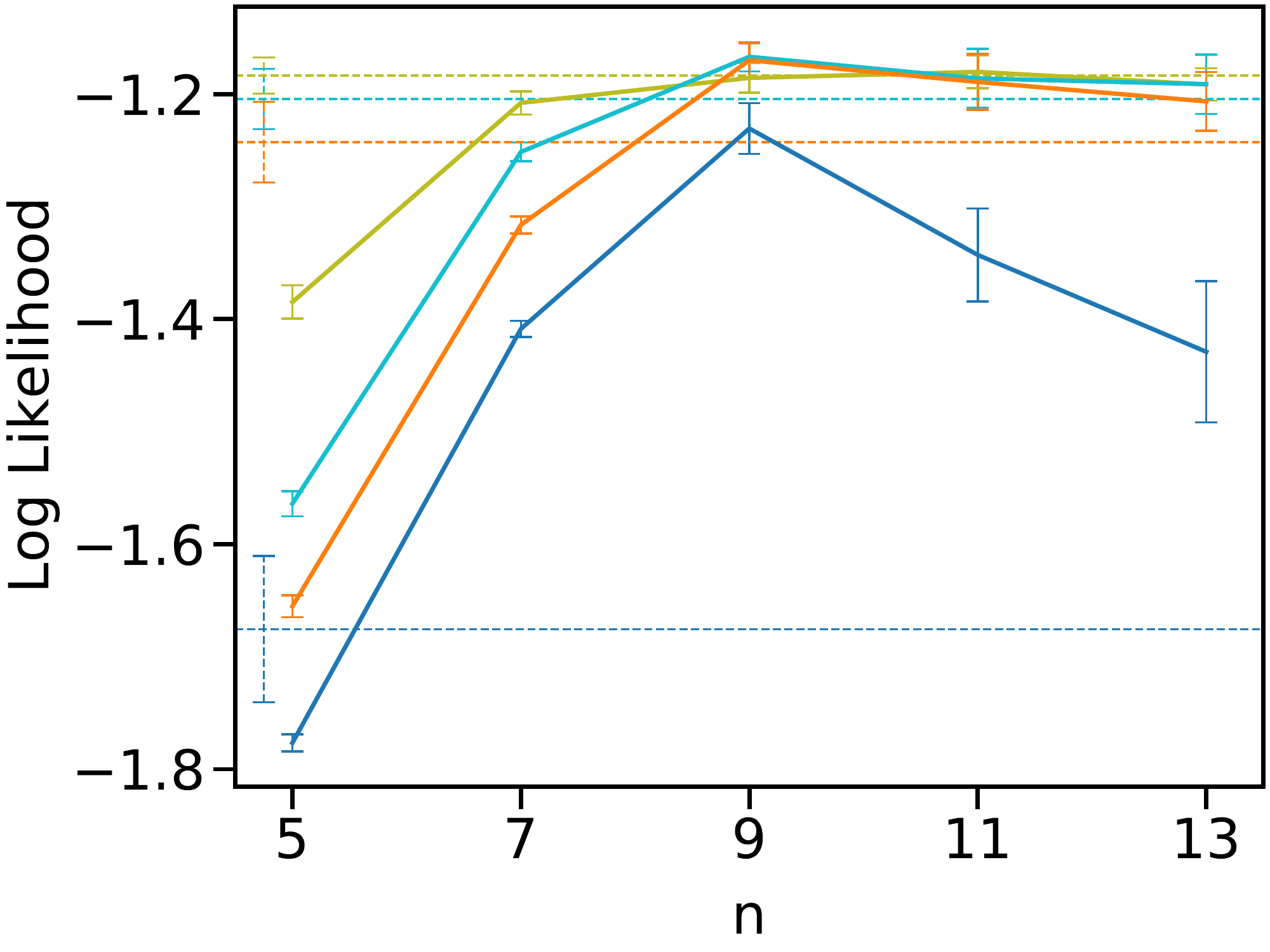}
        \caption{Wine}
        \label{fig:uci_wine}
    \end{subfigure}
    \hspace{15.5em}
    
\caption{UCI Data Regression with a \ac{bnn} Using Quantum \ac{ipe}. It shows the log-likelihood of classical prediction and quantum prediction for different qubit numbers $n$. Note that the y-axis is at different scales in the subplots.}
\label{fig:uci}
\end{center}
\end{figure}

\subsection{Linear Regression}

We see in \cref{fig:lr} that our expectation of greater precision for a higher qubit number $n$ holds. For $n=10$ qubits, the results are already comparable to the completely classical case. We also notice that quantum prediction seems to be more resilient to fewer qubits than quantum inference.
Specifically, if we compare the subfigures \cref{fig:lr_ciqp_10_0} and \cref{fig:lr_qicp_10_0}, the difference becomes apparent.
Further results are shown in \cref{fig:lr_add} in the Appendix.

In \cref{fig:lr_low_rank} in the Appendix, we see the difference between low-rank initialization (rank 3) and full-rank initialization (rank 5). We observe that the low-rank initialization still delivers acceptable results for 10 qubits when using quantum prediction. We also see that quantum inference is more susceptible to low-rank initialization. Increasing the qubit number likely makes the low-rank initialization almost equivalent to full-rank initialization because higher qubit number simulations approach the classical results. When looking at the completely classical simulations in \cref{fig:lr_cicp_0_fr} and \cref{fig:lr_cicp_0_lr}, we notice that low- and full-rank initialization are comparable.

\subsection{Binary Classification}

The binary classification in \cref{fig:bc_mean} confirms the results of the linear regression task. Again, there is greater precision with higher qubit numbers and quantum prediction is more resilient to fewer qubits than quantum inference.
These observations also qualitatively hold for the predictive uncertainties on this task, as shown in \cref{fig:bc_std} in the Appendix. 
Further results can be found \cref{fig:bc_mean_add} and \cref{fig:bc_std_add}.

Both \cref{fig:bc_mean_low_rank} and \cref{fig:bc_std_low_rank} in the Appendix confirm the observations regarding low-rank initialization. Here too, the low-rank initialization seems to provide sufficient results, especially if one would increase the number of qubits for the phase estimation.

\subsection{UCI Datasets}

In \cref{fig:uci}, we see the result of simulating the linear regression tasks in the four UCI datasets on our version of a quantum \ac{bnn}. The datasets are split 20 times into different training and prediction sets. The mean of the log-likelihood and the standard error are then used to create the plots. We simulate the linear regression for five different qubit numbers and 2000 weight samples of the posterior.

Both the Boston dataset in \cref{fig:uci_boston} and the Concrete dataset in \cref{fig:uci_concrete} show us expected results. The precision of the classical predictions increases with the number of hidden neurons. With the Boston dataset, the model stops improving after 10 hidden neurons, while the Concrete dataset shows improvements even after 20 hidden neurons per layer. As expected, the accuracy of the model increases with the number of qubits until it almost reaches classical prediction capabilities with 13 qubits.

For the Energy dataset in \cref{fig:uci_energy}, the behavior seems unexpected. The quantum prediction with the smallest number of neurons performs best. A possible explanation for this behavior is that the accuracy in the classical predictions does not increase with a larger number of hidden layer neurons. This suggests that the larger models do not capture more features of the data, while still incurring more inner products. The larger number of inner product calculations might lead to a larger overall error of these models when using quantum \ac{ipe}.

The Wine dataset in \cref{fig:uci_wine} seems to punish larger models even while using classical prediction. Thus it does not seem surprising that the quantum prediction also performs worse with the larger models.

\section{Related Work}
\label{sec:related_work}

\paragraph{Bayesian neural networks.}
Bayesian neural networks \citep{mackay1992practical, neal1992bayesian} have gained popularity recently \citep{wenzel2020good, fortuin2021bnnpriors, fortuin2021bayesian}.
They provide many benefits compared to their non-Bayesian counterparts, including calibrated uncertainties \citep{ovadia2019can}, principled inclusion of prior knowledge \citep{fortuin2021priors}, and automatic model selection \citep{immer2021scalable}.
While there has been work on \ac{bnn} inference using variational methods \citep{hernandez2015probabilistic, blundell2015weight, swiatkowski2020k, dusenberry2020efficient}, Laplace approximation \citep{daxberger2020expressive, immer2021improving}, and particle-based methods \citep{wang2018function, hu2019applying, ciosek2019conservative, dangelo2021stein, dangelo2021repulsive}, the gold-standard inference methods are still \ac{mcmc} methods \citep{izmailov2021bayesian}, such as SGLD \citep{welling2011bayesian}, GG-MC \citep{garriga2021exact}, HMC \citep{neal2011mcmc}, and NUTS \citep{hoffman2014no}.
We use NUTS sampling, but our algorithm is readily extensible to other inference settings.

\paragraph{Quantum machine learning.}
Quantum machine learning has gained a lot of interest in recent years, providing hope to enhance machine learning on a fault-tolerant universal quantum computer. It builds on several fundamental algorithms such as linear system solving \citep{HHL}, optimization \citep{kerenidis2020_gradient_descent}, recommendation systems \citep{kerenidis_recommendation_system}, dimensionality reduction \citep{Lloyd_PCA_quantum}, and many more. The quantum algorithms for neural networks by \citet{allcock2020quantum} and \citet{QCNN} were inspired by \citet{kerenidis2018qmeans}, who defined a fast quantum inner product estimation algorithm. We also build on this algorithm in our work. 
Recently, quantum Bayesian machine learning such as Gaussian Processes \citep{zhao2019bayesian} or Boltzmann machines \citep{wu2019bayesian} have also been explored. 
Moreover, Quantum Monte Carlo algorithms, such as the ones by \citet{montanaro2015quantum}, could be an interesting direction for use in \acp{bnn} in future work. 

\section{Conclusion}
\label{sec:conclusion}

We have shown that quantum deep learning techniques can be fruitfully combined with Bayesian neural networks.
In contrast to the standard point estimation setting, when sampling from Bayesian posteriors, one can achieve high fidelity of the samples even without repeating the quantum computations, already at small numbers of qubits.
The promised speedups of the quantum algorithms can thus be realized to their full extent in the Bayesian learning setting.
In future work, it will be exciting to extend these studies to more realistic prediction tasks and potentially speed up the inference even further through the use of quantum \ac{mcmc} techniques.

\bibliographystyle{plainnat}
\bibliography{references}

\begin{thebibliography}{38}
\providecommand{\natexlab}[1]{#1}
\providecommand{\url}[1]{\texttt{#1}}
\expandafter\ifx\csname urlstyle\endcsname\relax
  \providecommand{\doi}[1]{doi: #1}\else
  \providecommand{\doi}{doi: \begingroup \urlstyle{rm}\Url}\fi

\bibitem[Allcock et~al.(2020)Allcock, Hsieh, Kerenidis, and
  Zhang]{allcock2020quantum}
Jonathan Allcock, Chang-Yu Hsieh, Iordanis Kerenidis, and Shengyu Zhang.
\newblock Quantum algorithms for feedforward neural networks.
\newblock \emph{ACM Transactions on Quantum Computing}, 1\penalty0
  (1):\penalty0 1--24, 2020.

\bibitem[Blundell et~al.(2015)Blundell, Cornebise, Kavukcuoglu, and
  Wierstra]{blundell2015weight}
Charles Blundell, Julien Cornebise, Koray Kavukcuoglu, and Daan Wierstra.
\newblock Weight uncertainty in neural networks.
\newblock \emph{arXiv preprint arXiv:1505.05424}, 2015.

\bibitem[Brassard et~al.(2002)Brassard, Hoyer, Mosca, and
  Tapp]{brassard2002quantum}
Gilles Brassard, Peter Hoyer, Michele Mosca, and Alain Tapp.
\newblock Quantum amplitude amplification and estimation.
\newblock \emph{Contemporary Mathematics}, 305:\penalty0 53--74, 2002.

\bibitem[Cerezo et~al.(2020)Cerezo, Arrasmith, Babbush, Benjamin, Endo, Fujii,
  McClean, Mitarai, Yuan, Cincio, et~al.]{cerezo2020variational}
Marco Cerezo, Andrew Arrasmith, Ryan Babbush, Simon~C Benjamin, Suguru Endo,
  Keisuke Fujii, Jarrod~R McClean, Kosuke Mitarai, Xiao Yuan, Lukasz Cincio,
  et~al.
\newblock Variational quantum algorithms.
\newblock \emph{arXiv preprint arXiv:2012.09265}, 2020.

\bibitem[Ciosek et~al.(2019)Ciosek, Fortuin, Tomioka, Hofmann, and
  Turner]{ciosek2019conservative}
Kamil Ciosek, Vincent Fortuin, Ryota Tomioka, Katja Hofmann, and Richard
  Turner.
\newblock Conservative uncertainty estimation by fitting prior networks.
\newblock In \emph{International Conference on Learning Representations}, 2019.

\bibitem[D'Angelo and Fortuin(2021)]{dangelo2021repulsive}
Francesco D'Angelo and Vincent Fortuin.
\newblock Repulsive deep ensembles are bayesian.
\newblock \emph{arXiv preprint arXiv:2106.11642}, 2021.

\bibitem[D'Angelo et~al.(2021)D'Angelo, Fortuin, and Wenzel]{dangelo2021stein}
Francesco D'Angelo, Vincent Fortuin, and Florian Wenzel.
\newblock On stein variational neural network ensembles.
\newblock \emph{arXiv preprint arXiv:2106.10760}, 2021.

\bibitem[Daxberger et~al.(2020)Daxberger, Nalisnick, Allingham, Antor{\'a}n,
  and Hern{\'a}ndez-Lobato]{daxberger2020expressive}
Erik Daxberger, Eric Nalisnick, James~Urquhart Allingham, Javier Antor{\'a}n,
  and Jos{\'e}~Miguel Hern{\'a}ndez-Lobato.
\newblock Expressive yet tractable {B}ayesian deep learning via subnetwork
  inference.
\newblock \emph{arXiv preprint arXiv:2010.14689}, 2020.

\bibitem[Dusenberry et~al.(2020)Dusenberry, Jerfel, Wen, Ma, Snoek, Heller,
  Lakshminarayanan, and Tran]{dusenberry2020efficient}
Michael~W Dusenberry, Ghassen Jerfel, Yeming Wen, Yi-an Ma, Jasper Snoek,
  Katherine Heller, Balaji Lakshminarayanan, and Dustin Tran.
\newblock Efficient and scalable {B}ayesian neural nets with rank-1 factors.
\newblock \emph{arXiv preprint arXiv:2005.07186}, 2020.

\bibitem[Fortuin(2021)]{fortuin2021priors}
Vincent Fortuin.
\newblock Priors in bayesian deep learning: A review.
\newblock \emph{arXiv preprint arXiv:2105.06868}, 2021.

\bibitem[Fortuin et~al.(2021{\natexlab{a}})Fortuin, Garriga-Alonso, van~der
  Wilk, and Aitchison]{fortuin2021bnnpriors}
Vincent Fortuin, Adri{\`a} Garriga-Alonso, Mark van~der Wilk, and Laurence
  Aitchison.
\newblock Bnnpriors: A library for bayesian neural network inference with
  different prior distributions.
\newblock \emph{Software Impacts}, page 100079, 2021{\natexlab{a}}.

\bibitem[Fortuin et~al.(2021{\natexlab{b}})Fortuin, Garriga-Alonso, Wenzel,
  R{\"a}tsch, Turner, van~der Wilk, and Aitchison]{fortuin2021bayesian}
Vincent Fortuin, Adri{\`a} Garriga-Alonso, Florian Wenzel, Gunnar R{\"a}tsch,
  Richard Turner, Mark van~der Wilk, and Laurence Aitchison.
\newblock Bayesian neural network priors revisited.
\newblock \emph{arXiv preprint arXiv:2102.06571}, 2021{\natexlab{b}}.

\bibitem[Garriga-Alonso and Fortuin(2021)]{garriga2021exact}
Adri{\`a} Garriga-Alonso and Vincent Fortuin.
\newblock Exact langevin dynamics with stochastic gradients.
\newblock \emph{arXiv preprint arXiv:2102.01691}, 2021.

\bibitem[Harrow et~al.(2009)Harrow, Hassidim, and Lloyd]{HHL}
Aram~W Harrow, Avinatan Hassidim, and Seth Lloyd.
\newblock Quantum algorithm for linear systems of equations.
\newblock \emph{Physical review letters}, 103\penalty0 (15):\penalty0 150502,
  2009.

\bibitem[Hern{\'a}ndez-Lobato and Adams(2015)]{hernandez2015probabilistic}
Jos{\'e}~Miguel Hern{\'a}ndez-Lobato and Ryan Adams.
\newblock Probabilistic backpropagation for scalable learning of bayesian
  neural networks.
\newblock In \emph{International Conference on Machine Learning}, pages
  1861--1869. PMLR, 2015.

\bibitem[Hoffman and Gelman(2014)]{hoffman2014no}
Matthew~D Hoffman and Andrew Gelman.
\newblock The no-u-turn sampler: adaptively setting path lengths in hamiltonian
  monte carlo.
\newblock \emph{J. Mach. Learn. Res.}, 15\penalty0 (1):\penalty0 1593--1623,
  2014.

\bibitem[Hu et~al.(2019)Hu, Szerlip, Karaletsos, and Singh]{hu2019applying}
Xinyu Hu, Paul Szerlip, Theofanis Karaletsos, and Rohit Singh.
\newblock Applying svgd to bayesian neural networks for cyclical time-series
  prediction and inference.
\newblock \emph{arXiv preprint arXiv:1901.05906}, 2019.

\bibitem[Immer et~al.(2021{\natexlab{a}})Immer, Bauer, Fortuin, R{\"a}tsch, and
  Khan]{immer2021scalable}
Alexander Immer, Matthias Bauer, Vincent Fortuin, Gunnar R{\"a}tsch, and
  Mohammad~Emtiyaz Khan.
\newblock Scalable marginal likelihood estimation for model selection in deep
  learning.
\newblock \emph{arXiv preprint arXiv:2104.04975}, 2021{\natexlab{a}}.

\bibitem[Immer et~al.(2021{\natexlab{b}})Immer, Korzepa, and
  Bauer]{immer2021improving}
Alexander Immer, Maciej Korzepa, and Matthias Bauer.
\newblock Improving predictions of bayesian neural nets via local
  linearization.
\newblock In \emph{International Conference on Artificial Intelligence and
  Statistics}, pages 703--711. PMLR, 2021{\natexlab{b}}.

\bibitem[Izmailov et~al.(2021)Izmailov, Vikram, Hoffman, and
  Wilson]{izmailov2021bayesian}
Pavel Izmailov, Sharad Vikram, Matthew~D Hoffman, and Andrew~Gordon Wilson.
\newblock What are bayesian neural network posteriors really like?
\newblock \emph{arXiv preprint arXiv:2104.14421}, 2021.

\bibitem[Kerenidis and Prakash(2016)]{kerenidis_recommendation_system}
Iordanis Kerenidis and Anupam Prakash.
\newblock Quantum recommendation systems.
\newblock \emph{arXiv preprint arXiv:1603.08675}, 2016.

\bibitem[Kerenidis and Prakash(2020)]{kerenidis2020_gradient_descent}
Iordanis Kerenidis and Anupam Prakash.
\newblock Quantum gradient descent for linear systems and least squares.
\newblock \emph{Physical Review A}, 101\penalty0 (2):\penalty0 022316, 2020.

\bibitem[Kerenidis et~al.(2018)Kerenidis, Landman, Luongo, and
  Prakash]{kerenidis2018qmeans}
Iordanis Kerenidis, Jonas Landman, Alessandro Luongo, and Anupam Prakash.
\newblock q-means: A quantum algorithm for unsupervised machine learning, 2018.

\bibitem[Kerenidis et~al.(2020)Kerenidis, Landman, and Prakash]{QCNN}
Iordanis Kerenidis, Jonas Landman, and Anupam Prakash.
\newblock Quantum algorithms for deep convolutional neural networks.
\newblock In \emph{Proceedings of the International Conference on Learning
  Representations ({ICLR})}, 2020.

\bibitem[Lloyd et~al.(2014)Lloyd, Mohseni, and Rebentrost]{Lloyd_PCA_quantum}
Seth Lloyd, Masoud Mohseni, and Patrick Rebentrost.
\newblock Quantum principal component analysis.
\newblock \emph{Nature Physics}, 10\penalty0 (9):\penalty0 631--633, 2014.

\bibitem[MacKay(1992)]{mackay1992practical}
David~J.C. MacKay.
\newblock A practical {B}ayesian framework for backpropagation networks.
\newblock \emph{Neural computation}, 4\penalty0 (3):\penalty0 448--472, 1992.

\bibitem[Montanaro(2015)]{montanaro2015quantum}
Ashley Montanaro.
\newblock Quantum speedup of monte carlo methods.
\newblock \emph{Proceedings of the Royal Society A: Mathematical, Physical and
  Engineering Sciences}, 471\penalty0 (2181):\penalty0 20150301, 2015.

\bibitem[Neal(1992)]{neal1992bayesian}
Radford~M. Neal.
\newblock {B}ayesian training of backpropagation networks by the {H}ybrid
  {M}onte {C}arlo method.
\newblock Technical report, University of Toronto, 1992.

\bibitem[Neal et~al.(2011)]{neal2011mcmc}
Radford~M Neal et~al.
\newblock Mcmc using hamiltonian dynamics.
\newblock \emph{Handbook of markov chain monte carlo}, 2\penalty0
  (11):\penalty0 2, 2011.

\bibitem[Nielsen and Chuang(2002)]{NC02}
Michael~A Nielsen and Isaac Chuang.
\newblock Quantum computation and quantum information, 2002.

\bibitem[Ovadia et~al.(2019)Ovadia, Fertig, Ren, Nado, Sculley, Nowozin,
  Dillon, Lakshminarayanan, and Snoek]{ovadia2019can}
Yaniv Ovadia, Emily Fertig, Jie Ren, Zachary Nado, David Sculley, Sebastian
  Nowozin, Joshua Dillon, Balaji Lakshminarayanan, and Jasper Snoek.
\newblock Can you trust your model's uncertainty? {E}valuating predictive
  uncertainty under dataset shift.
\newblock In \emph{Advances in Neural Information Processing Systems}, pages
  13991--14002, 2019.

\bibitem[Preskill(2018)]{NISQpreskill}
John Preskill.
\newblock Quantum computing in the nisq era and beyond.
\newblock \emph{Quantum}, 2:\penalty0 79, 2018.

\bibitem[Swiatkowski et~al.(2020)Swiatkowski, Roth, Veeling, Tran, Dillon,
  Mandt, Snoek, Salimans, Jenatton, and Nowozin]{swiatkowski2020k}
Jakub Swiatkowski, Kevin Roth, Bastiaan~S Veeling, Linh Tran, Joshua~V Dillon,
  Stephan Mandt, Jasper Snoek, Tim Salimans, Rodolphe Jenatton, and Sebastian
  Nowozin.
\newblock The k-tied normal distribution: A compact parameterization of
  {G}aussian mean field posteriors in {B}ayesian neural networks.
\newblock \emph{arXiv preprint arXiv:2002.02655}, 2020.

\bibitem[Wang et~al.(2018)Wang, Ren, Zhu, and Zhang]{wang2018function}
Ziyu Wang, Tongzheng Ren, Jun Zhu, and Bo~Zhang.
\newblock Function space particle optimization for bayesian neural networks.
\newblock In \emph{International Conference on Learning Representations}, 2018.

\bibitem[Welling and Teh(2011)]{welling2011bayesian}
Max Welling and Yee~Whye Teh.
\newblock Bayesian learning via stochastic gradient langevin dynamics.
\newblock In \emph{Proceedings of the 28th International Conference on
  International Conference on Machine Learning}, pages 681--688, 2011.

\bibitem[Wenzel et~al.(2020)Wenzel, Roth, Veeling, {\'S}wi{\,{a}}tkowski, Tran,
  Mandt, Snoek, Salimans, Jenatton, and Nowozin]{wenzel2020good}
Florian Wenzel, Kevin Roth, Bastiaan~S Veeling, Jakub {\'S}wi{\,{a}}tkowski,
  Linh Tran, Stephan Mandt, Jasper Snoek, Tim Salimans, Rodolphe Jenatton, and
  Sebastian Nowozin.
\newblock How good is the {B}ayes posterior in deep neural networks really?
\newblock In \emph{International Conference on Machine Learning}, 2020.

\bibitem[Wu et~al.(2019)Wu, Yu, Qin, Wen, and Gao]{wu2019bayesian}
Yusen Wu, Chao-hua Yu, Sujuan Qin, Qiaoyan Wen, and Fei Gao.
\newblock Bayesian machine learning for boltzmann machine in quantum-enhanced
  feature spaces.
\newblock \emph{arXiv preprint arXiv:1912.10857}, 2019.

\bibitem[Zhao et~al.(2019)Zhao, Pozas-Kerstjens, Rebentrost, and
  Wittek]{zhao2019bayesian}
Zhikuan Zhao, Alejandro Pozas-Kerstjens, Patrick Rebentrost, and Peter Wittek.
\newblock Bayesian deep learning on a quantum computer.
\newblock \emph{Quantum Machine Intelligence}, 1\penalty0 (1):\penalty0 41--51,
  2019.

\end{thebibliography}

\newpage

\onecolumn

\appendix
\counterwithin{figure}{section}
\counterwithin{table}{section}

\section{Source Code and Data}
\label{sec:source_code}
The source code and additional data from the simulations presented in this paper can be found on GitHub: \url{https://github.com/NoahBerner/quantum_bayesian_neural_networks}.

\section{Method details}

\subsection{\emph{R} Terms in Quantum Inference and Prediction Algorithm}

The \emph{R} terms appearing in the runtime of both quantum algorithms are a new phenomenon, not observed in classical algorithms. The variable $R$ in \cref{eq:quantum_inference_runtime} is defined as

\begin{equation}
\begin{split}
    R &= R_a + R_\delta + R_W,\\
    R_a &= \frac{1}{KN (\Omega - n_1)} \sum_{k, n} \sum_{\ell=2}^L \sum_{j=1}^{n_\ell} \|X^{[k,\ell,j]}\|_F \|a^{k, n, \ell-1}\|,\\
    R_\delta &= \frac{1}{KN (\Omega - n_1)} \sum_{k, n}\sum_{\ell=1}^{L-1} \sum_{j=1}^{n_\ell} \|\tilde{X}^{[k,\ell+1,j]}\|_F \|\delta^{k, n, \ell+1}\|,\\
    R_W &= \frac{1}{KN (\Omega - n_1)} \sum_{\ell=2}^L \sum_{j=1}^{n_\ell} \left(\frac{\|X^{[k,\ell,j]}\|_F}{\|W^{k,\ell}_j\|} + \frac{\|\tilde{X}^{[k,\ell,j]}\|_F}{\|(W^{k,\ell})_j^\top\|}\right),
\end{split}
\label{eq:R}
\end{equation}
where $n_\ell$ is the number of neurons in the $\ell$-th layer, the \ac{bnn} consists of $L$ layers, the weight matrix $W^{\ell}$ is associated between layers $\ell-1$ and $\ell$, $W^{k,\ell}_j$ is the $k$-th sample of the $j$-th row of the weight matrix $W^{\ell}$, $X^{[k,\ell,j]}$ is the implicitly stored version of $W^{k,\ell}_j$, $a^{k, n, \ell}$ is the output for the $n$-th datapoint of the $\ell$-th layer using the $k$-th weight sample,  $\delta^{k, n, \ell}$ is the output for the $n$-th datapoint of the $\ell$-th layer using the $k$-th weight sample in the backward pass and $\|\|_F$ is the Frobenius norm. 

The variable $R_e$ used in \cref{eq:quantum_prediction_runtime} is defined as

\begin{equation}
    R_e = \frac{1}{(\Omega-n_1)} \sum_{\ell = 2}^L \sum_{j = 1}^{n_\ell} \|W_j^\ell\| \|a^{\ell-1}\|.
\label{eq:R_e}
\end{equation}

As argued by \citet{allcock2020quantum}, both these values are expected to be small for practical parameter regimes, which we also expect to hold for \acp{bnn}.

\subsection{Jacobian-Vector Product of an Inner Product}
\label{sec:jvp}

The \ac{jvp} of an inner product contains itself two inner products between vectors:
\begin{equation}
    \nabla (v_i^\top v_j) \cdot (t_1, t_2) = v_j \cdot t_1 + v_i \cdot t_2,
\end{equation}
where $t_1$ and $t_2$ are the tangent vectors. In our simulation, we replace the exact calculation of these inner products with the estimate of the \ac{ipe} routine. This gives us an estimate of the \ac{jvp} (and thus of the gradient), instead of the true \ac{jvp} value. This allows for a faster runtime of the backpropagation algorithm.

\section{Preliminaries in Quantum Computing}\label{preliminariesquantum}
We present a succinct broad-audience quantum information background necessary for this work. See the book by \citet{NC02} for a detailed course.

\paragraph{Qubits:} In classical computing, a bit can be either 0 or 1. From a quantum information perspective, a quantum bit or \emph{qubit} can be in state $\ket{0}$ or $\ket{1}$. We use the \emph{braket} notation $\ket{\cdot}$ to specify the quantum nature of the bit. The qubits can be in superposition of both states $\alpha\ket{0}+\beta\ket{1}$ where $\alpha,\beta \in \mathbb{C}$ such that $|\alpha|^2 + |\beta|^2 = 1$. The coefficients $\alpha$ and $\beta$ are called \emph{amplitudes}.  The probabilities of observing either 0 or 1 when \emph{measuring} the qubit are linked to the amplitudes:
\begin{equation}
    p(0)=|\alpha|^2, \quad p(1)=|\beta|^2
\end{equation}

As quantum physics teaches us, any superposition is possible before the measurement, which gives special abilities in terms of computation. With $n$ qubits, $2^n$ possible binary combinations (e.g.  $\ket{01\cdots1001}$) can exist simultaneously, each with its own amplitude.

A $n$ qubits system can be represented as a normalized vector in a $2^n$ dimensional Hilbert space. A multi-qubit system is called a quantum \emph{register}. If $\ket{p}$ and $\ket{q}$ are two quantum states or quantum registers, the whole system can be represented as a tensor product $\ket{p}\otimes\ket{q}$, also written as $\ket{p}\ket{q}$ or $\ket{p,q}$.

\paragraph{Quantum Computation:} As logical gates in classical circuits, qubits or quantum registers are processed using quantum gates. A quantum gate is a \emph{unitary} mapping in the Hilbert space, preserving the unit norm of the quantum state vector. Therefore, a quantum gate acting on $n$ qubits is a matrix $U \in \mathbb{C}^{2^n}$ such that $UU^{\dagger}=U^{\dagger}U=I$, with $U^{\dagger}$ being the adjoint, or conjugate transpose, of $U$. 

Common single qubit gates include 
the Hadamard gate 
$\frac{1}{\sqrt{2}}\begin{pmatrix}
1 & 1 \\
1 & -1 \\
\end{pmatrix}$ that maps $\ket{0} \mapsto \frac{1}{\sqrt{2}}(\ket{0}+\ket{1})$ and $\ket{1} \mapsto \frac{1}{\sqrt{2}}(\ket{0}-\ket{1})$, creating a quantum superposition,
the NOT gate 
$\begin{pmatrix}
0 & 1 \\
1 & 0 \\
\end{pmatrix}$ 
that permutes $\ket{0}$ and $\ket{1}$, 
or $R_y$ rotation gate parametrized by an angle $\theta
$, given by
$\begin{pmatrix}
\cos(\theta/2) & -\sin(\theta/2) \\
\sin(\theta/2) & \cos(\theta/2) \\
\end{pmatrix}$.

Common two-qubits gates include 
the CNOT gate
$\begin{pmatrix}
1 & 0 & 0 & 0 \\
0 & 1 & 0 & 0 \\
0 & 0 & 0 & 1 \\
0 & 0 & 1 & 0 \\
\end{pmatrix}$
which is a NOT gate applied on the second qubit only if the first one is in state $\ket{1}$.

The main advantage of quantum gates is their ability to be applied to a superposition of inputs. Indeed, given a gate $U$ such that $U\ket{x} \mapsto \ket{f(x)}$, we can apply it to all possible combinations of $x$ at once $U(\frac{1}{C}\sum_{x}\ket{x}) \mapsto \frac{1}{C}\sum_{x}\ket{f(x)}$.

\section{Error Analysis of the Quantum Inner Product Estimation Algorithm}
\label{sec:ipe_error_analysis}

The \acl{ipe} routine takes as input two vectors as quantum states $\ket{v_i}$ and $\ket{v_j}$. It outputs an estimate of the inner product $\langle v_i | v_j \rangle$. The vectors are amplitude encoded, meaning that $\ket{v_i}$ is defined as 
\begin{equation}
    \ket{v_i} = \sum_{l=0}^d v_{i,l} \ket{l},
\end{equation}
where $v_{i,l}$ is the $l$-th element of the vector $v_i$ and for the dimension $d=2^n$ has to hold, where $n$ is the number of qubits of state $\ket{v_i}$. Note, that amplitude encoding enforces that the vectors $v_i$ and $v_j$ are normalized. Thus also for the \ac{ipe} the input vectors are normalized. Their norms are stored during the quantum inference and prediction algorithms to unnormalize the inner product estimate after the \ac{ipe} computation. This means that the error of the \ac{ipe} is also multiplied by the norms $\|v_i\|$ and $\|v_j\|$.

\ac{ipe} calculates the inner product estimate by preparing a state $\ket{\psi}$. $\ket{\psi}$ has an amplitude on one of the measurable basis states that is proportional to the inner product $\langle v_i | v_j \rangle$. It then uses amplitude estimation, which was introduced by \citet{brassard2002quantum}, as a subroutine to compute the value of this amplitude. Amplitude estimation uses phase estimation as a subroutine, so we will first look at phase estimation and cascade the error backward.

\subsection{Phase Estimation}
\label{sec:phase_estimation}

Phase estimation receives as input the following quantum state:

\begin{equation}
    \frac{1}{\sqrt{2^n}} \sum_{y=0}^{2^n-1} e^{2 \pi i \omega y} |y\rangle,
    \label{phase_estimation_input}
\end{equation}
where $n$ is the number of qubits. The desired output is a good estimate of the phase parameter $\omega$.

In general, the output of the phase estimation algorithm will be a superposition
\begin{equation}
    |\tilde{\omega}\rangle = \sum_x \alpha_x(\omega) |x\rangle
\end{equation}
of all possible integer states $|x\rangle$, where $x \in \{0,\dots,2^n-1\}$. We are interested in the amplitudes $|\alpha_x(\omega)|^2$ which define the output distribution of the phase estimation algorithm.

Let $b$ be the integer in the range 0 to $2^n - 1$ such that $\frac{b}{2^n} = 0.b_1 \dots b_n$ is the best $n$ bit approximation of $\omega$ which is less than $\omega$. Then, the difference $\delta \equiv \omega - \frac{b}{2^n}$ satisfies $0 \leq \omega \leq 2^{-n}$. Applying the phase estimation algorithm, also known as the inverse Quantum Fourier Transform, yields the state  

\begin{equation}
    \frac{1}{2^n} \sum_{k, l = 0}^{2^n-1} e^{\frac{-2 \pi i k l}{2^n}} e^{2 \pi i \omega k} |l\rangle
\end{equation}

The amplitude $\alpha_l$ of the state $|(b+l)(\text{mod}\ 2^n)\rangle$ is
\begin{equation}
\begin{split}
    \alpha_l &= \frac{1}{2^n} \sum_{k = 0}^{2^n-1} \left( e^{2 \pi i (\omega - (b+l)/2^n)} \right)^k\\
    &= \frac{1}{2^n} \left(\frac{1 - e^{2 \pi i (2^n \omega - (b+l))}}{1 - e^{2 \pi i (\omega - (b+l)/2^n)}}\right)\\
    &= \frac{1}{2^n} \left(\frac{1 - e^{2 \pi i (2^n \delta - l)}}{1 - e^{2 \pi i (\delta - l/2^n)}}\right)
\end{split}
\end{equation}
Here, the second equality stems from the closed-form formula for the geometric series. 

The probability to measure the integer $b+l\mod2^n$ is 
\begin{equation}
\begin{split}
    |\alpha_l|^2 &= \frac{1}{2^{2n}} \left|\left(\frac{1 - e^{2 \pi i (2^n \delta - l)}}{1 - e^{2 \pi i (\delta - l/2^n)}}\right)\right|^2\\
    &= \frac{1}{2^{2n}} \left| \frac{2 \sin(\pi (2^n \delta - l))}{2 \sin(\pi (\delta - l/2^n))} \right|^2\\
    &= \frac{1}{2^{2n}} \frac{\sin^2(\pi (2^n \delta - l))}{\sin^2(\pi (\delta - l/2^n))},
\end{split}
\label{eq:phase_estimation_distribution}
\end{equation}
where we used the fact that $|1 - e^{2ix}|^2 = 4 |\sin(x)|^2$. Note that the distribution in \cref{eq:phase_estimation_distribution} depends (through the variable $\delta$) on the phase $\omega$, which is the variable that the phase estimation algorithm is trying to estimate. This means that it will not be possible to predict the exact output distribution of the phase estimation algorithm, since that would require knowledge of $\omega$.

\subsection{Amplitude Estimation}
\label{sec:amplitude_estimation}

The amplitude of the state $\ket{\psi}$ given by the \ac{ipe} as an input to the amplitude estimation algorithm is

\begin{equation}
    a = \frac{1}{2}\left(\frac{\braket{v_i}{v_j}}{\norm{v_i}\norm{v_j}}+1\right)
\end{equation}

Phase estimation can obtain an estimate of $\omega$ for an amplitude of the form $\sin^2(\pi \omega)$. Thus the phase we try to estimate using phase estimation is 

\begin{equation}
    \overline{\omega} = \frac{1}{\pi}\arcsin(\sqrt{a})
\end{equation}

Using the results from \cref{sec:phase_estimation}, we know that the phase estimate will be

\begin{equation}
    \omega_l = \frac{(b+l) \mod 2^n}{2^n}, l \in \{0,1, \dots, {2^n-1}\},
\end{equation}
where $b = \text{argmin}_i |\frac{i}{2^n}-\overline{\omega}|, i \in \{0,1, \dots, {2^n}\}$, with a probability of

\begin{equation}
    p_l =  \frac{1}{2^{2n}} \frac{\sin^2( \pi  (2^n \delta - l))}{\sin^2( \pi (\delta - l / 2^n))}, \delta = \overline{\omega} - b/2^n.
\end{equation}

The output of the amplitude estimation will be 

\begin{equation}
    a_l = \sin^2(\pi \omega_l)
\end{equation}

and the final output of the \ac{ipe} algorithm is

\begin{equation}
    (2a_l-1) \|v_i\| \|v_j\|.
\end{equation}

as an estimate of the inner product $\braket{v_i}{v_j}$ between the vectors $v_i$ and $v_j$.

\section{Additional results}
\label{sec:add_results}

\begin{figure}[!b]
\vskip 0.2in
\begin{center}
    \begin{subfigure}[b]{0.24\linewidth}
        \includegraphics[width=\textwidth]{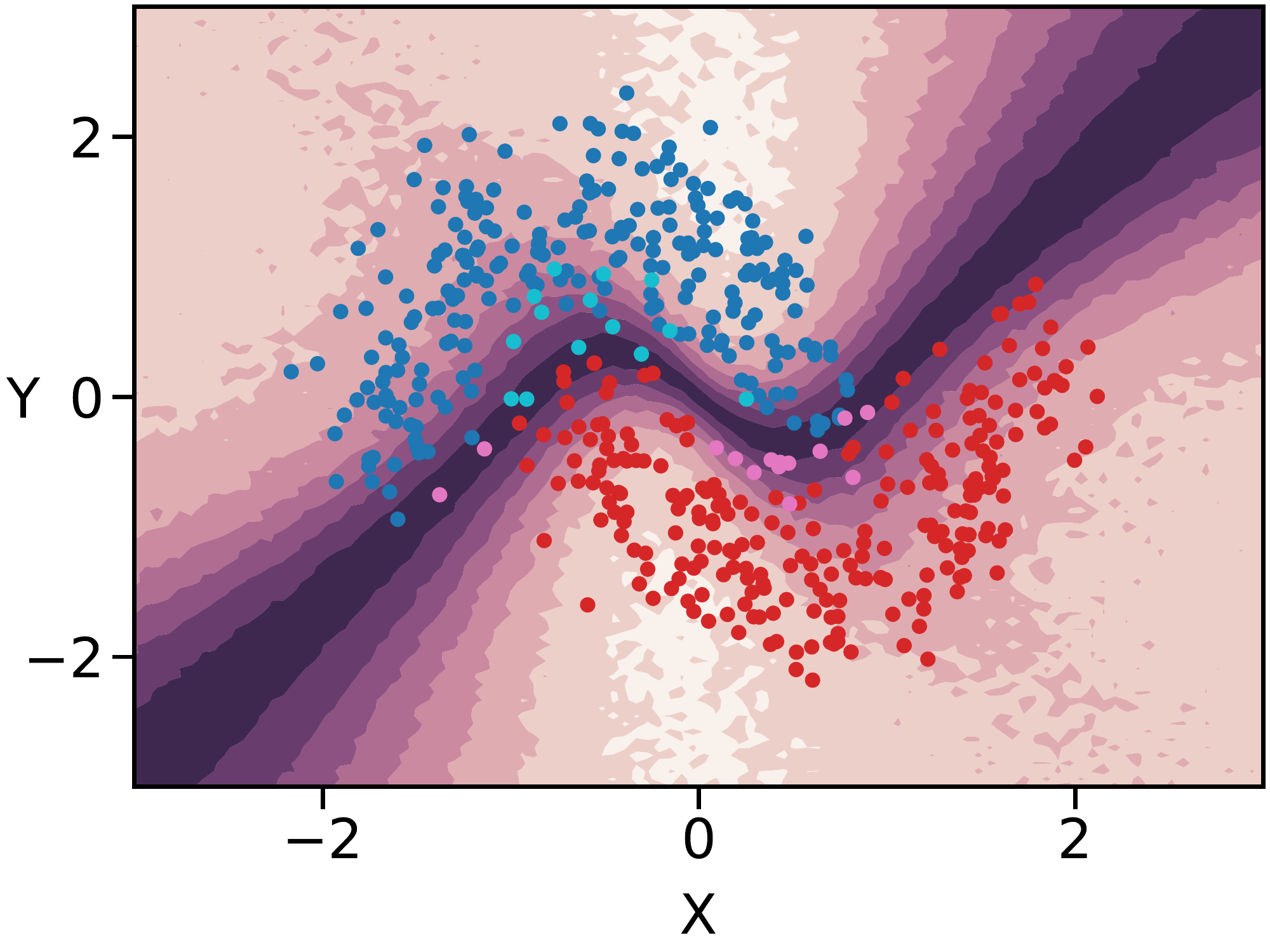}
        \caption{CICP (Reference)}
        \label{fig:bc_std_cicp_0}
    \end{subfigure}
    \begin{subfigure}[b]{0.24\linewidth}
        \includegraphics[width=\textwidth]{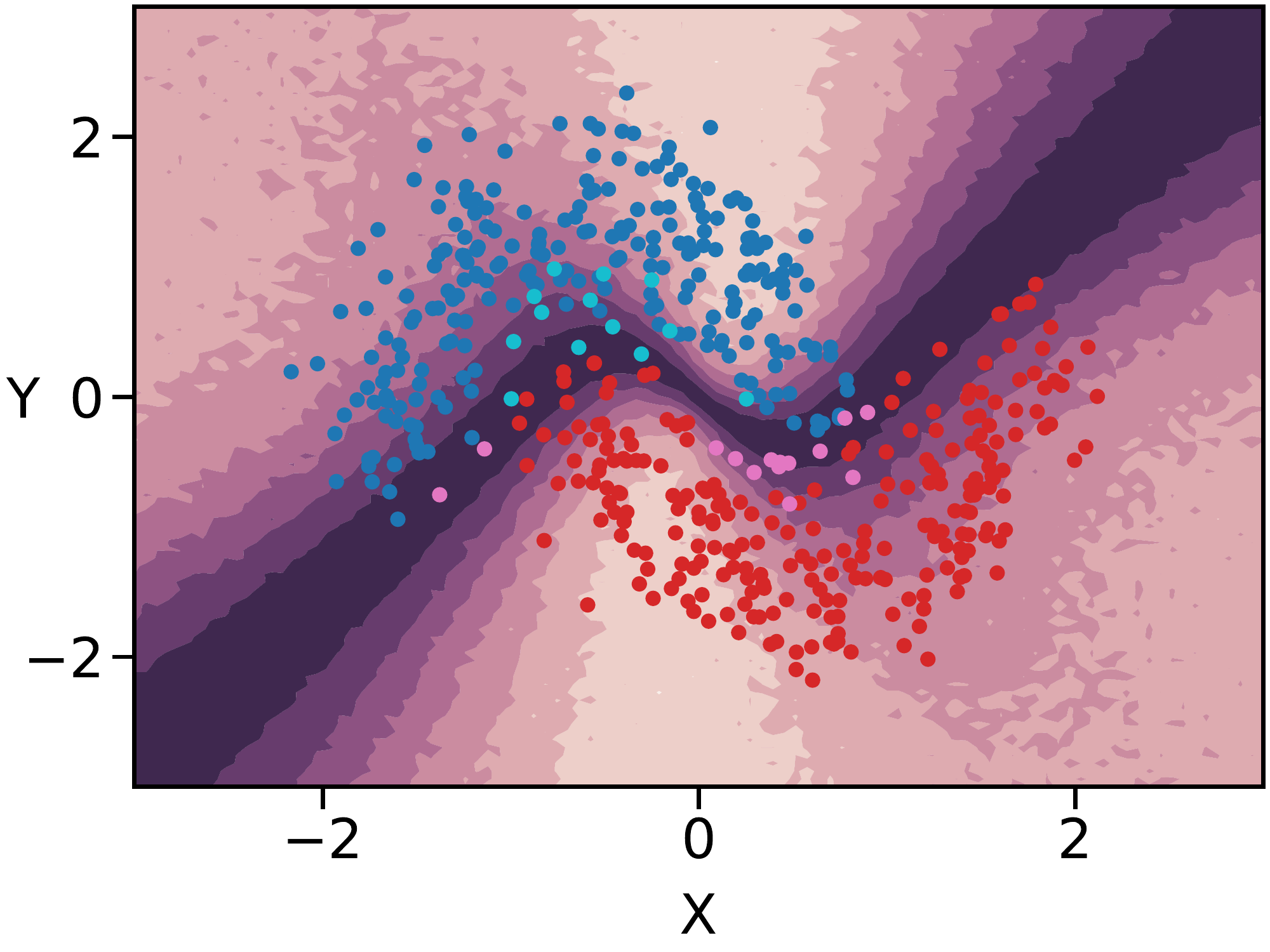}
        \caption{CIQP, $n=5$}
        \label{fig:bc_std_ciqp_5_0}
    \end{subfigure}
    \begin{subfigure}[b]{0.24\linewidth}
        \includegraphics[width=\textwidth]{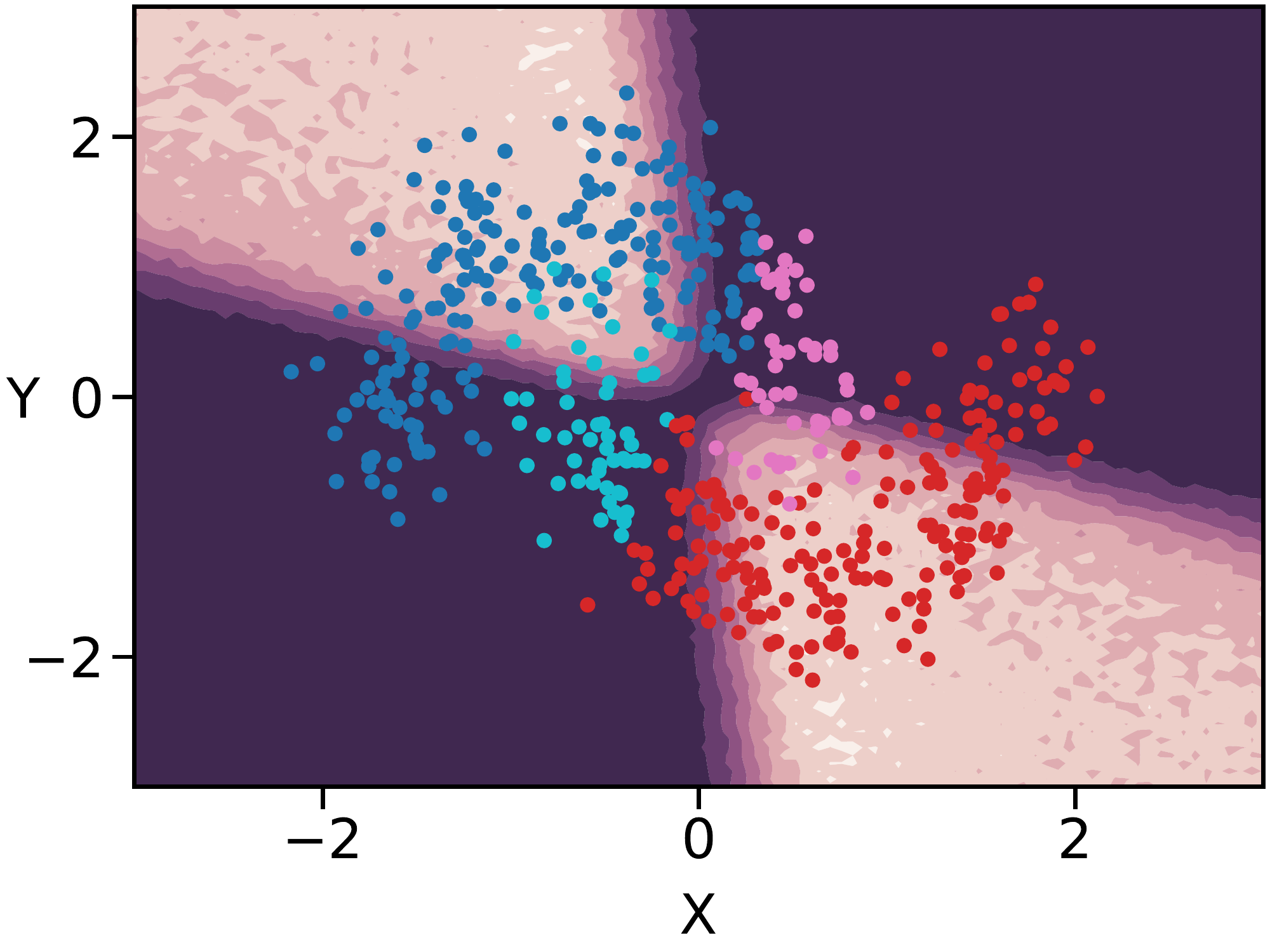}
        \caption{QICP, $n=5$}
        \label{fig:bc_std_qicp_5_0}
    \end{subfigure}
    \begin{subfigure}[b]{0.24\linewidth}
        \includegraphics[width=\textwidth]{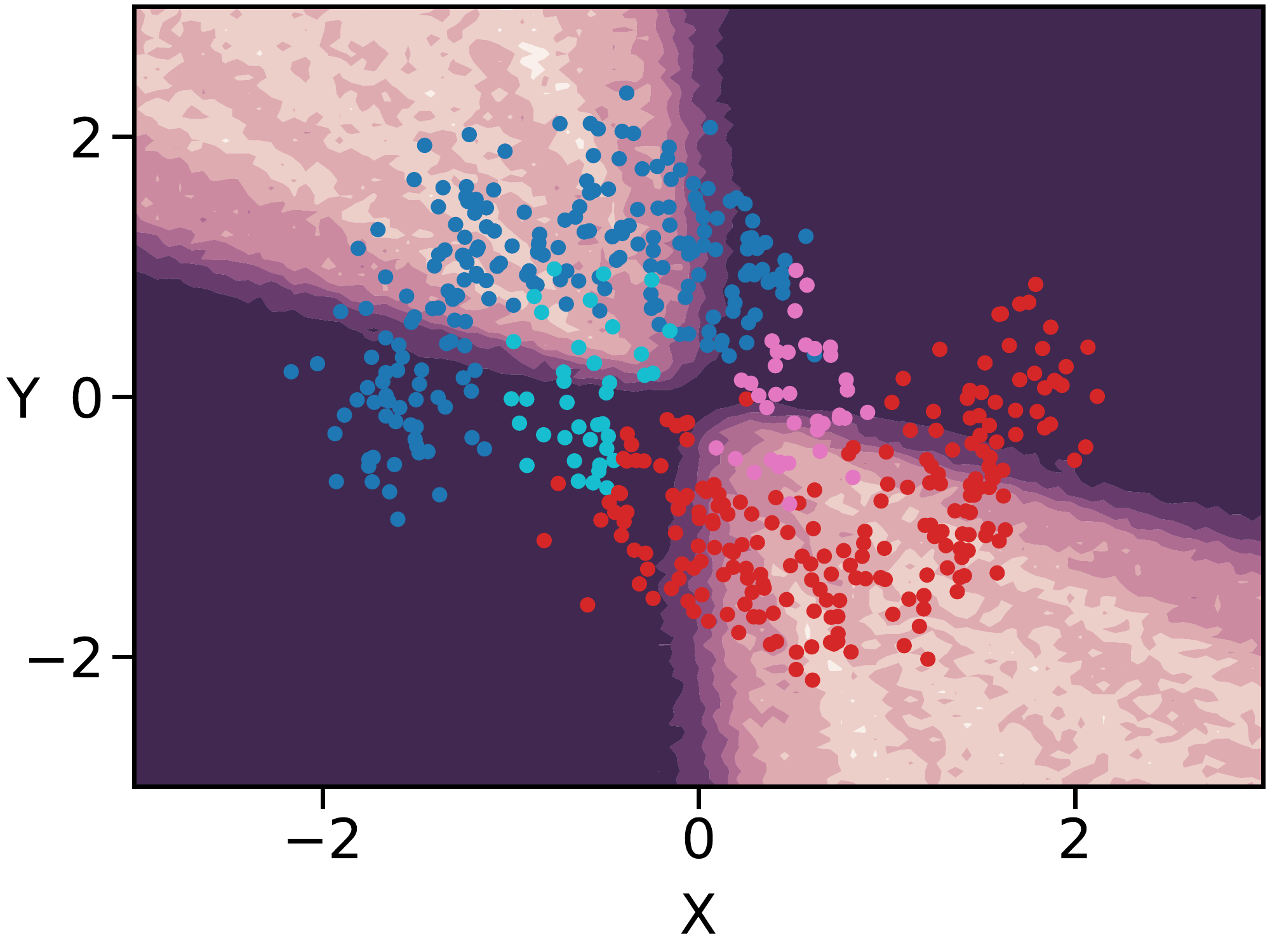}
        \caption{QIQP, $n=5$}
        \label{fig:bc_std_qiqp_5_0}
    \end{subfigure}
    \vskip 1.0em
    \begin{subfigure}[b]{0.24\linewidth}
        \includegraphics[width=\textwidth]{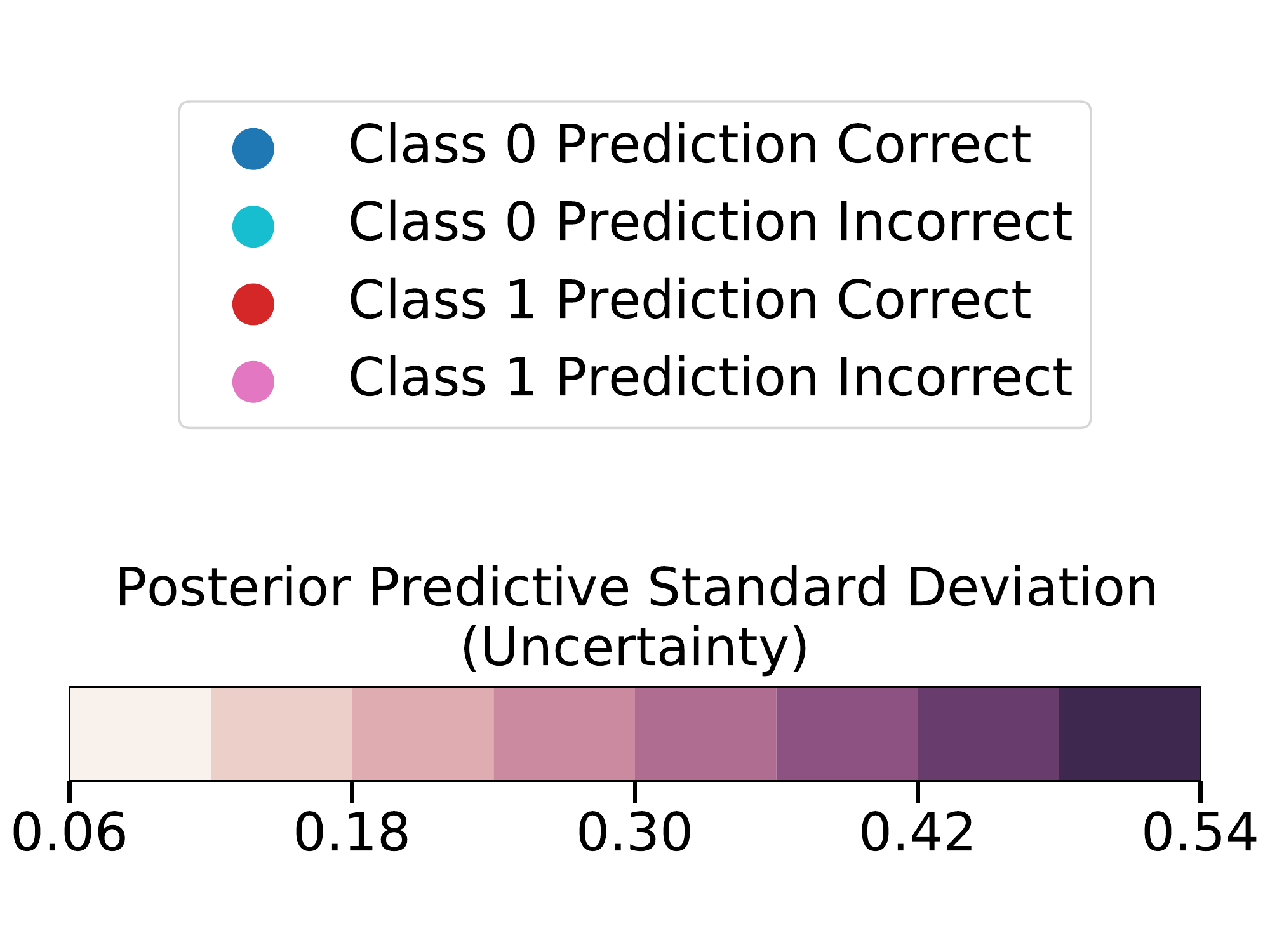}
        \caption{Legend}
    \end{subfigure}
    \begin{subfigure}[b]{0.24\linewidth}
        \includegraphics[width=\textwidth]{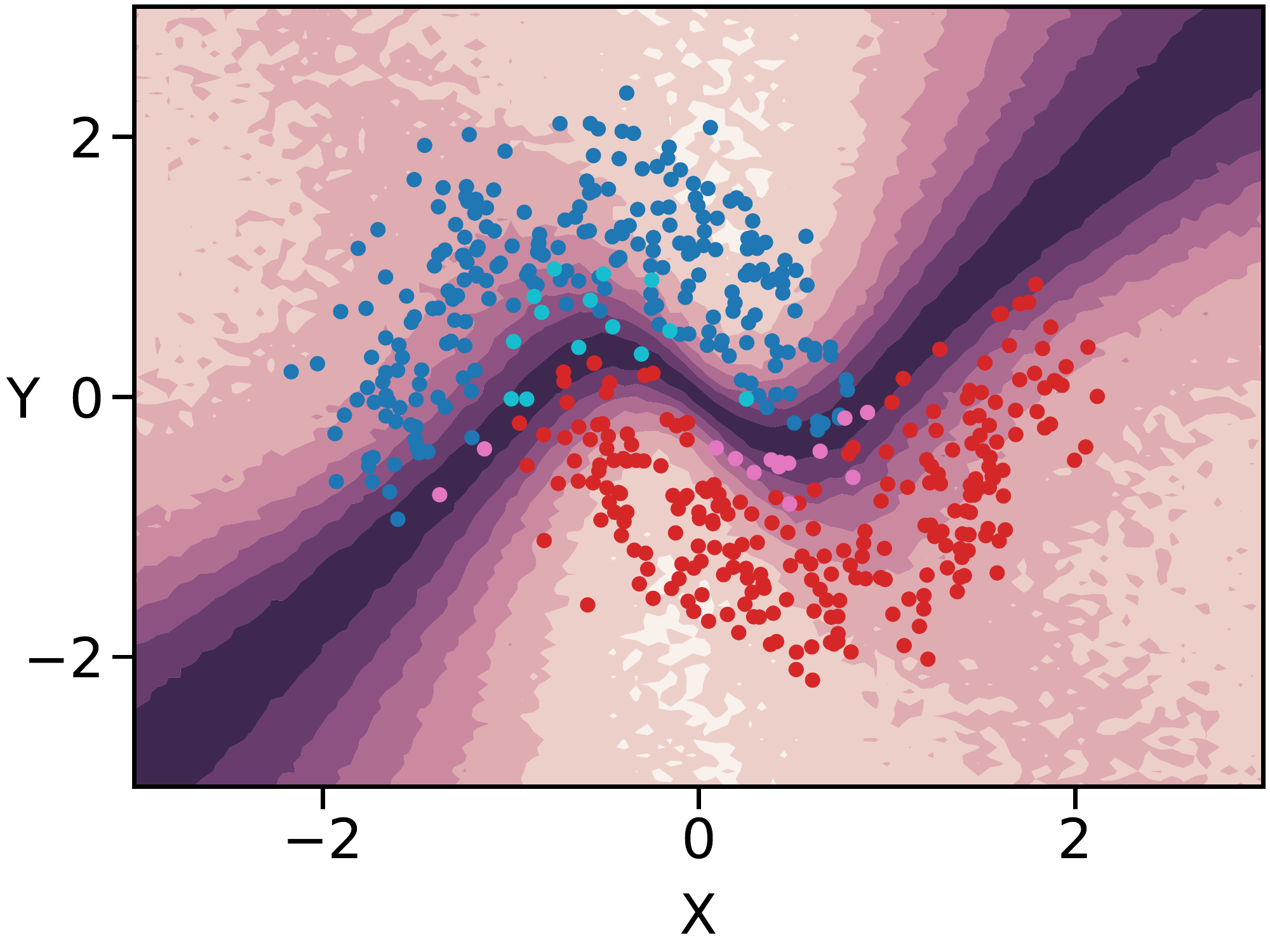}
        \caption{CIQP, $n=10$}
        \label{fig:bc_std_ciqp_10_0}
    \end{subfigure}
    \begin{subfigure}[b]{0.24\linewidth}
        \includegraphics[width=\textwidth]{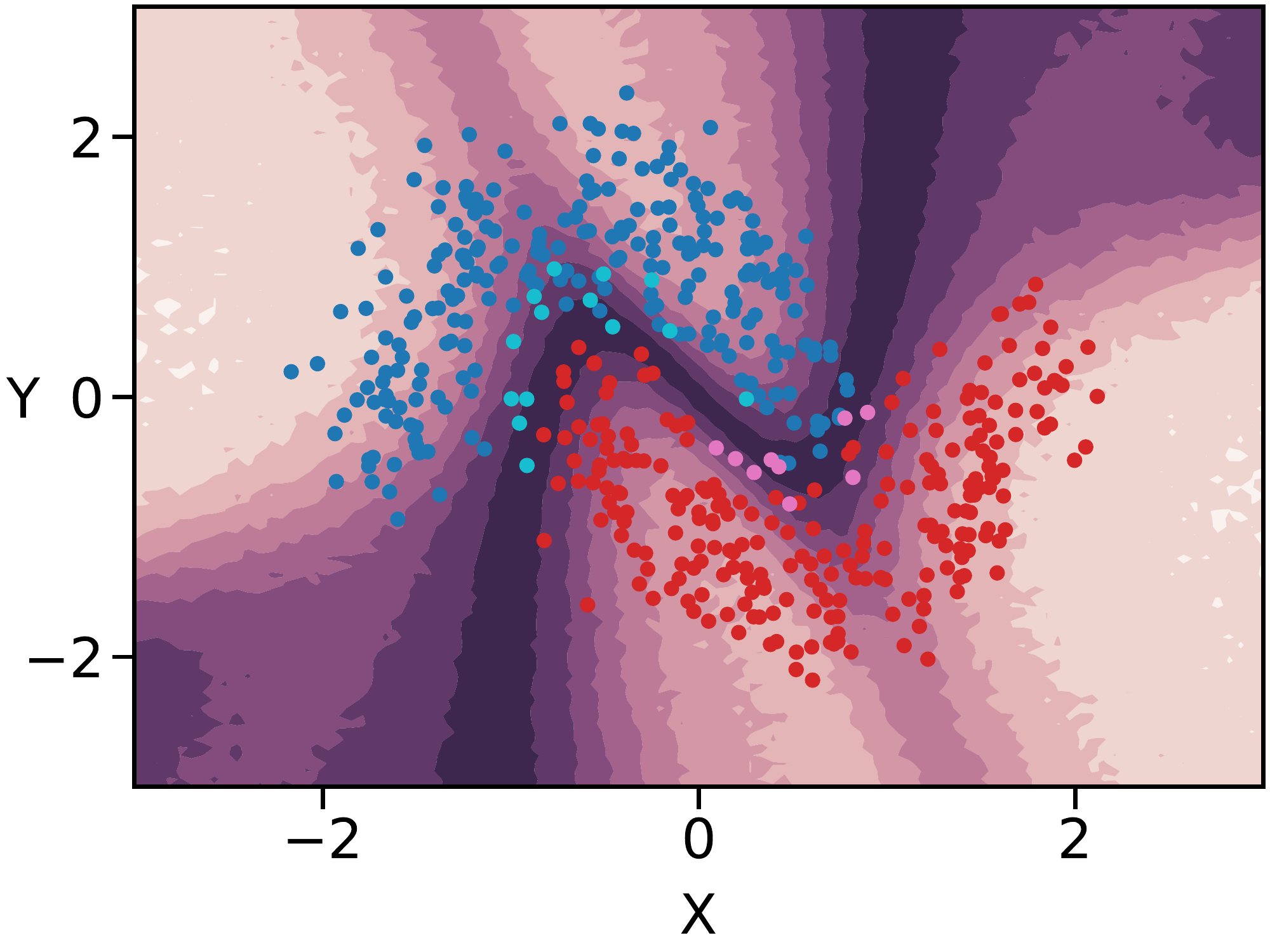}
        \caption{QICP, $n=10$}
        \label{fig:bc_std_qicp_10_0}
    \end{subfigure}
    \begin{subfigure}[b]{0.24\linewidth}
        \includegraphics[width=\textwidth]{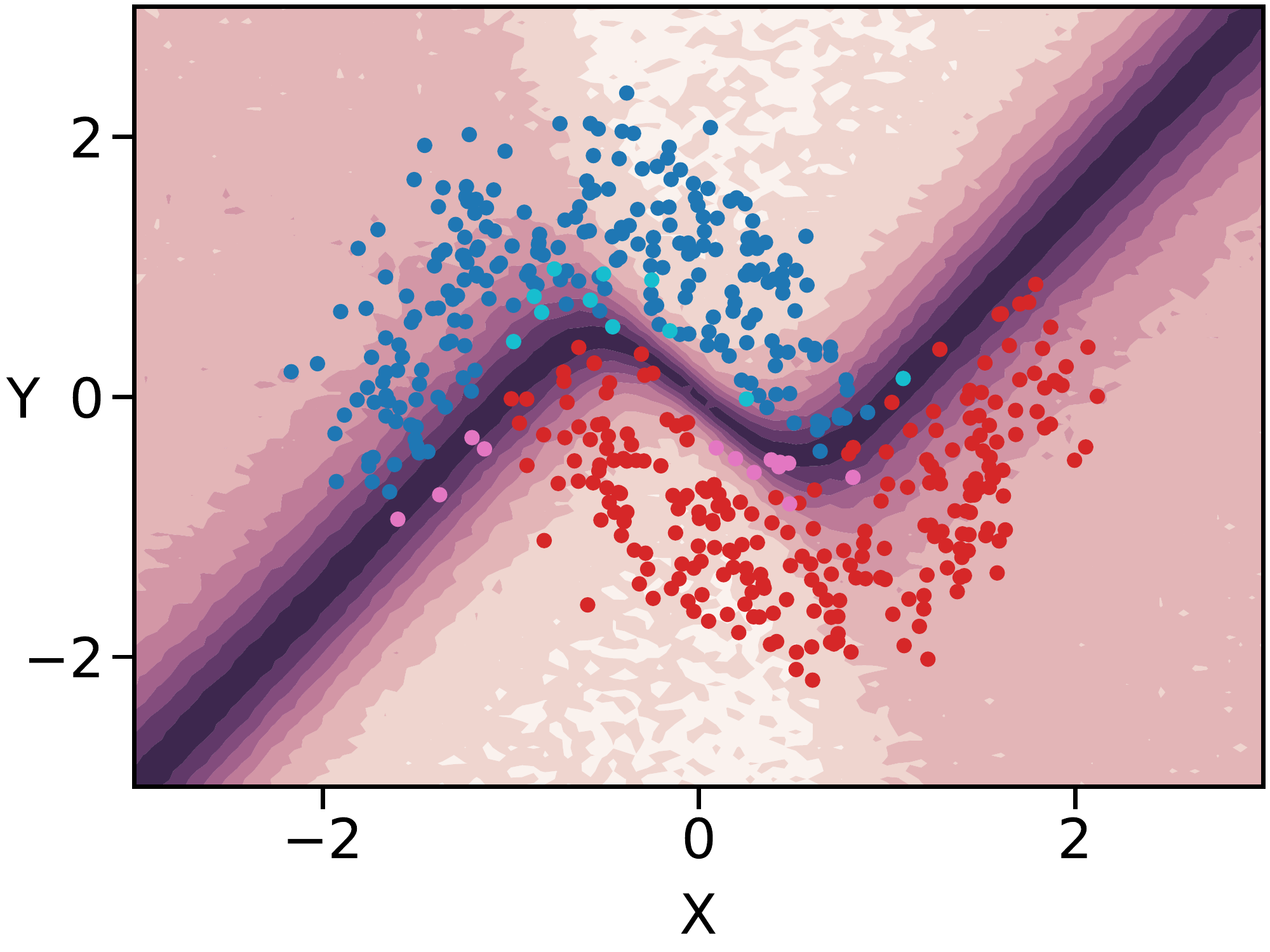}
        \caption{QIQP, $n=10$}
        \label{fig:bc_std_qiqp_10_0}
    \end{subfigure}
    
\caption{Posterior Predictive Standard Deviation of Binary Classification with \ac{bnn}: \emph{C} and \emph{Q} stand for \emph{Classical} and \emph{Quantum} respectively. \emph{I} and \emph{P} stand for \emph{Inference} and \emph{Prediction}. The Figure shows the expected increase in accuracy for higher qubit numbers $n$.}
\label{fig:bc_std}
\end{center}
\vskip -0.2in
\end{figure}

The additional results from the simulation of both the linear regression and binary classification task can be found in this section. \cref{fig:bc_std} shows the standard deviation of the \ac{bnn} in the binary classification task. 

\begin{figure}[t]
\begin{center}
    \begin{subfigure}[b]{0.24\linewidth}
        \includegraphics[width=\textwidth]{figs/linear_regression/cicp_nh_5_seed_0_fr.pdf}
        \caption{CICP, FR}
        \label{fig:lr_cicp_0_fr}
    \end{subfigure}
    \begin{subfigure}[b]{0.24\linewidth}
        \includegraphics[width=\textwidth]{figs/linear_regression/ciqp_n_10_nh_5_seed_0_fr.pdf}
        \caption{CIQP, FR}
        \label{fig:lr_ciqp_10_0_fr}
    \end{subfigure}
    \begin{subfigure}[b]{0.24\linewidth}
        \includegraphics[width=\textwidth]{figs/linear_regression/qicp_n_10_nh_5_seed_0_fr.pdf}
        \caption{QICP, FR}
        \label{fig:lr_qicp_10_0_fr}
    \end{subfigure}
    \begin{subfigure}[b]{0.24\linewidth}
        \includegraphics[width=\textwidth]{figs/linear_regression/qiqp_n_10_nh_5_seed_0_fr.pdf}
        \caption{QIQP, FR}
        \label{fig:lr_qiqp_10_0_fr}
    \end{subfigure}
    \vskip 1.0em
    \begin{subfigure}[b]{0.24\linewidth}
        \includegraphics[width=\textwidth]{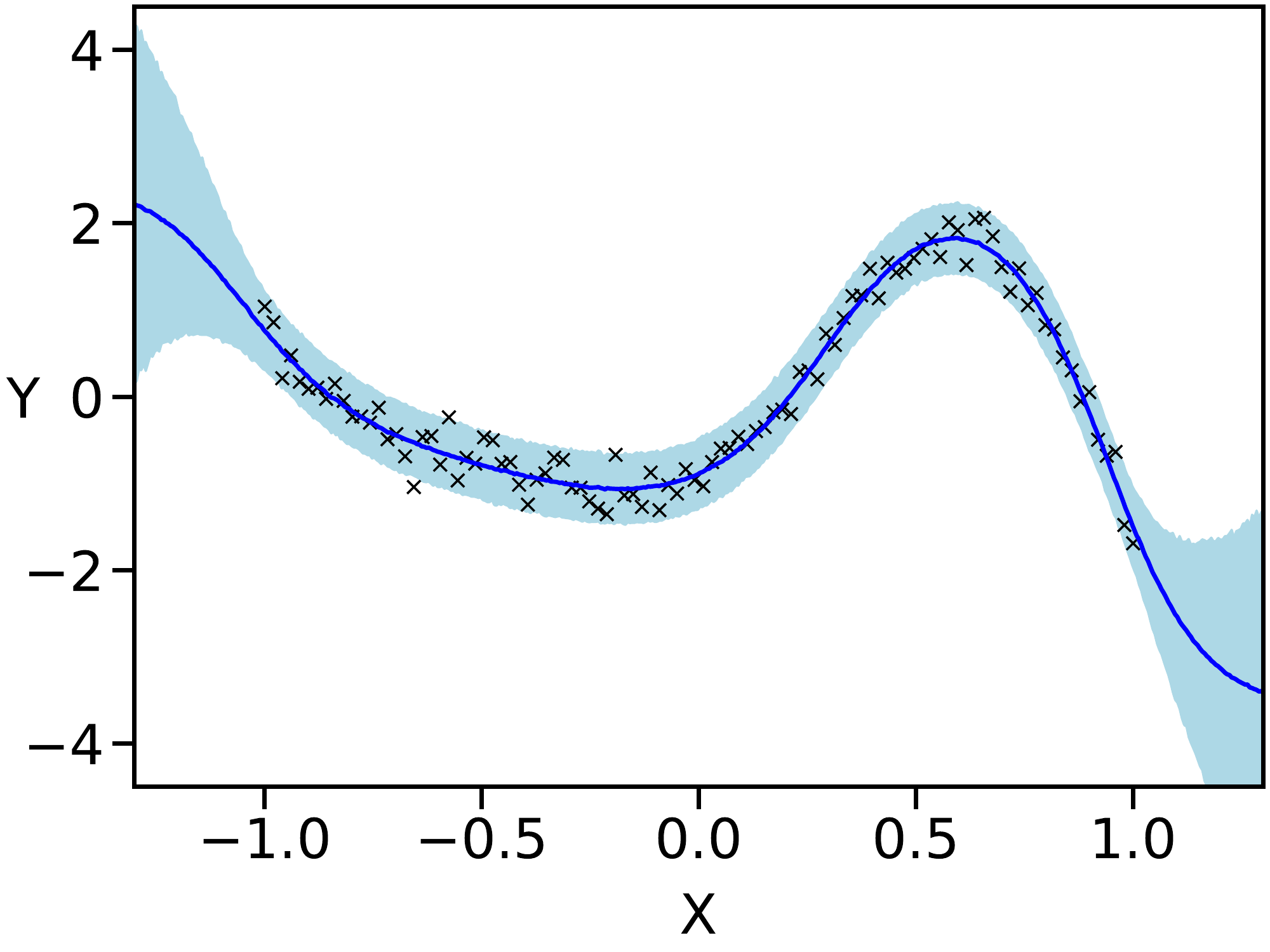}
        \caption{CICP, LR}
        \label{fig:lr_cicp_0_lr}
    \end{subfigure}
    \begin{subfigure}[b]{0.24\linewidth}
        \includegraphics[width=\textwidth]{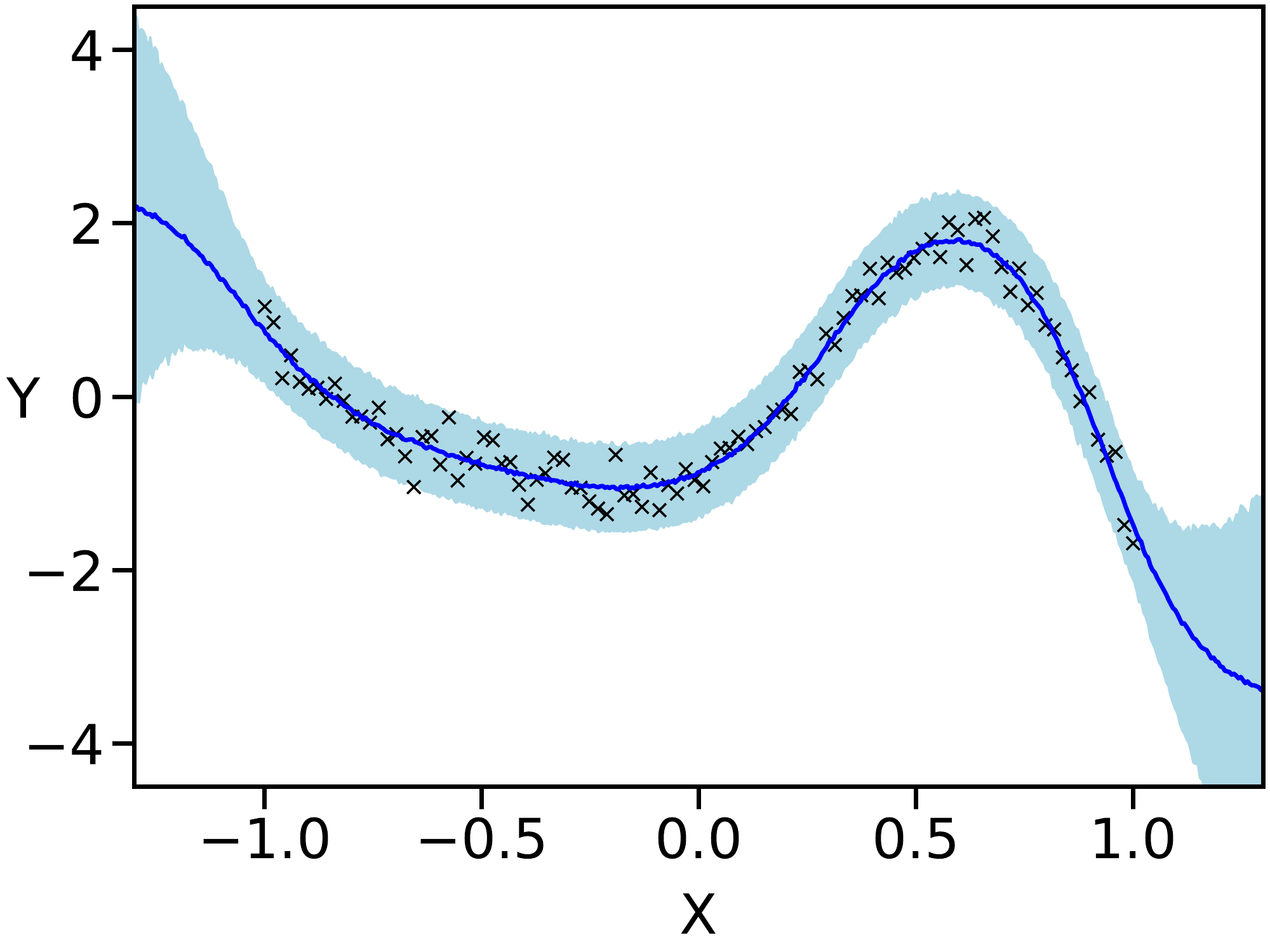}
        \caption{CIQP, LR}
        \label{fig:lr_ciqp_10_0_lr}
    \end{subfigure}
    \begin{subfigure}[b]{0.24\linewidth}
        \includegraphics[width=\textwidth]{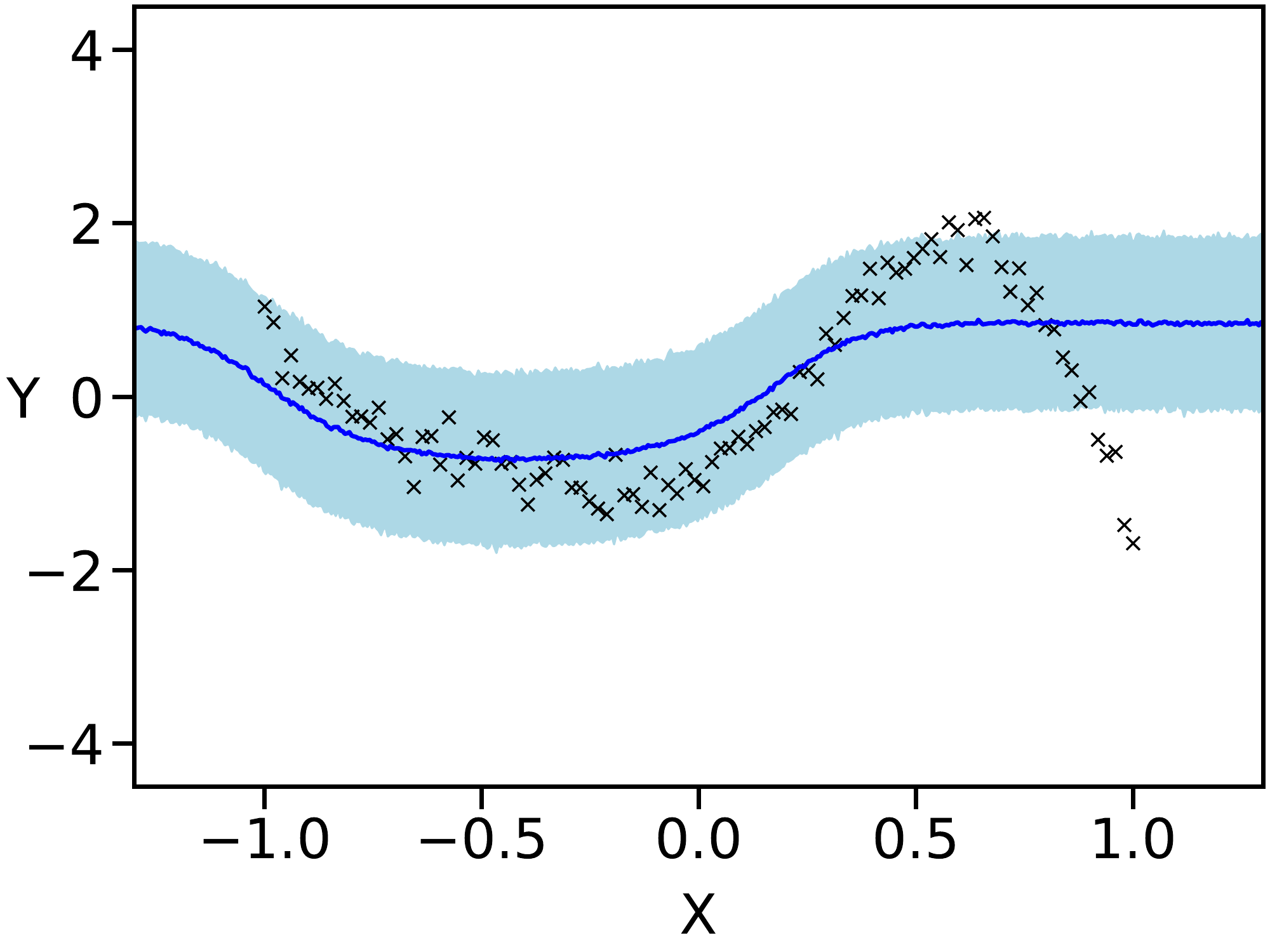}
        \caption{QICP, LR}
        \label{fig:lr_qicp_10_0_lr}
    \end{subfigure}
    \begin{subfigure}[b]{0.24\linewidth}
        \includegraphics[width=\textwidth]{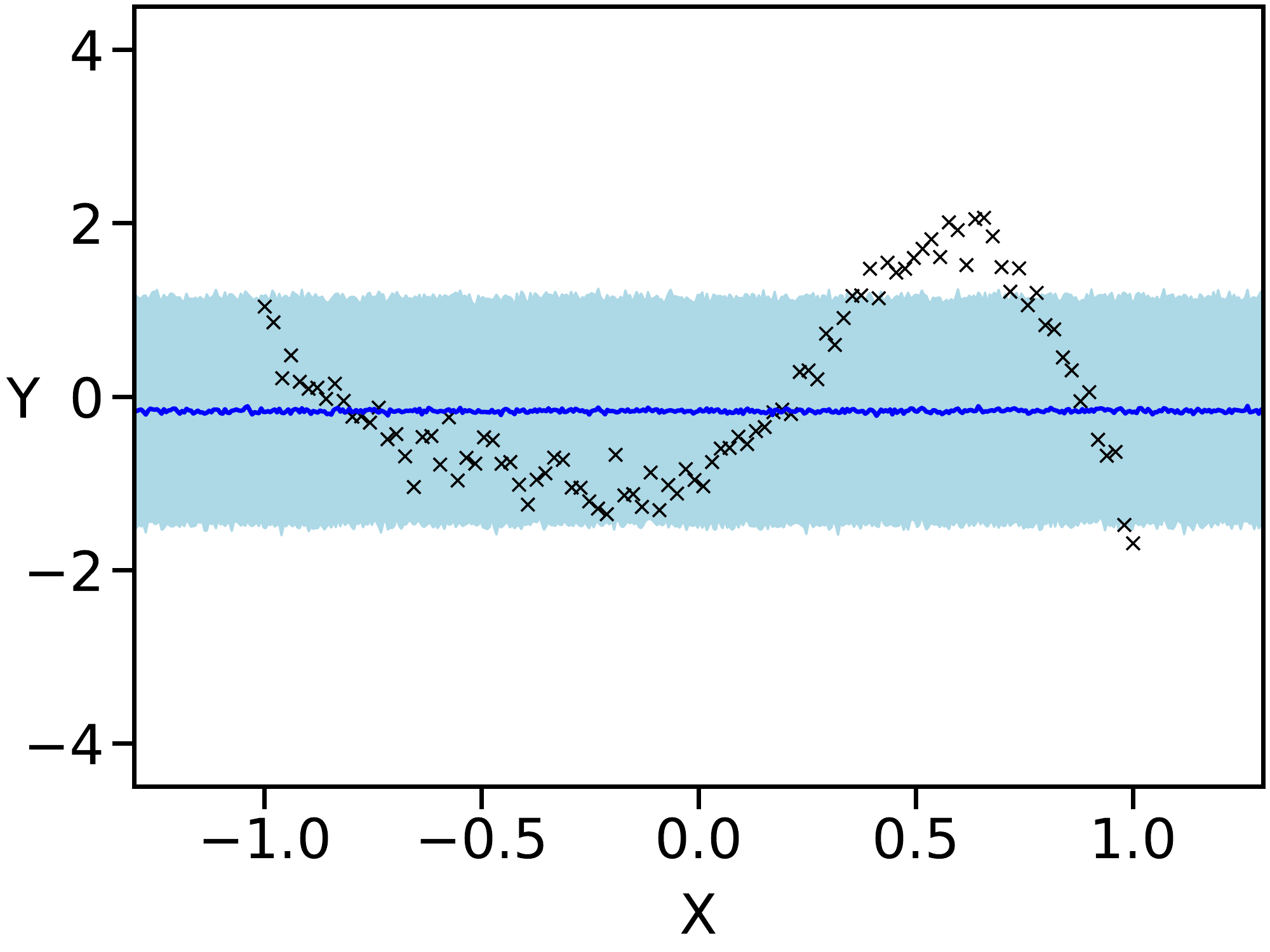}
        \caption{QIQP, LR}
        \label{fig:lr_qiqp_10_0_lr}
    \end{subfigure}
    
\caption{Comparison between Full-Rank Initialization (Rank 5) and Low-Rank Initialization (Rank 3) for the Linear Regression Task on a \ac{bnn} for 10 qubits. FR stands for Full-Rank Initialization and LR stands for Low-Rank Initialization.}
\label{fig:lr_low_rank}
\end{center}
\end{figure}

\begin{figure}[t]
\begin{center}
    \begin{subfigure}[b]{0.24\linewidth}
        \includegraphics[width=\textwidth]{figs/binary_classification/mean/cicp_nh_5_seed_0_fr.pdf}
        \caption{CICP, FR}
        \label{fig:bc_mean_cicp_0_fr}
    \end{subfigure}
    \begin{subfigure}[b]{0.24\linewidth}
        \includegraphics[width=\textwidth]{figs/binary_classification/mean/ciqp_n_10_nh_5_seed_0_fr.pdf}
        \caption{CIQP, FR}
        \label{fig:bc_mean_ciqp_10_0_fr}
    \end{subfigure}
    \begin{subfigure}[b]{0.24\linewidth}
        \includegraphics[width=\textwidth]{figs/binary_classification/mean/qicp_n_10_nh_5_seed_0_fr.pdf}
        \caption{QICP, FR}
        \label{fig:bc_mean_qicp_10_0_fr}
    \end{subfigure}
    \begin{subfigure}[b]{0.24\linewidth}
        \includegraphics[width=\textwidth]{figs/binary_classification/mean/qiqp_n_10_nh_5_seed_0_fr.pdf}
        \caption{QIQP, FR}
        \label{fig:bc_mean_qiqp_10_0_fr}
    \end{subfigure}
    \vskip 1.0em
    \begin{subfigure}[b]{0.24\linewidth}
        \includegraphics[width=\textwidth]{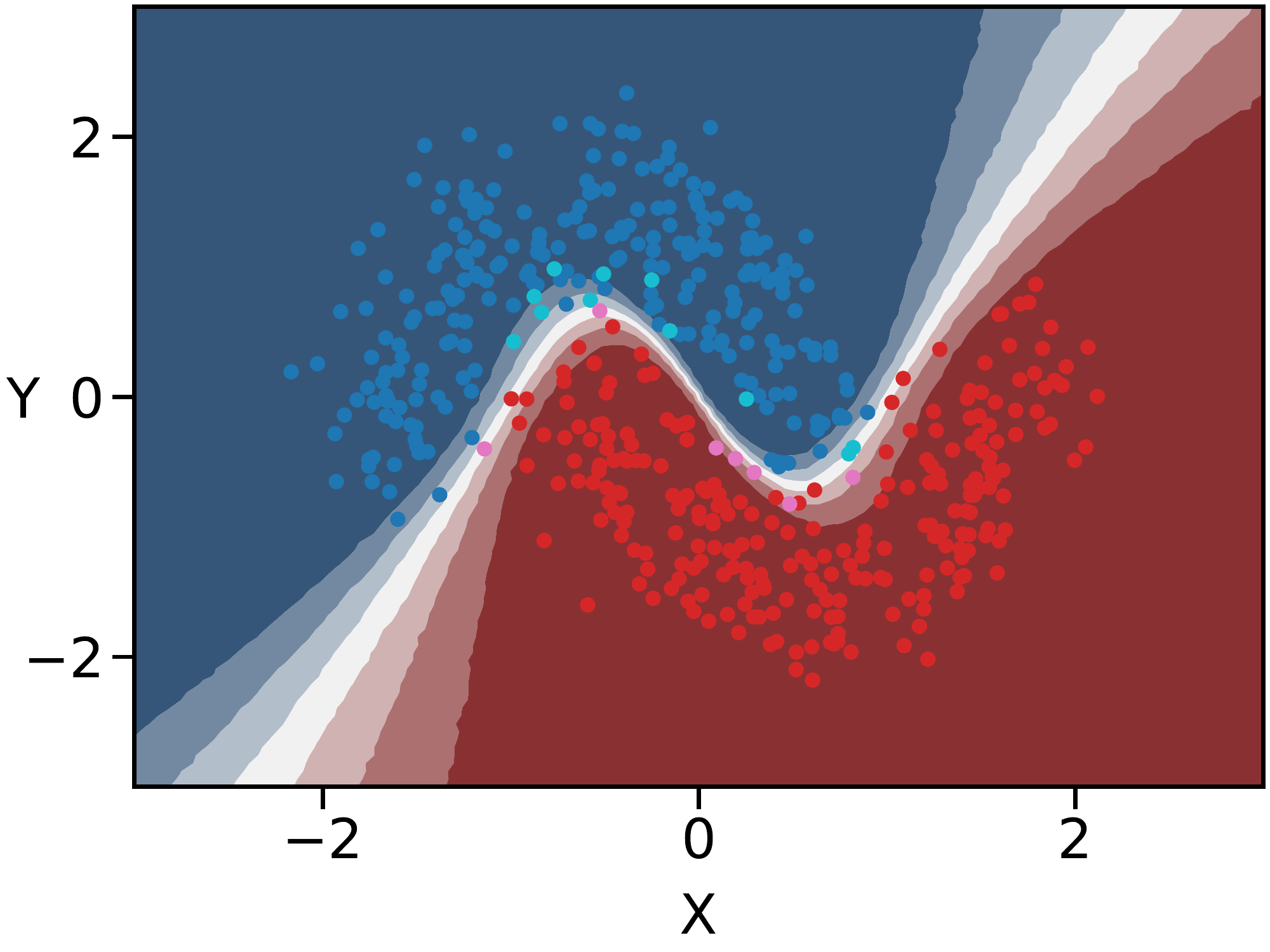}
        \caption{CICP, LR}
        \label{fig:bc_mean_cicp_0_lr}
    \end{subfigure}
    \begin{subfigure}[b]{0.24\linewidth}
        \includegraphics[width=\textwidth]{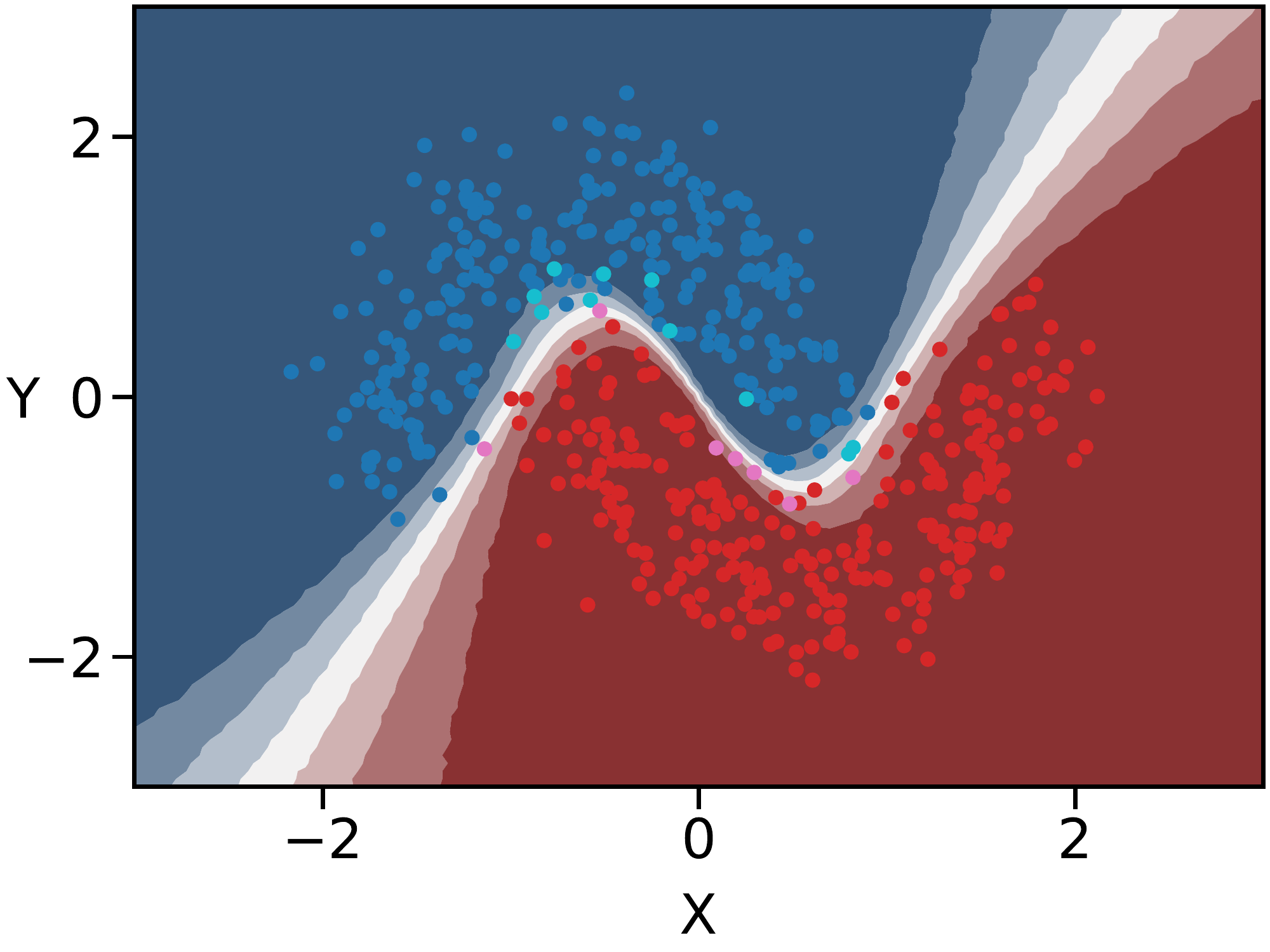}
        \caption{CIQP, LR}
        \label{fig:bc_mean_ciqp_10_0_lr}
    \end{subfigure}
    \begin{subfigure}[b]{0.24\linewidth}
        \includegraphics[width=\textwidth]{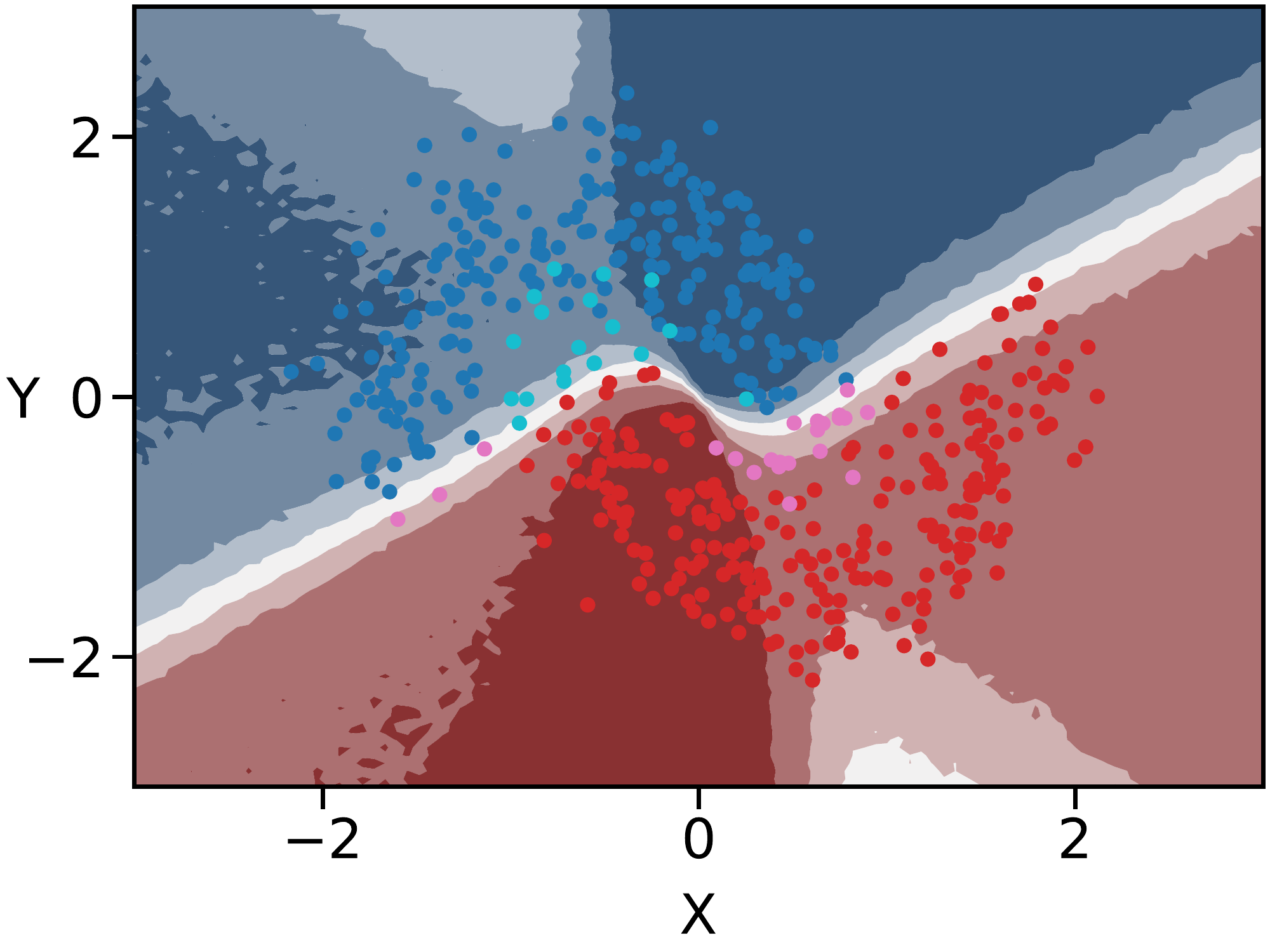}
        \caption{QICP, LR}
        \label{fig:bc_mean_qicp_10_0_lr}
    \end{subfigure}
    \begin{subfigure}[b]{0.24\linewidth}
        \includegraphics[width=\textwidth]{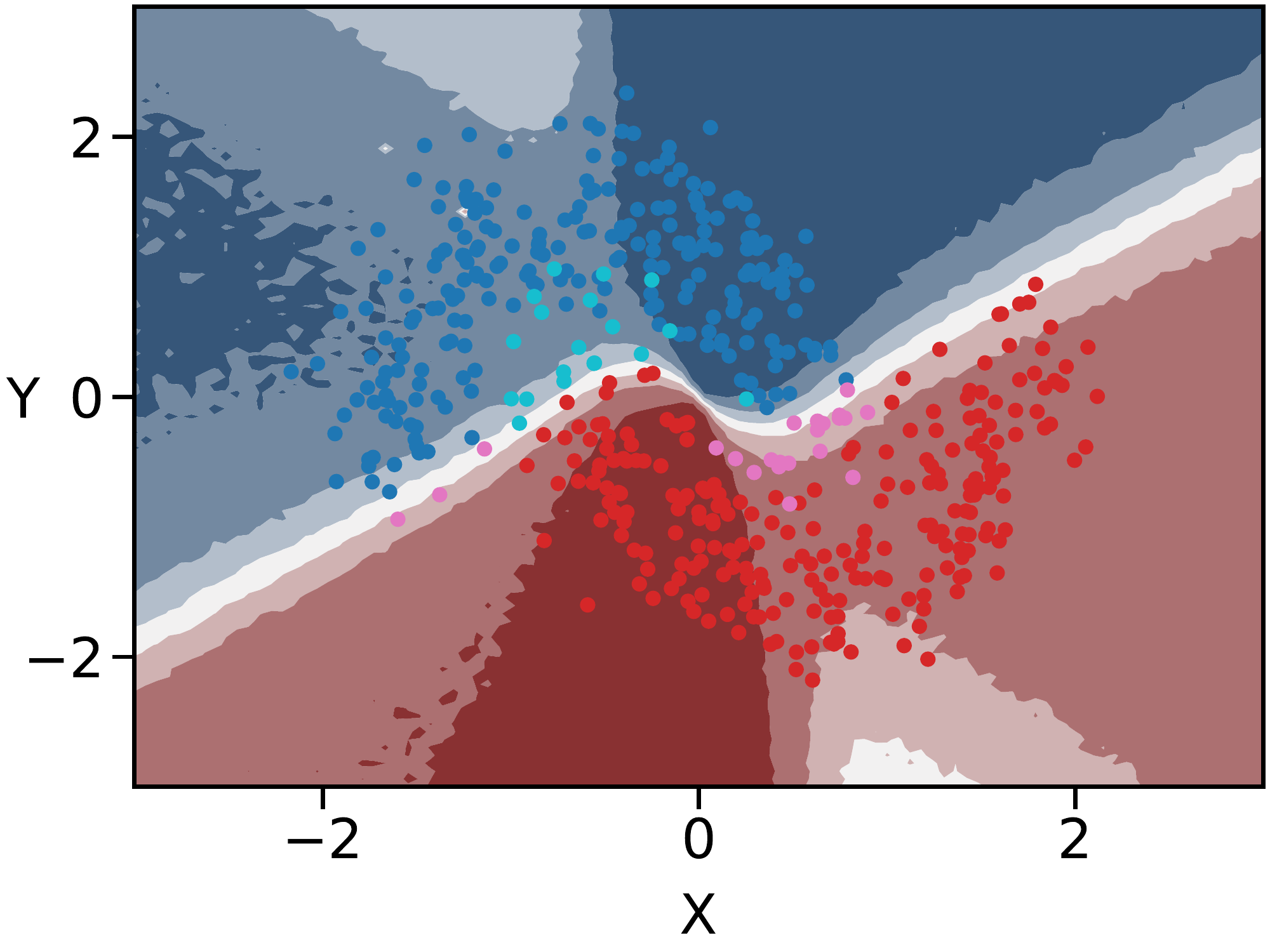}
        \caption{QIQP, LR}
        \label{fig:bc_mean_qiqp_10_0_lr}
    \end{subfigure}
    
\caption{Comparison between Full-Rank Initialization (Rank 5) and Low-Rank Initialization (Rank 3) for the Binary Classification Task on a \ac{bnn} for 10 qubits. FR stands for Full-Rank Initialization and LR stands for Low-Rank Initialization.}
\label{fig:bc_mean_low_rank}
\end{center}
\end{figure}

\begin{figure}[t]
\begin{center}
    \begin{subfigure}[b]{0.24\linewidth}
        \includegraphics[width=\textwidth]{figs/binary_classification/std/cicp_nh_5_seed_0_fr.pdf}
        \caption{CICP, FR}
        \label{fig:bc_std_cicp_0_fr}
    \end{subfigure}
    \begin{subfigure}[b]{0.24\linewidth}
        \includegraphics[width=\textwidth]{figs/binary_classification/std/ciqp_n_10_nh_5_seed_0_fr.pdf}
        \caption{CIQP, FR}
        \label{fig:bc_std_ciqp_10_0_fr}
    \end{subfigure}
    \begin{subfigure}[b]{0.24\linewidth}
        \includegraphics[width=\textwidth]{figs/binary_classification/std/qicp_n_10_nh_5_seed_0_fr.pdf}
        \caption{QICP, FR}
        \label{fig:bc_std_qicp_10_0_fr}
    \end{subfigure}
    \begin{subfigure}[b]{0.24\linewidth}
        \includegraphics[width=\textwidth]{figs/binary_classification/std/qiqp_n_10_nh_5_seed_0_fr.pdf}
        \caption{QIQP, FR}
        \label{fig:bc_std_qiqp_10_0_fr}
    \end{subfigure}
    \vskip 1.0em
    \begin{subfigure}[b]{0.24\linewidth}
        \includegraphics[width=\textwidth]{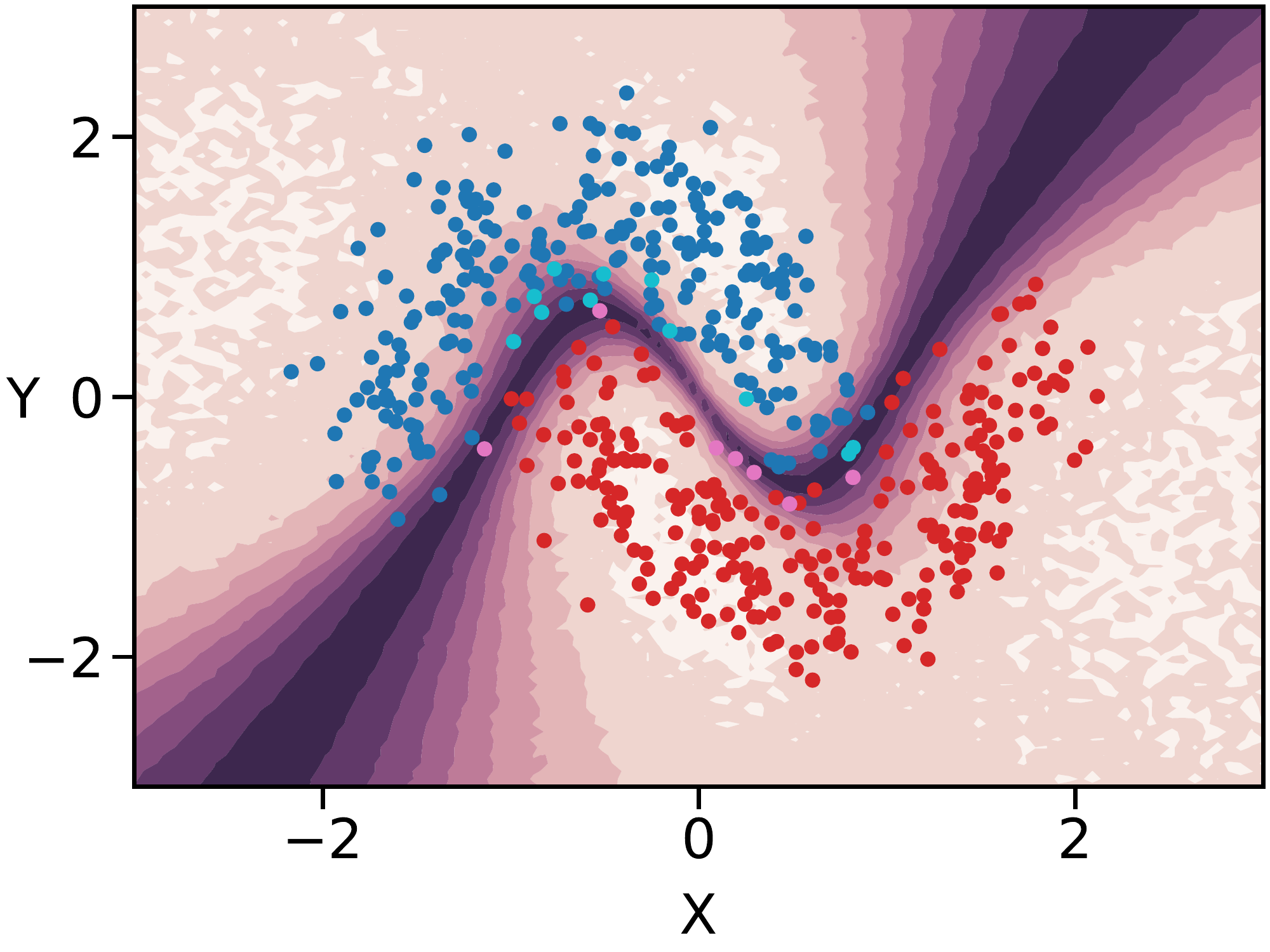}
        \caption{CICP, LR}
        \label{fig:bc_std_cicp_0_lr}
    \end{subfigure}
    \begin{subfigure}[b]{0.24\linewidth}
        \includegraphics[width=\textwidth]{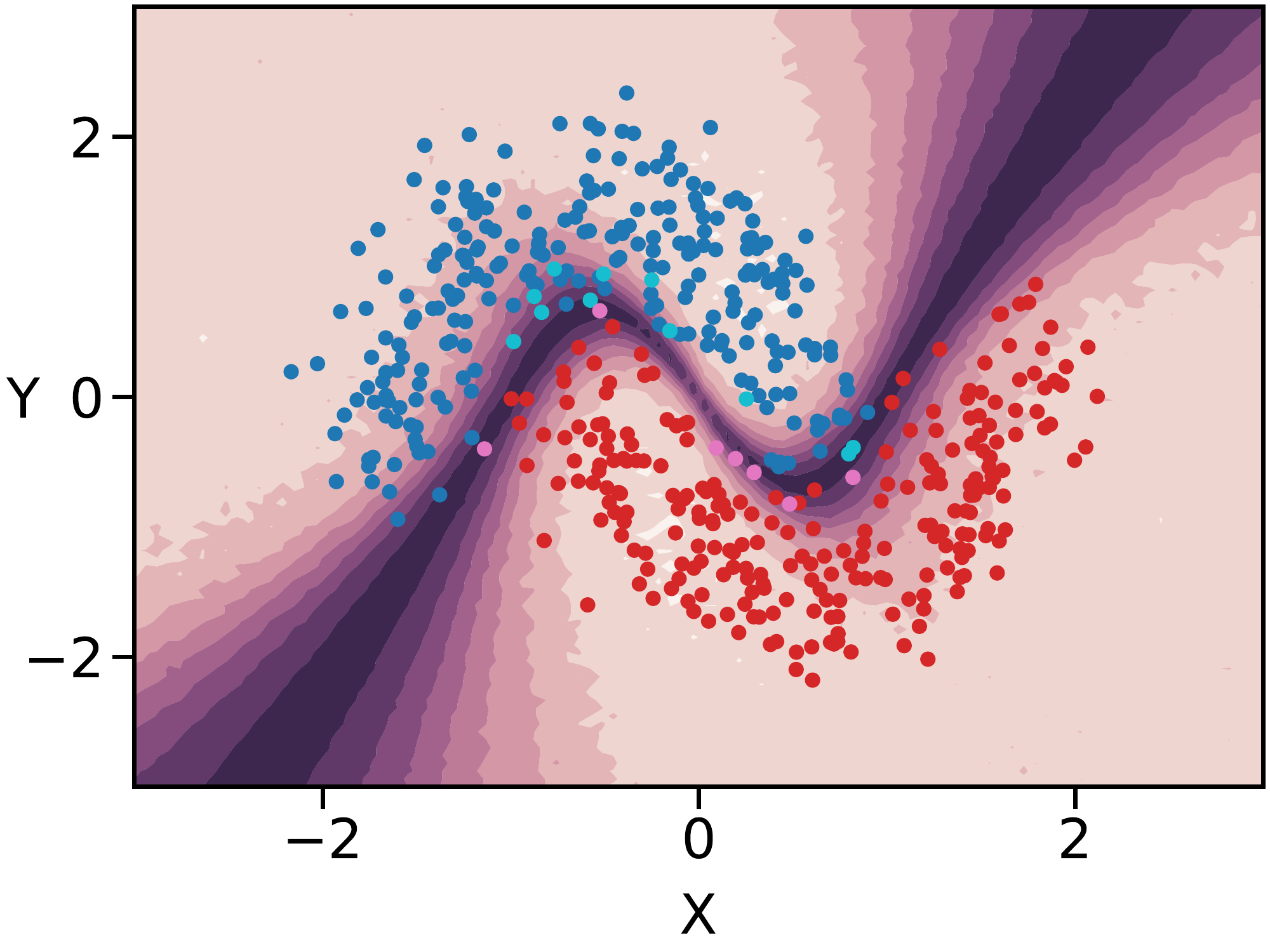}
        \caption{CIQP, LR}
        \label{fig:bc_std_ciqp_10_0_lr}
    \end{subfigure}
    \begin{subfigure}[b]{0.24\linewidth}
        \includegraphics[width=\textwidth]{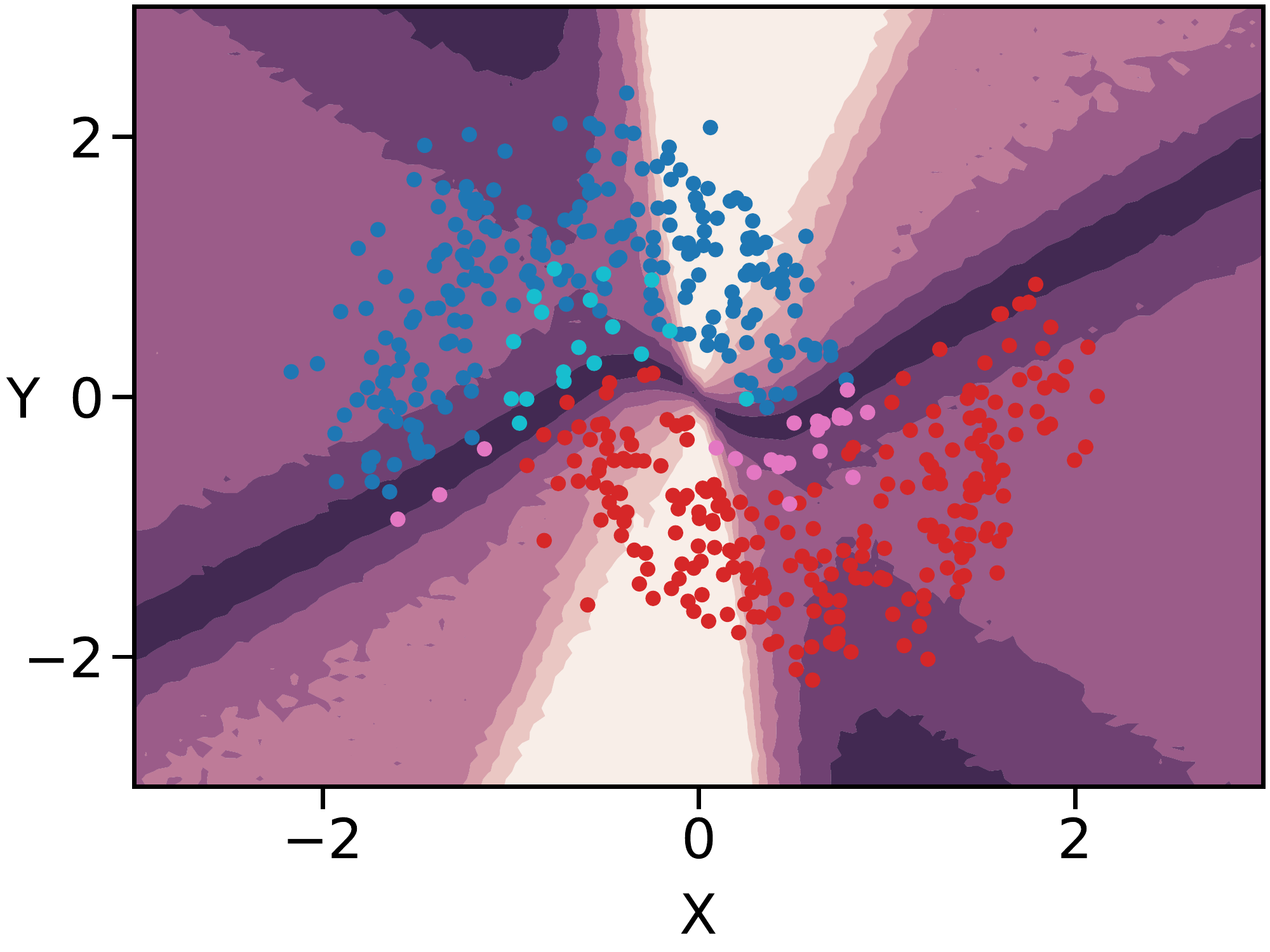}
        \caption{QICP, LR}
        \label{fig:bc_std_qicp_10_0_lr}
    \end{subfigure}
    \begin{subfigure}[b]{0.24\linewidth}
        \includegraphics[width=\textwidth]{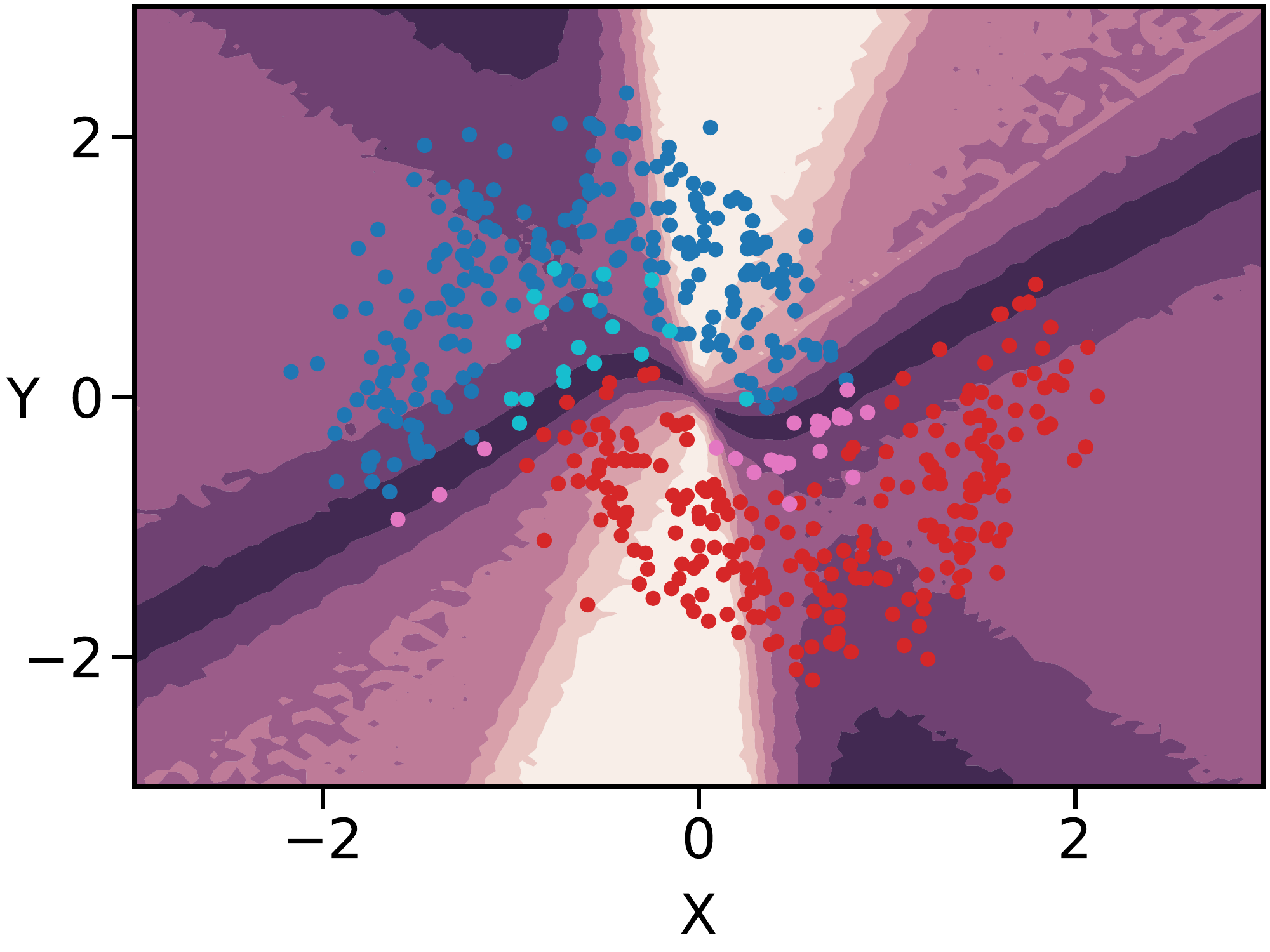}
        \caption{QIQP, LR}
        \label{fig:bc_std_qiqp_10_0_lr}
    \end{subfigure}
    
\caption{Comparison between Full-Rank Initialization (Rank 5) and Low-Rank Initialization (Rank 3) for the Posterior Predictive Standard Deviation of Binary Classification with \ac{bnn} for 10 qubits. FR stands for Full-Rank Initialization and LR stands for Low-Rank Initialization.}
\label{fig:bc_std_low_rank}
\end{center}
\end{figure}

\cref{fig:lr_low_rank}, \cref{fig:bc_mean_low_rank} and \cref{fig:bc_std_low_rank} show a comparison between low-rank and full-rank initialization of the linear regression and binary classification task.

\begin{figure}[t]
\vskip 0.2in
\begin{center}
    \begin{subfigure}[b]{0.24\linewidth}
        \includegraphics[width=\textwidth]{figs/linear_regression/cicp_nh_5_seed_0_fr.pdf}
        \caption{CICP (Reference)}
    \end{subfigure}
    \begin{subfigure}[b]{0.24\linewidth}
        \includegraphics[width=\textwidth]{figs/linear_regression/ciqp_n_5_nh_5_seed_0_fr.pdf}
        \caption{CIQP, $n=5$}
    \end{subfigure}
    \begin{subfigure}[b]{0.24\linewidth}
        \includegraphics[width=\textwidth]{figs/linear_regression/qicp_n_5_nh_5_seed_0_fr.pdf}
        \caption{QICP, $n=5$}
    \end{subfigure}
    \begin{subfigure}[b]{0.24\linewidth}
        \includegraphics[width=\textwidth]{figs/linear_regression/qiqp_n_5_nh_5_seed_0_fr.pdf}
        \caption{QIQP, $n=5$}
    \end{subfigure}
    \vskip 1.0em
    \begin{subfigure}[b]{0.24\linewidth}
        \includegraphics[width=\textwidth]{figs/linear_regression/legend.pdf}
        \caption{Legend}
    \end{subfigure}
    \begin{subfigure}[b]{0.24\linewidth}
        \includegraphics[width=\textwidth]{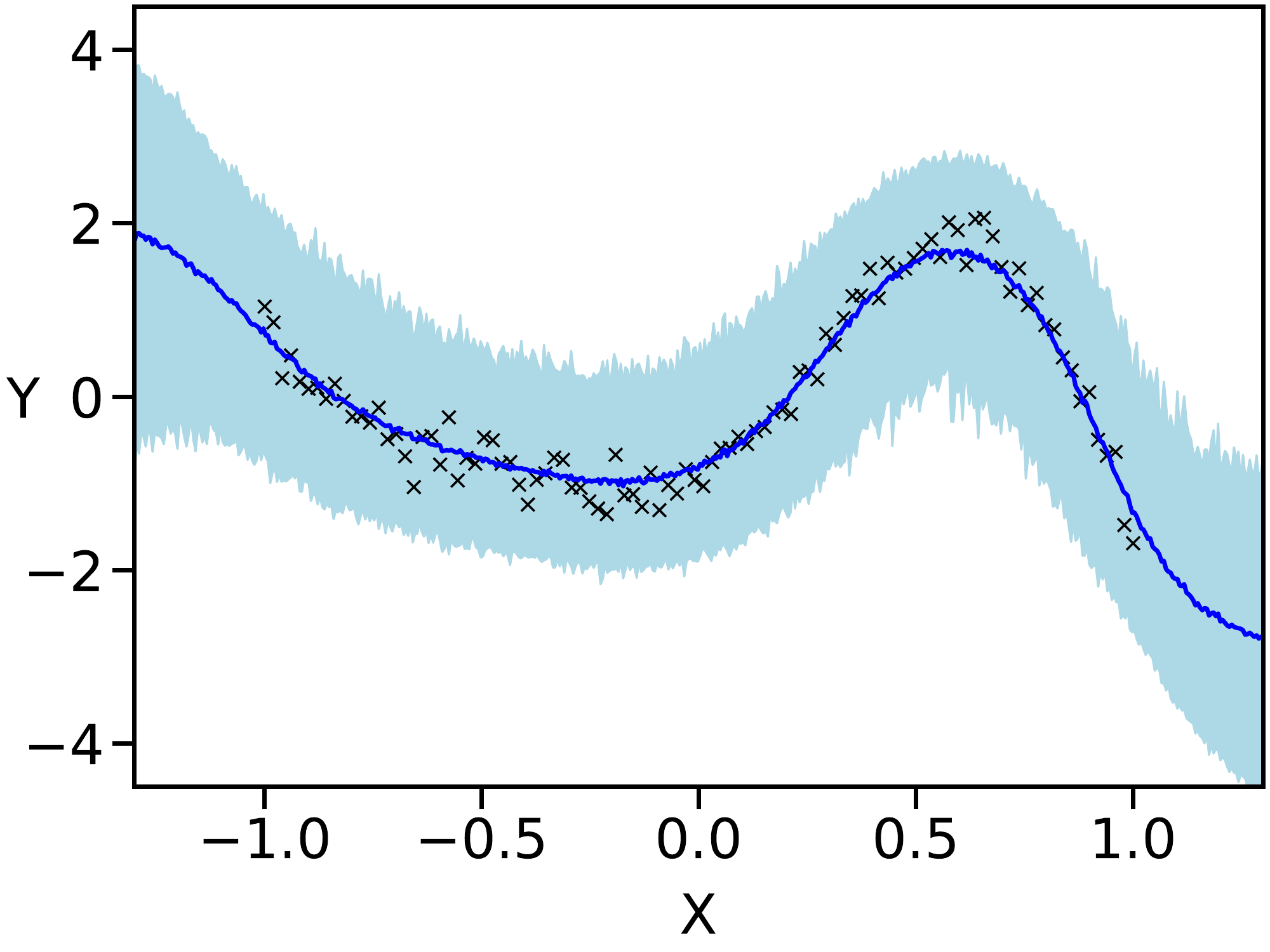}
        \caption{CIQP, $n=7$}
    \end{subfigure}
    \begin{subfigure}[b]{0.24\linewidth}
        \includegraphics[width=\textwidth]{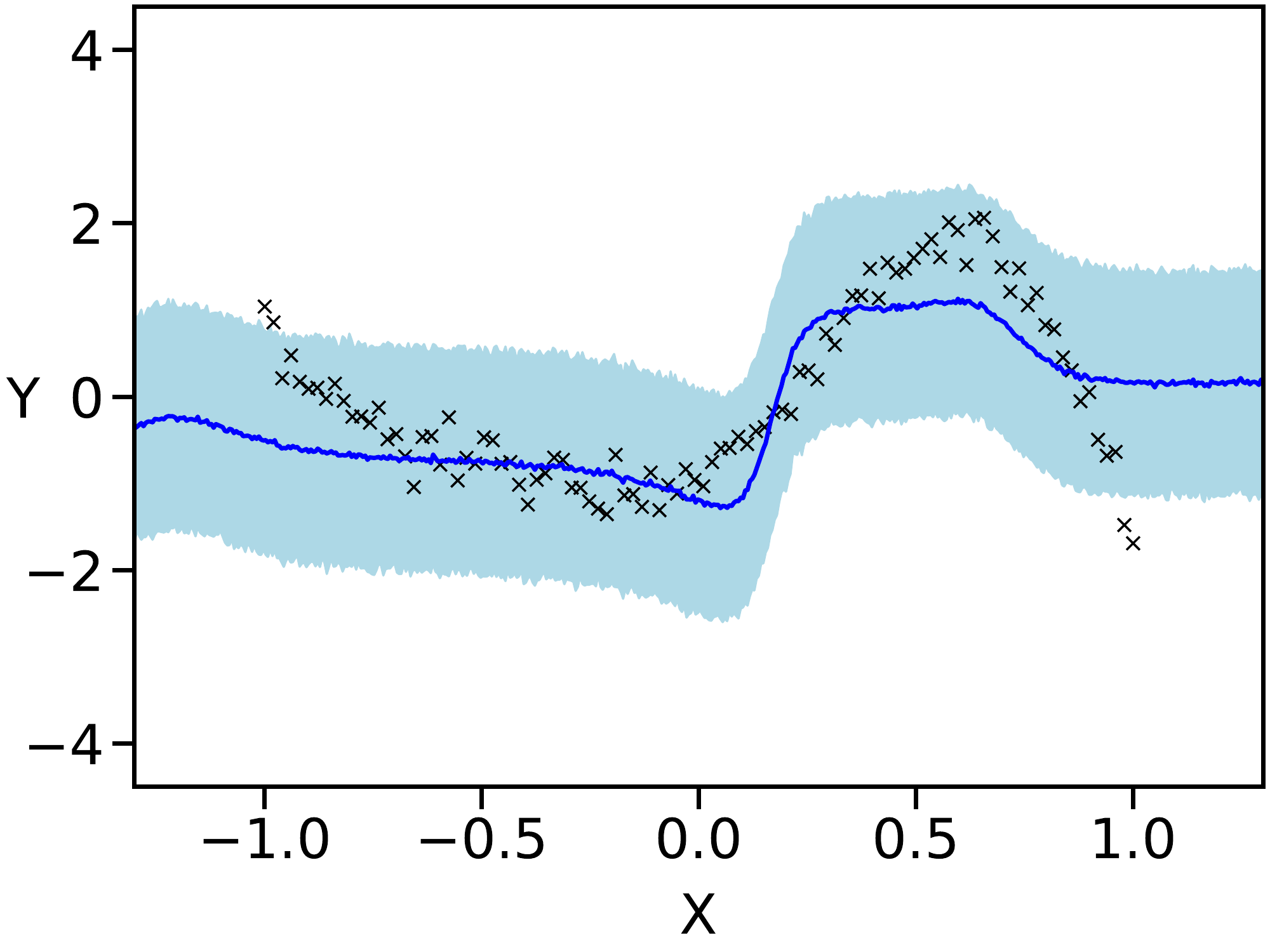}
        \caption{QICP, $n=7$}
    \end{subfigure}
    \begin{subfigure}[b]{0.24\linewidth}
        \includegraphics[width=\textwidth]{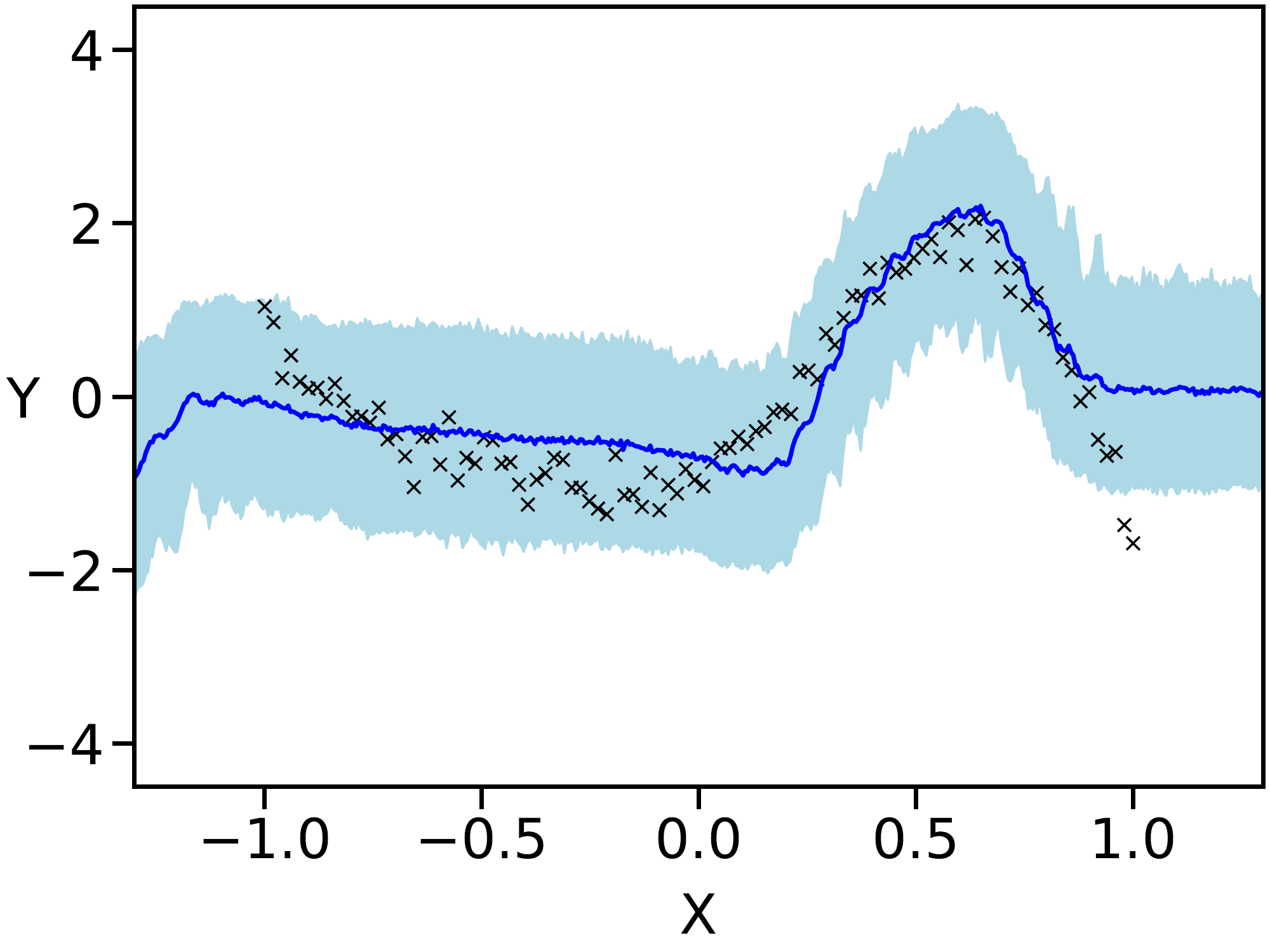}
        \caption{QIQP, $n=7$}
    \end{subfigure}
    \vskip 1.0em    
    \hspace{11.3em}
    \begin{subfigure}[b]{0.24\linewidth}
        \includegraphics[width=\textwidth]{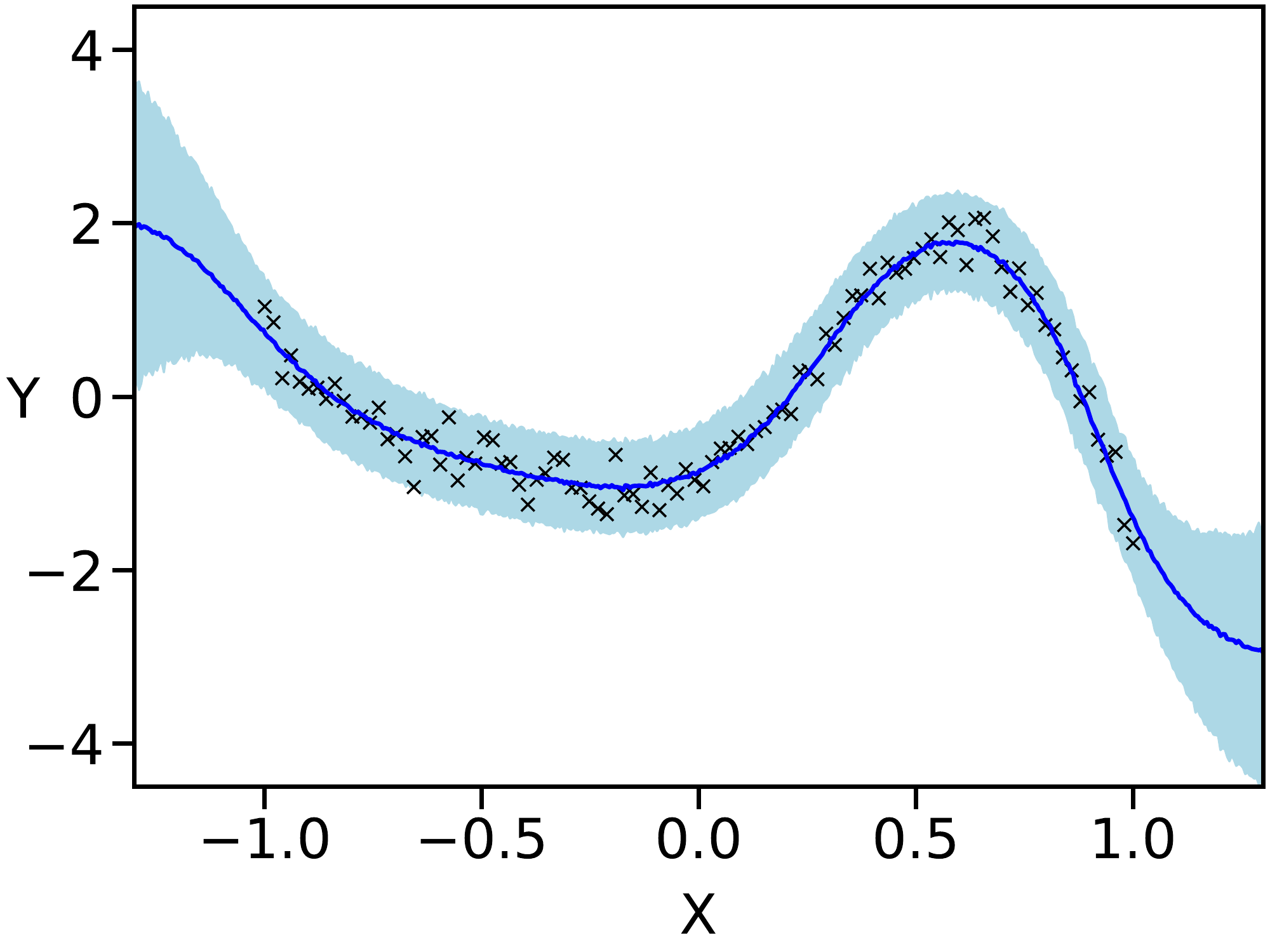}
        \caption{CIQP, $n=9$}
    \end{subfigure}
    \begin{subfigure}[b]{0.24\linewidth}
        \includegraphics[width=\textwidth]{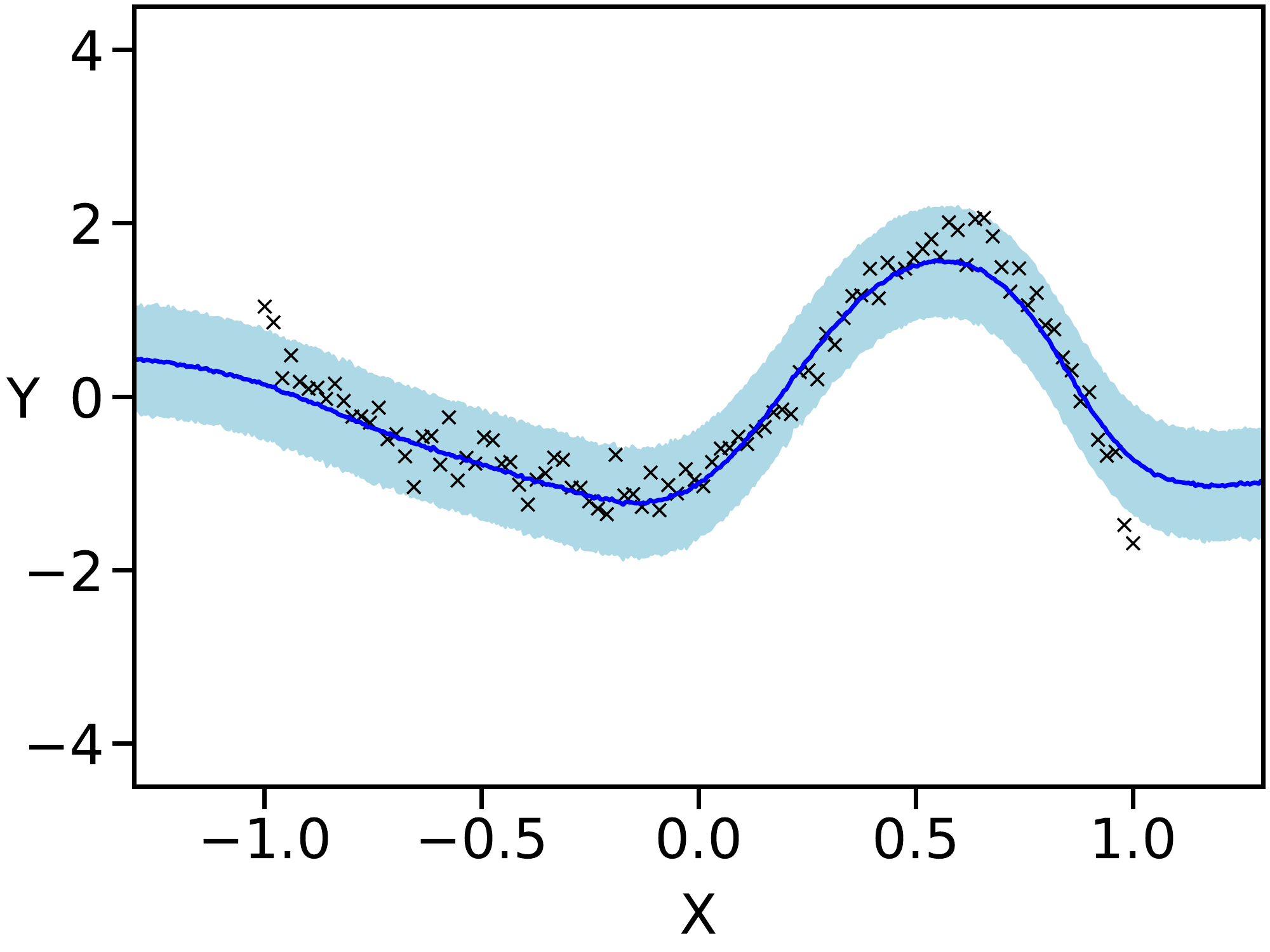}
        \caption{QICP, $n=9$}
    \end{subfigure}
    \begin{subfigure}[b]{0.24\linewidth}
        \includegraphics[width=\textwidth]{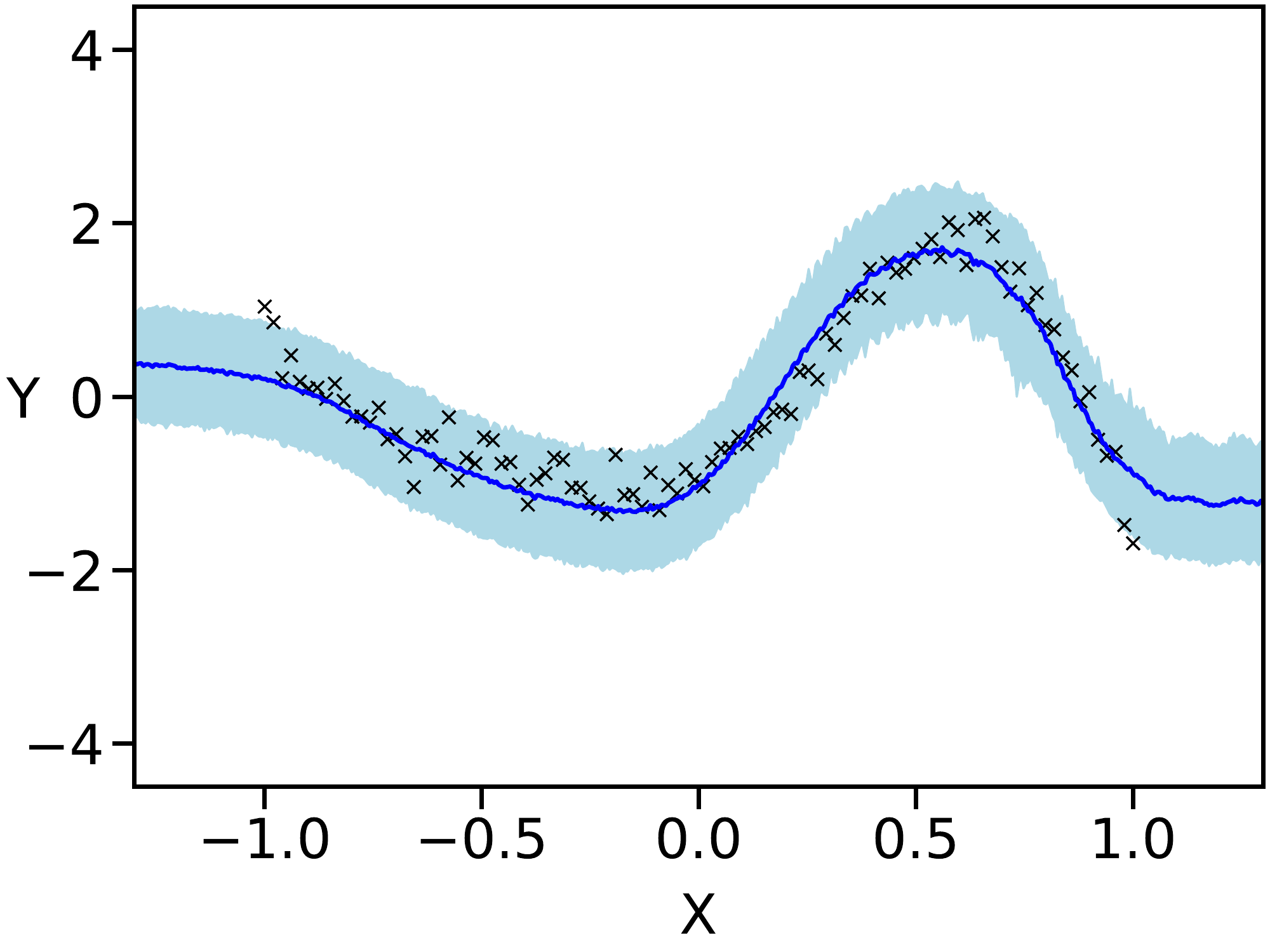}
        \caption{QIQP, $n=9$}
    \end{subfigure}
    \vskip 1.0em    
    \hspace{11.3em}
    \begin{subfigure}[b]{0.24\linewidth}
        \includegraphics[width=\textwidth]{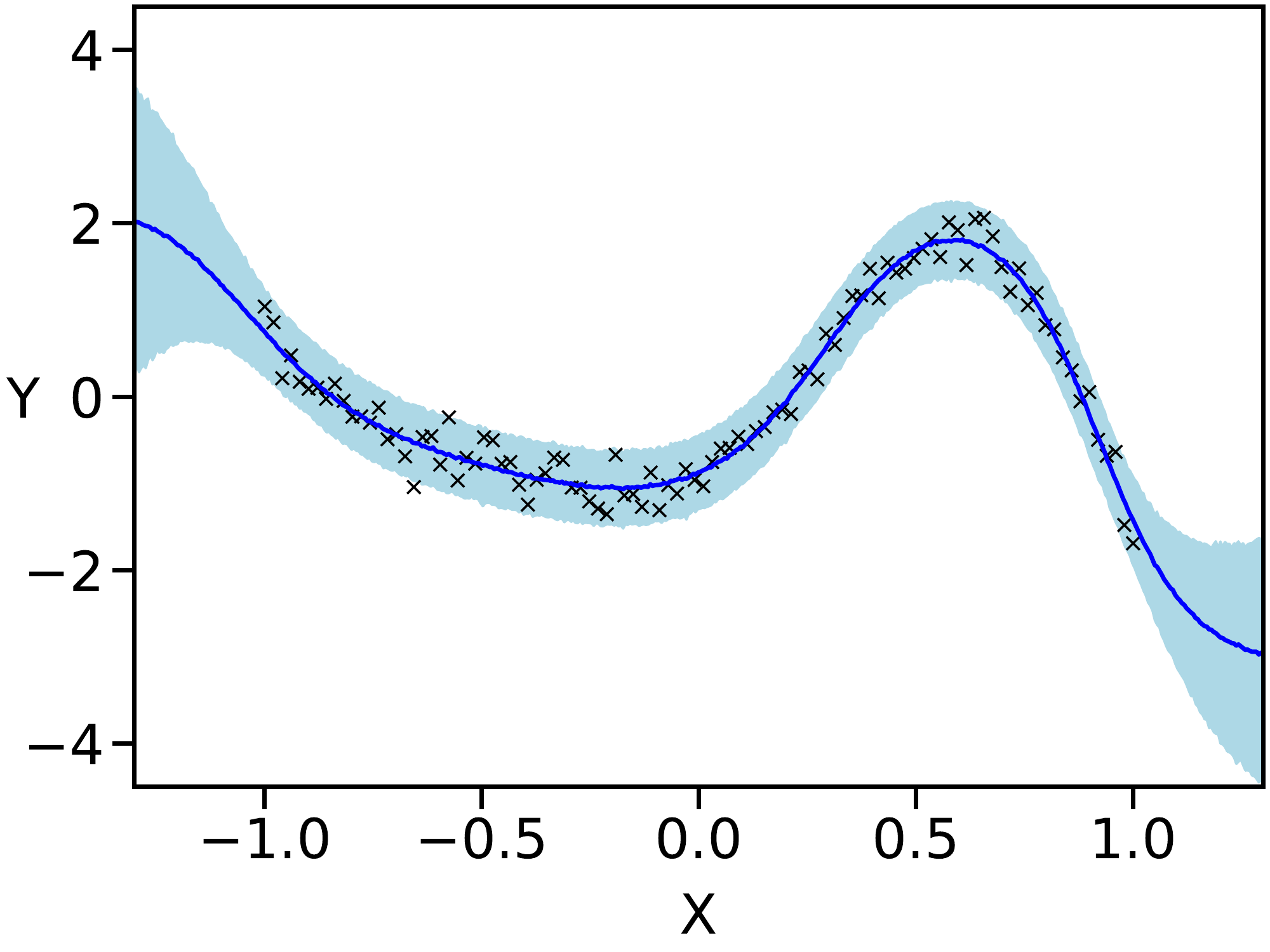}
        \caption{CIQP, $n=11$}
    \end{subfigure}
    \begin{subfigure}[b]{0.24\linewidth}
        \includegraphics[width=\textwidth]{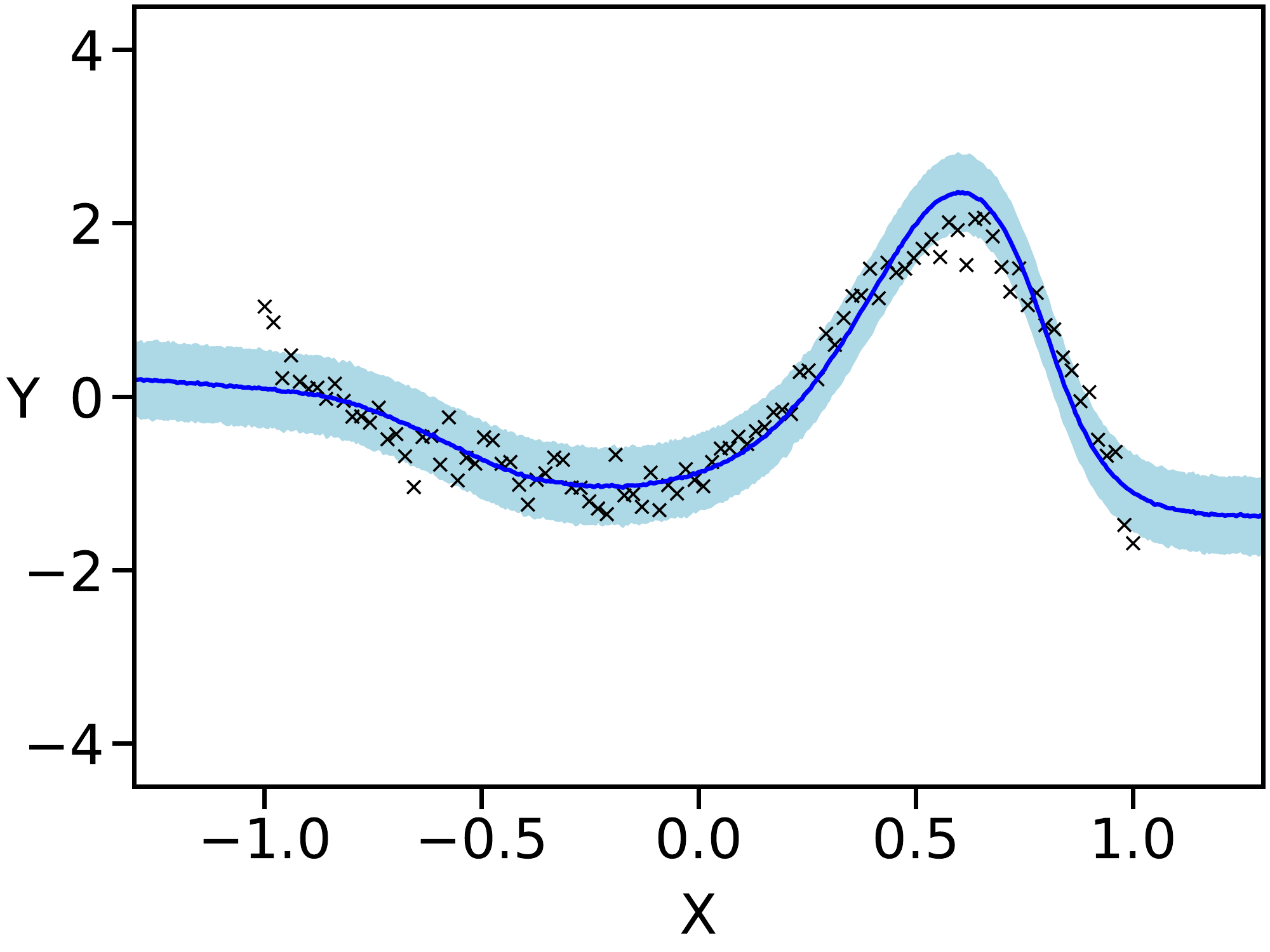}
        \caption{QICP, $n=11$}
    \end{subfigure}
    \begin{subfigure}[b]{0.24\linewidth}
        \includegraphics[width=\textwidth]{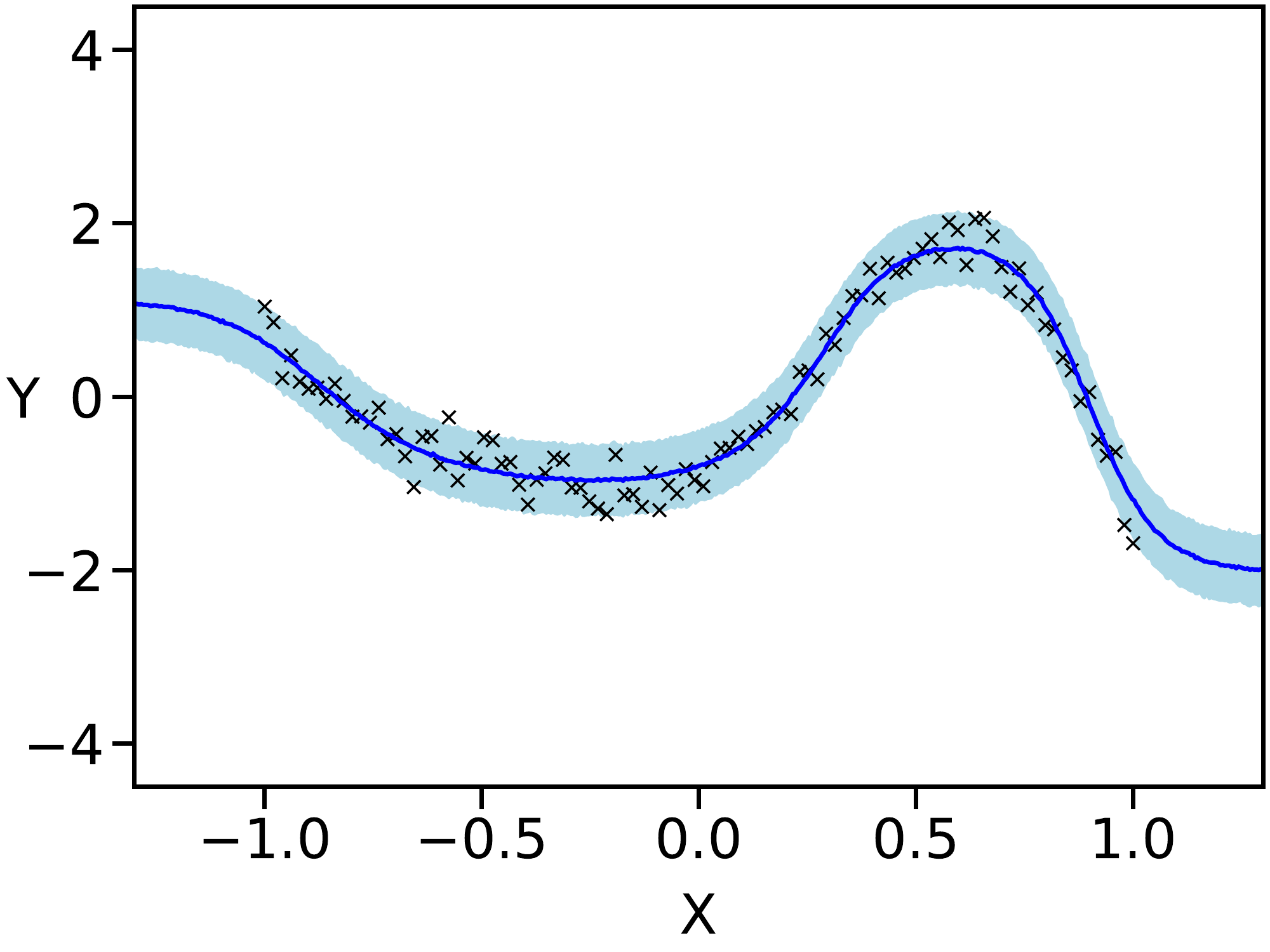}
        \caption{QIQP, $n=11$}
    \end{subfigure}
    \vskip 1.0em    
    \hspace{11.3em}
    \begin{subfigure}[b]{0.24\linewidth}
        \includegraphics[width=\textwidth]{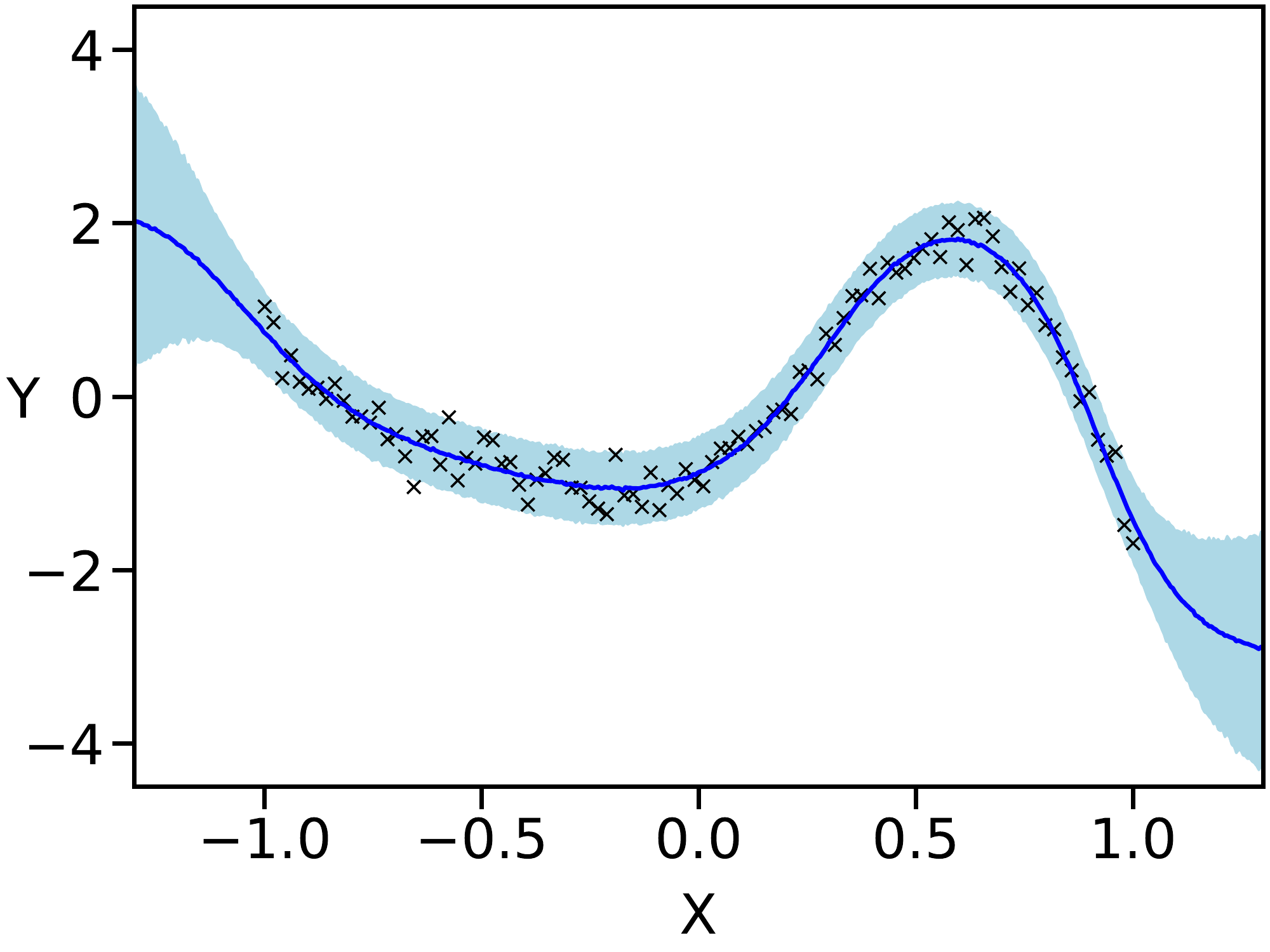}
        \caption{CIQP, $n=13$}
    \end{subfigure}
    \begin{subfigure}[b]{0.24\linewidth}
        \includegraphics[width=\textwidth]{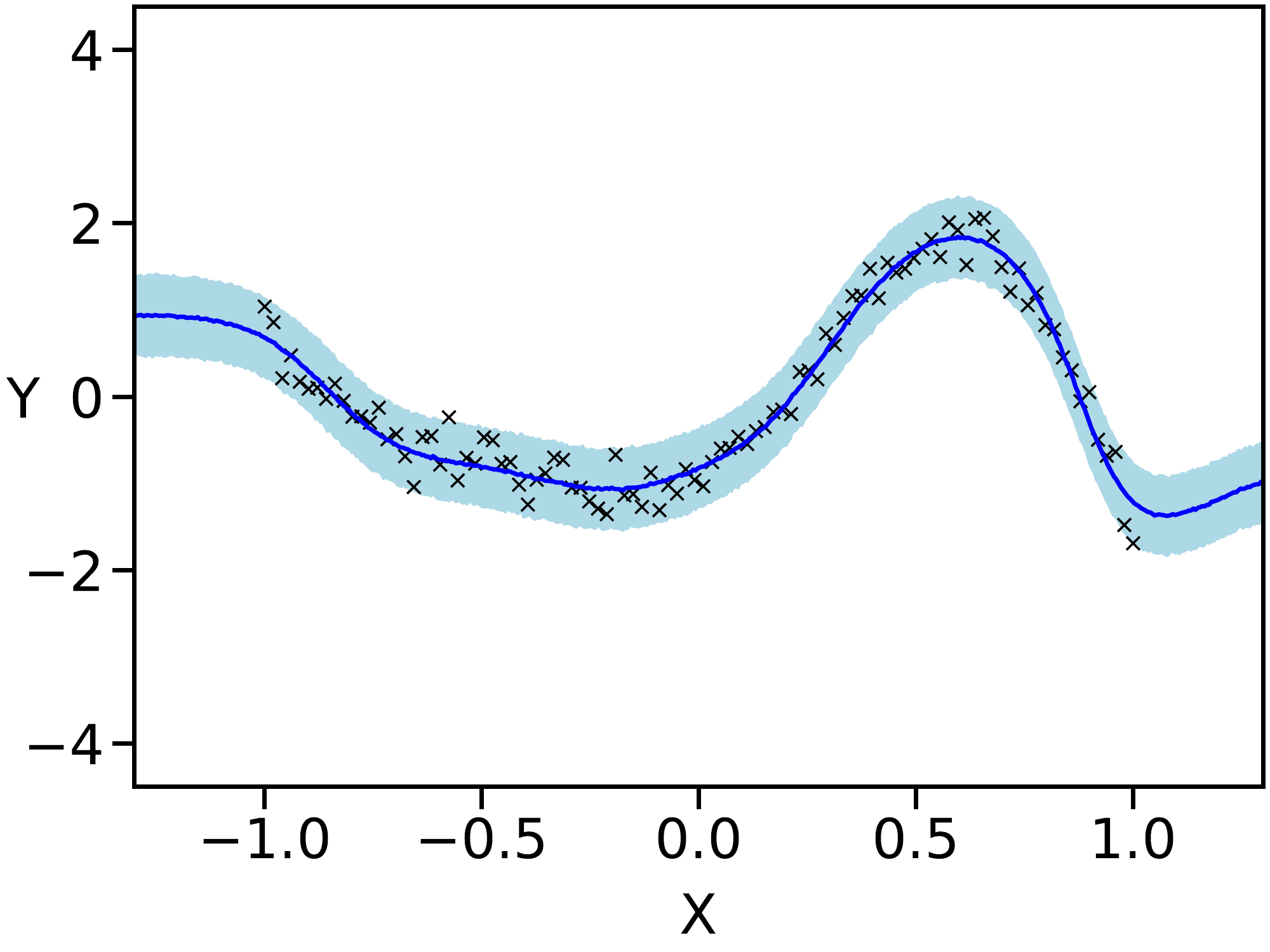}
        \caption{QICP, $n=13$}
    \end{subfigure}
    \begin{subfigure}[b]{0.24\linewidth}
        \includegraphics[width=\textwidth]{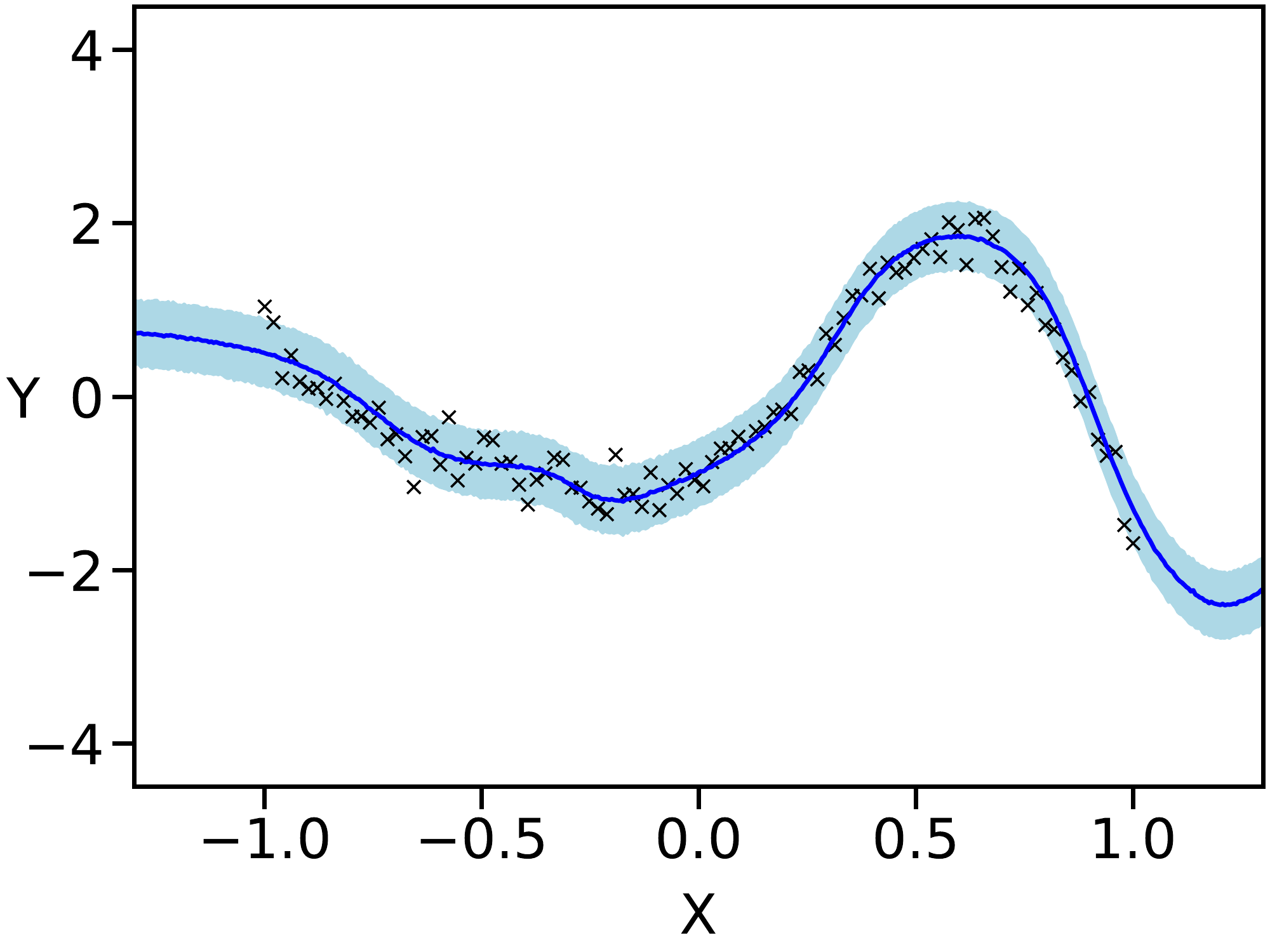}
        \caption{QIQP, $n=13$}
    \end{subfigure}
    
\caption{Additional Results for Linear Regression with \ac{bnn}: \emph{C} and \emph{Q} stand for \emph{Classical} and \emph{Quantum} respectively. \emph{I} and \emph{P} stand for \emph{Inference} and \emph{Prediction}. The Figure shows the expected increase in accuracy for higher qubit numbers $n$.}
\label{fig:lr_add}
\end{center}
\vskip -0.2in
\end{figure}

\begin{figure}[t]
\vskip 0.2in
\begin{center}
    \begin{subfigure}[b]{0.24\linewidth}
        \includegraphics[width=\textwidth]{figs/binary_classification/mean/cicp_nh_5_seed_0_fr.pdf}
        \caption{CICP (Reference)}
    \end{subfigure}
    \begin{subfigure}[b]{0.24\linewidth}
        \includegraphics[width=\textwidth]{figs/binary_classification/mean/ciqp_n_5_nh_5_seed_0_fr.pdf}
        \caption{CIQP, $n=5$}
    \end{subfigure}
    \begin{subfigure}[b]{0.24\linewidth}
        \includegraphics[width=\textwidth]{figs/binary_classification/mean/qicp_n_5_nh_5_seed_0_fr.pdf}
        \caption{QICP, $n=5$}
    \end{subfigure}
    \begin{subfigure}[b]{0.24\linewidth}
        \includegraphics[width=\textwidth]{figs/binary_classification/mean/qiqp_n_5_nh_5_seed_0_fr.pdf}
        \caption{QIQP, $n=5$}
    \end{subfigure}
    \vskip 1.0em
    \begin{subfigure}[b]{0.24\linewidth}
        \includegraphics[width=\textwidth]{figs/binary_classification/legend_mean.pdf}
        \caption{Legend}
    \end{subfigure}
    \begin{subfigure}[b]{0.24\linewidth}
        \includegraphics[width=\textwidth]{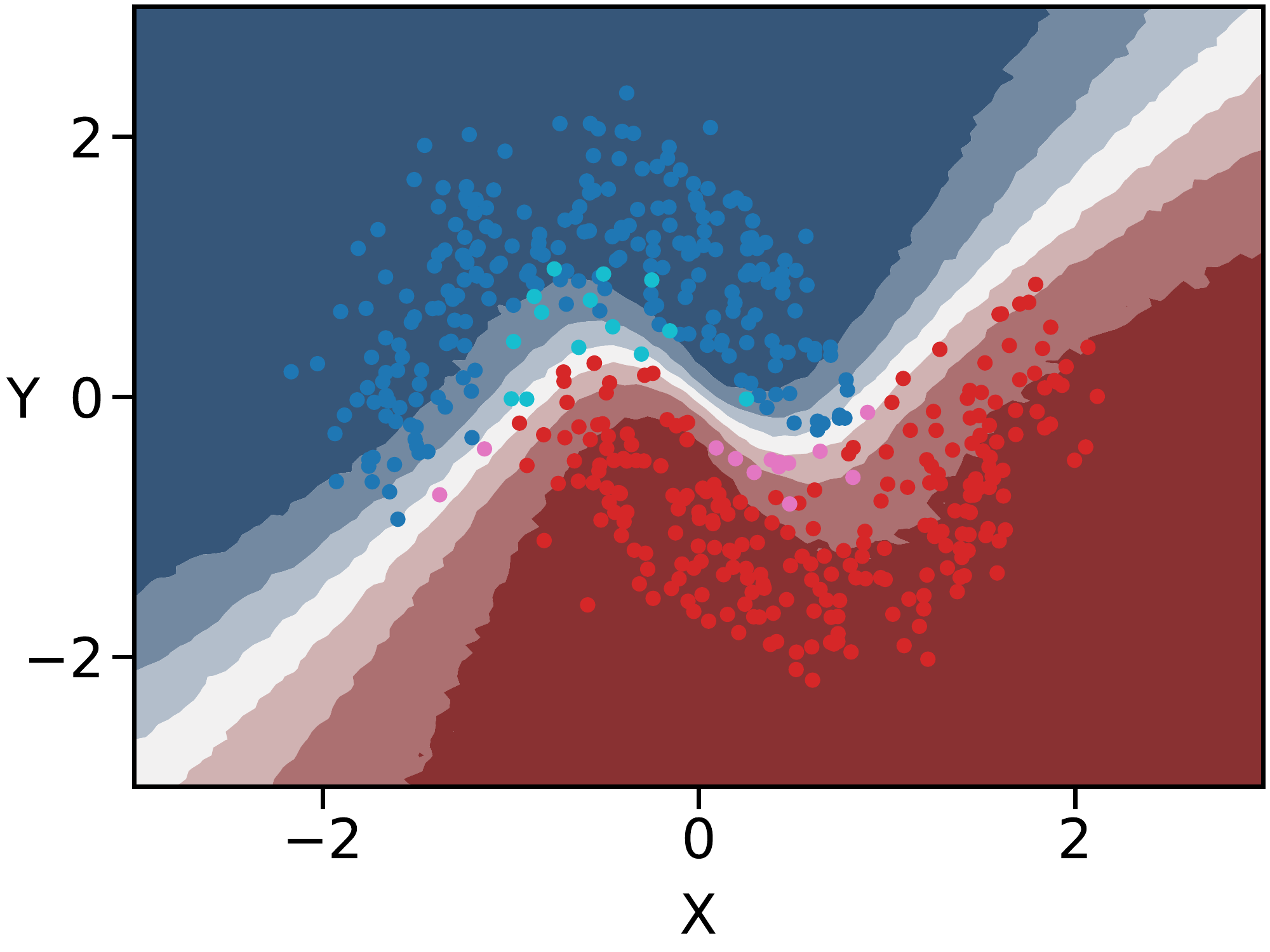}
        \caption{CIQP, $n=7$}
    \end{subfigure}
    \begin{subfigure}[b]{0.24\linewidth}
        \includegraphics[width=\textwidth]{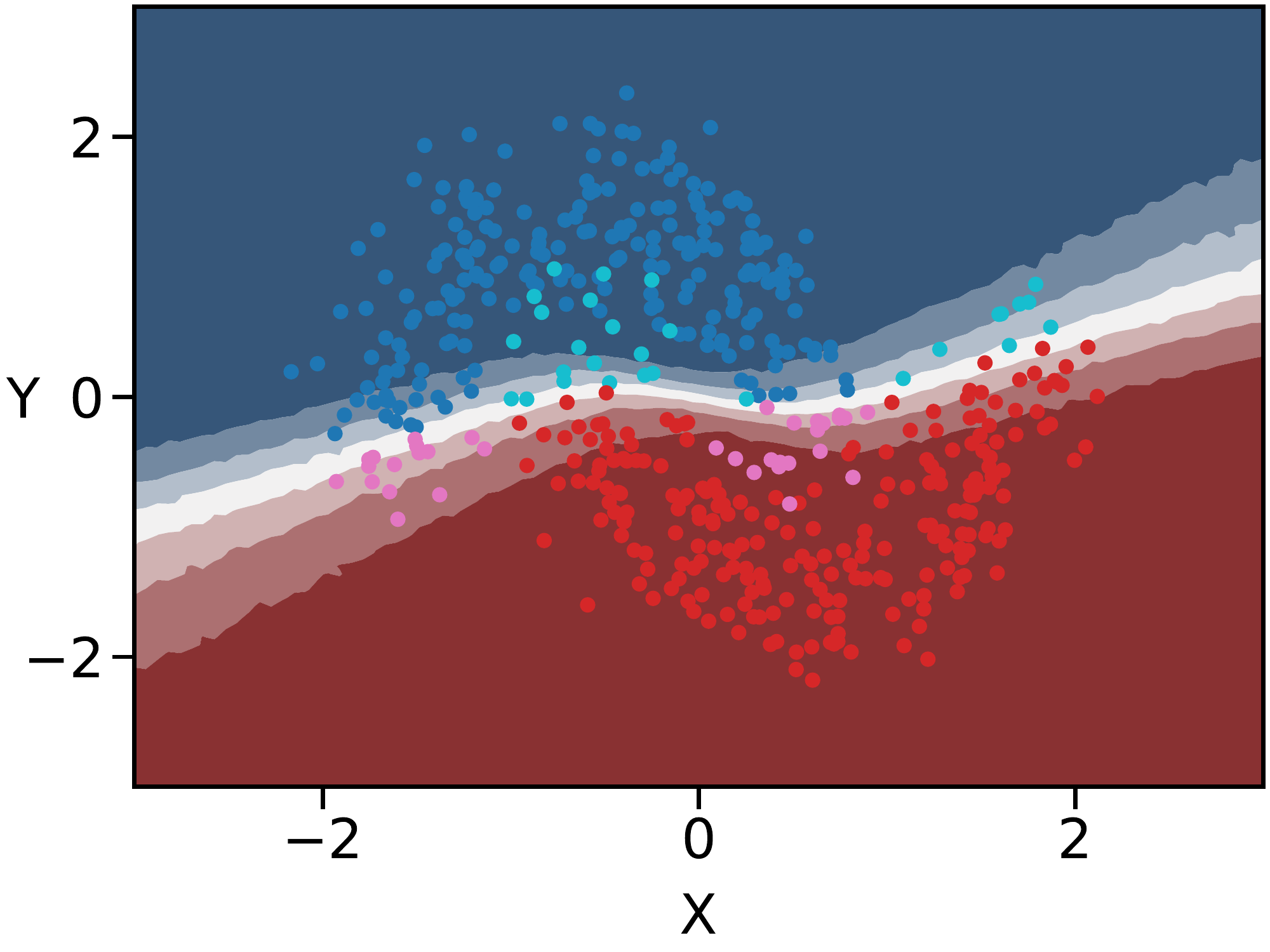}
        \caption{QICP, $n=7$}
    \end{subfigure}
    \begin{subfigure}[b]{0.24\linewidth}
        \includegraphics[width=\textwidth]{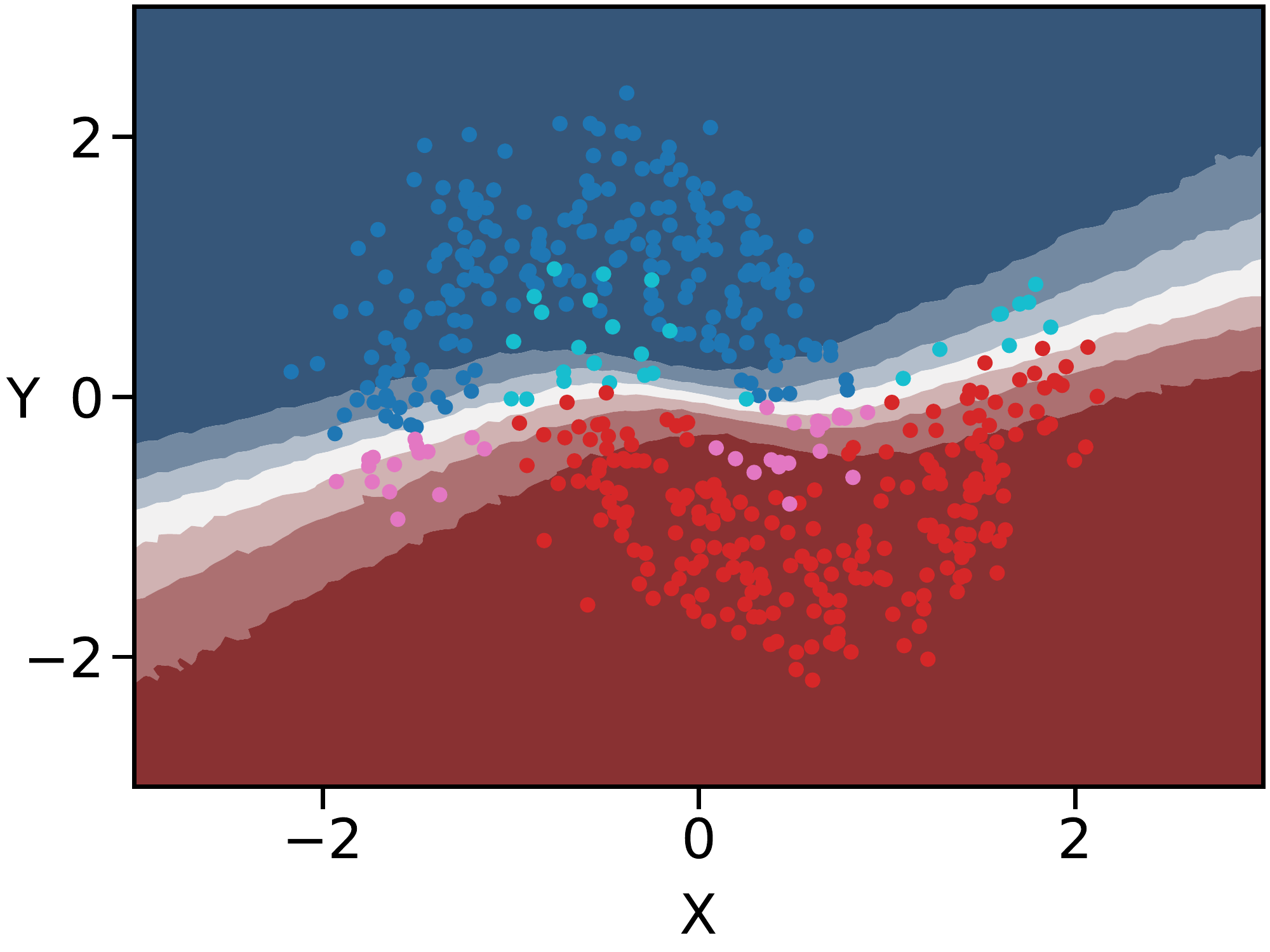}
        \caption{QIQP, $n=7$}
    \end{subfigure}
    \vskip 1.0em    
    \hspace{11.3em}
    \begin{subfigure}[b]{0.24\linewidth}
        \includegraphics[width=\textwidth]{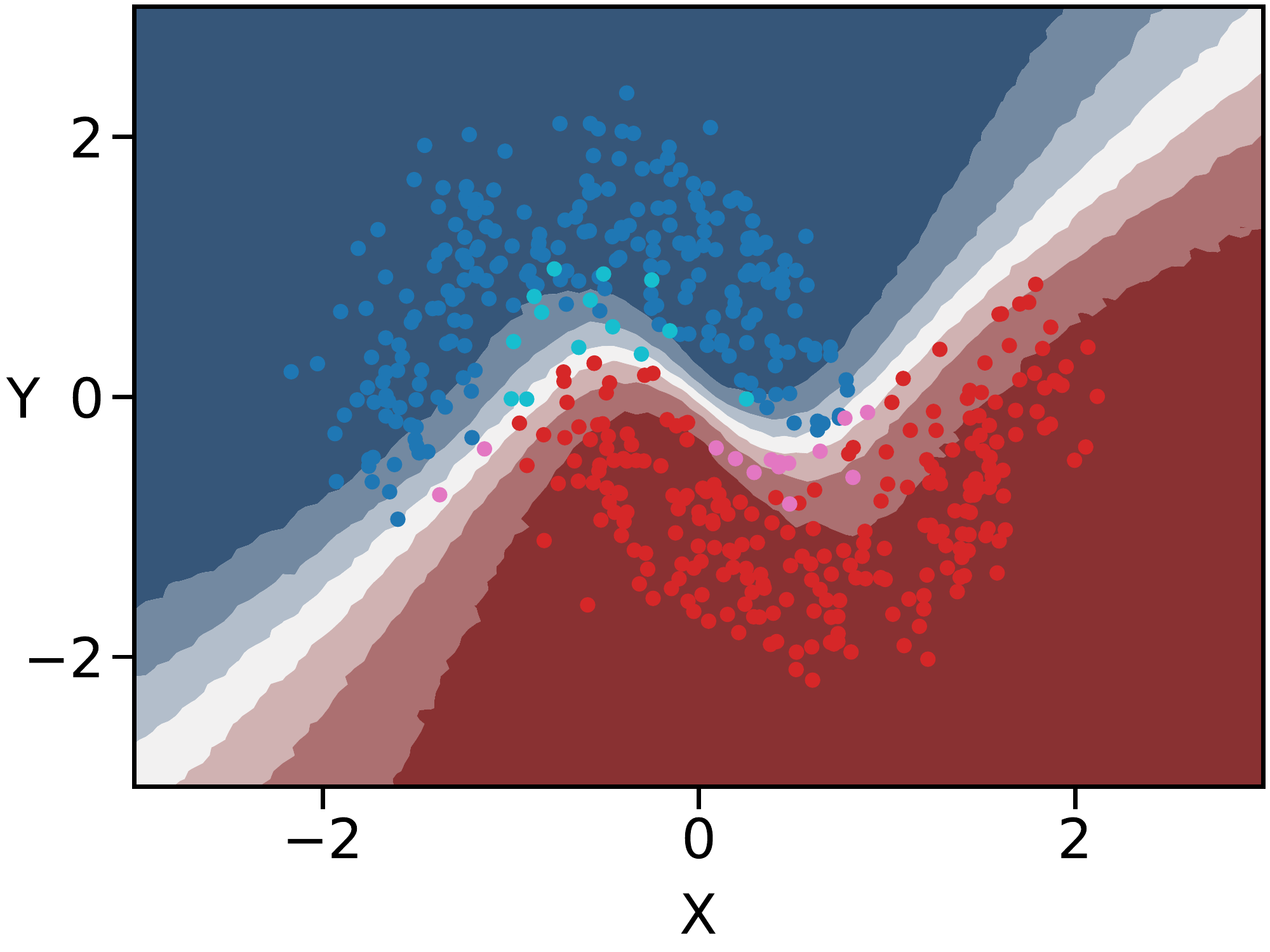}
        \caption{CIQP, $n=9$}
    \end{subfigure}
    \begin{subfigure}[b]{0.24\linewidth}
        \includegraphics[width=\textwidth]{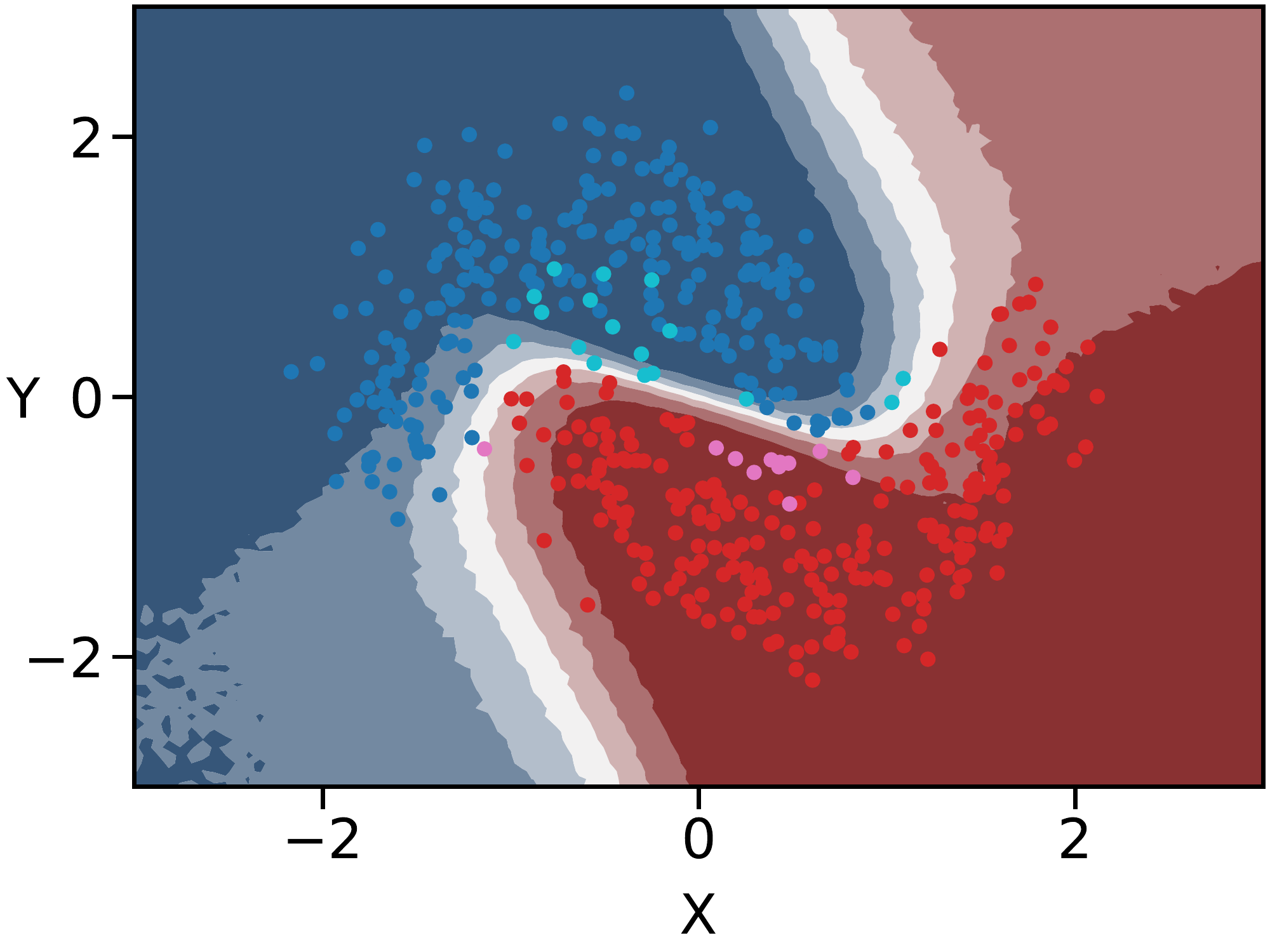}
        \caption{QICP, $n=9$}
    \end{subfigure}
    \begin{subfigure}[b]{0.24\linewidth}
        \includegraphics[width=\textwidth]{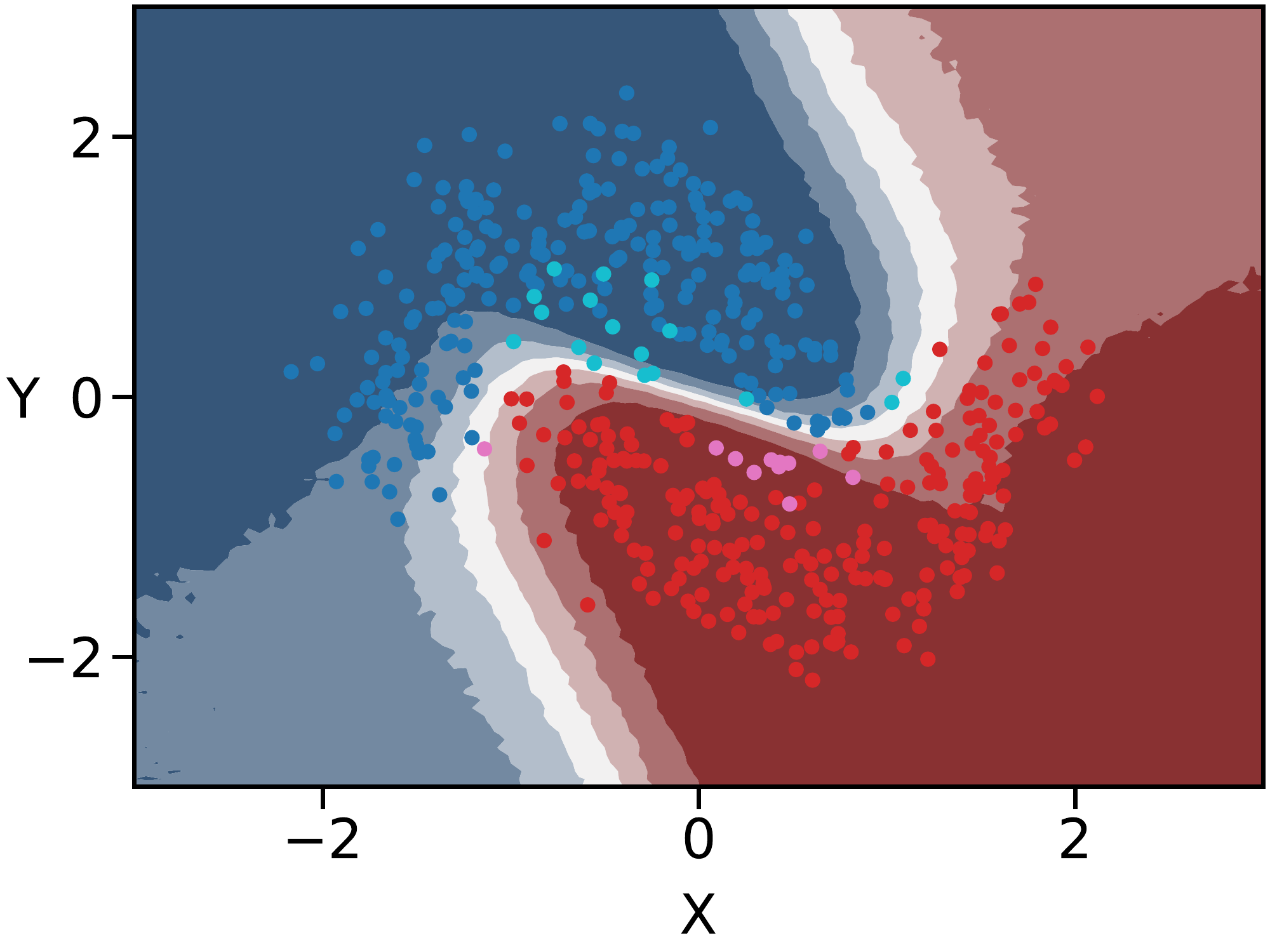}
        \caption{QIQP, $n=9$}
    \end{subfigure}
    \vskip 1.0em    
    \hspace{11.3em}
    \begin{subfigure}[b]{0.24\linewidth}
        \includegraphics[width=\textwidth]{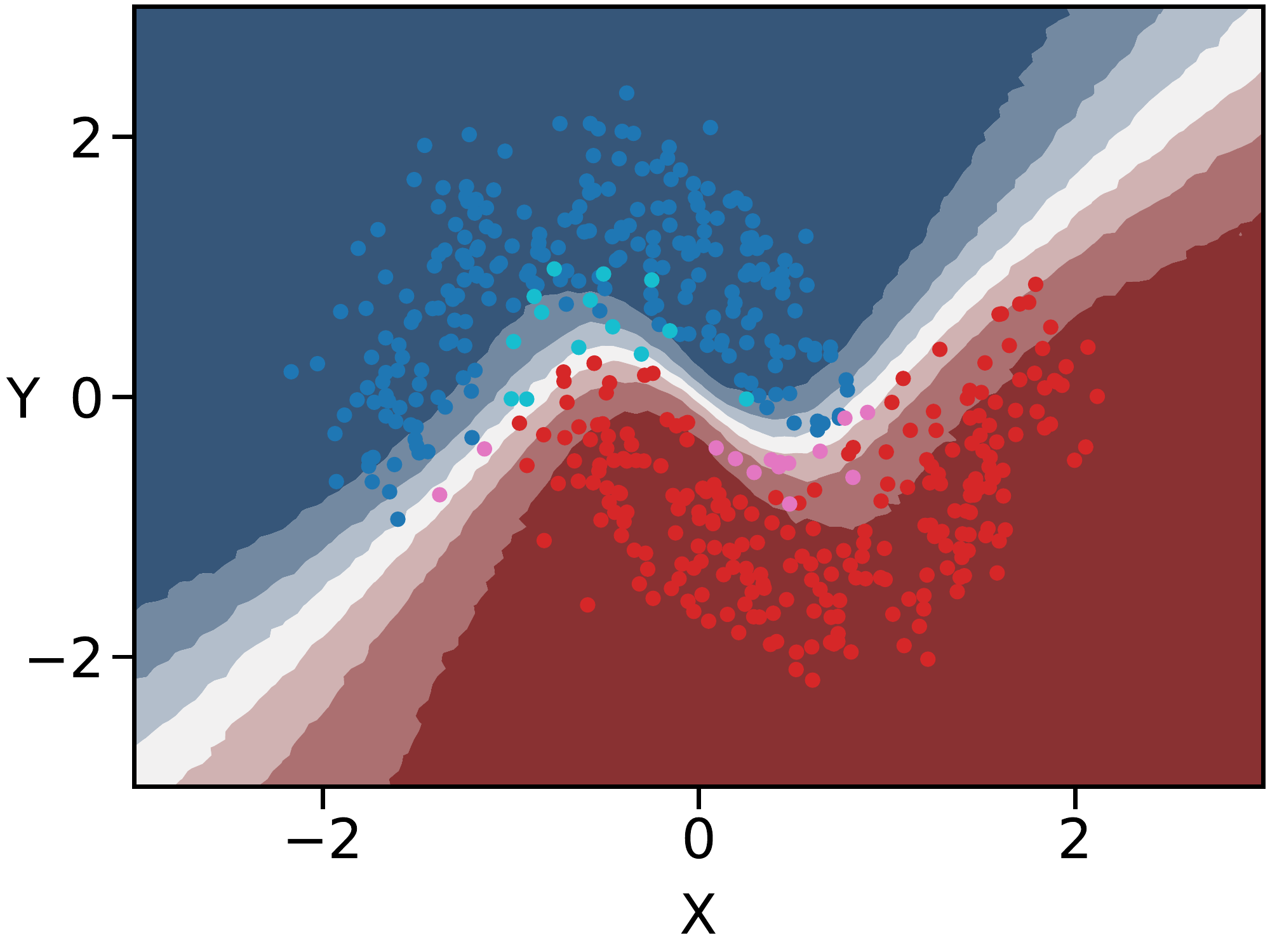}
        \caption{CIQP, $n=11$}
    \end{subfigure}
    \begin{subfigure}[b]{0.24\linewidth}
        \includegraphics[width=\textwidth]{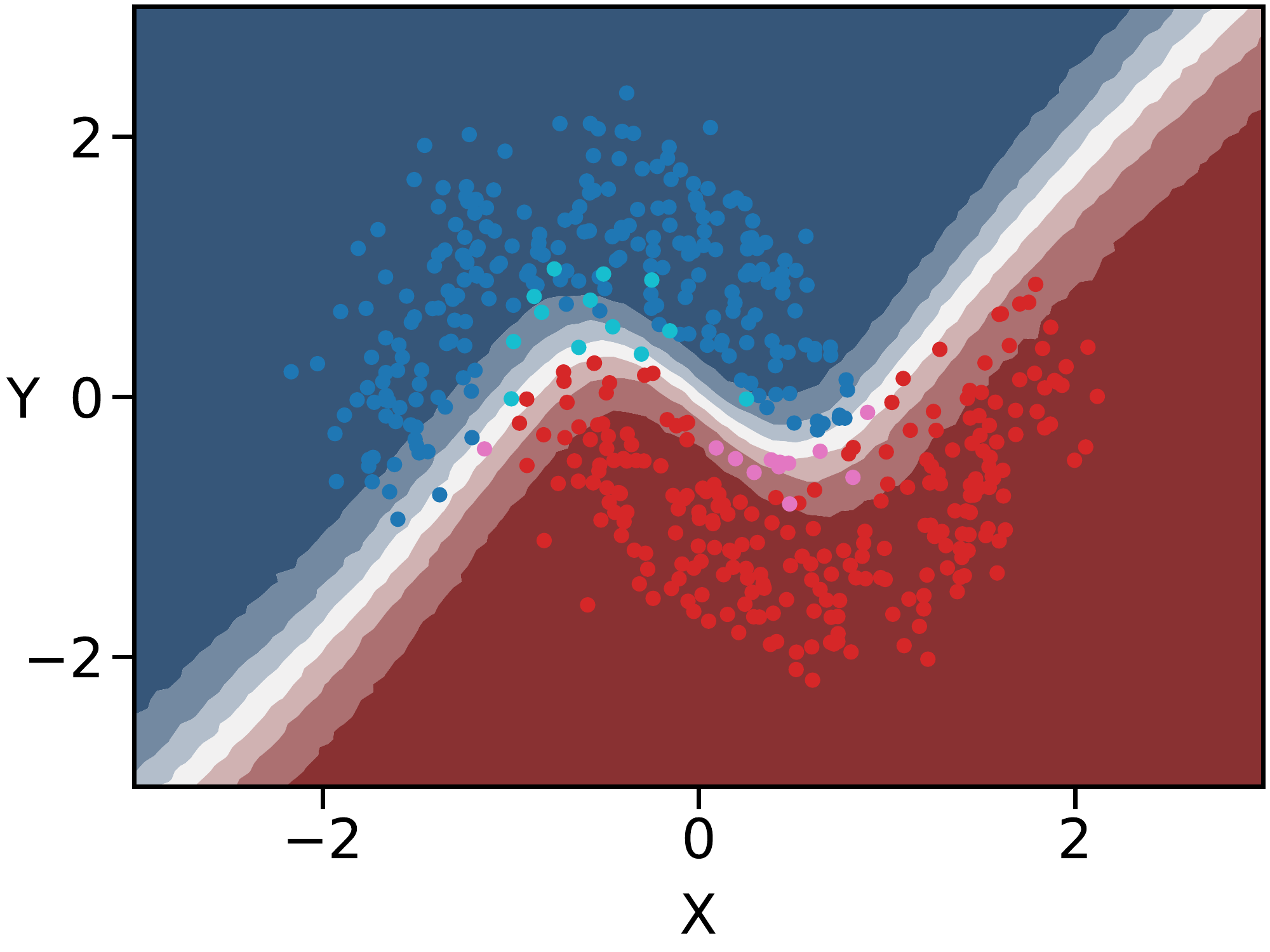}
        \caption{QICP, $n=11$}
    \end{subfigure}
    \begin{subfigure}[b]{0.24\linewidth}
        \includegraphics[width=\textwidth]{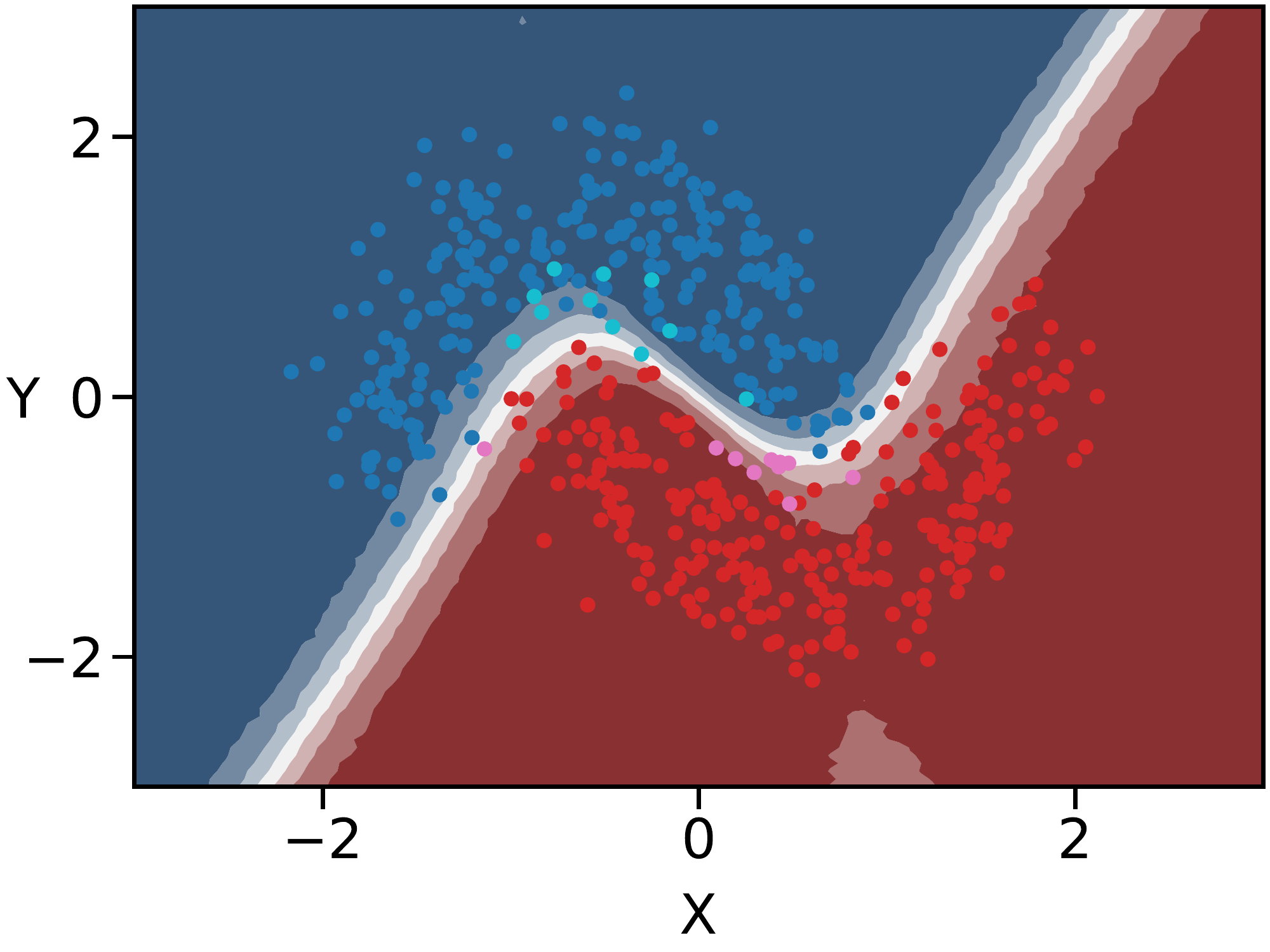}
        \caption{QIQP, $n=11$}
    \end{subfigure}
    \vskip 1.0em    
    \hspace{11.3em}
    \begin{subfigure}[b]{0.24\linewidth}
        \includegraphics[width=\textwidth]{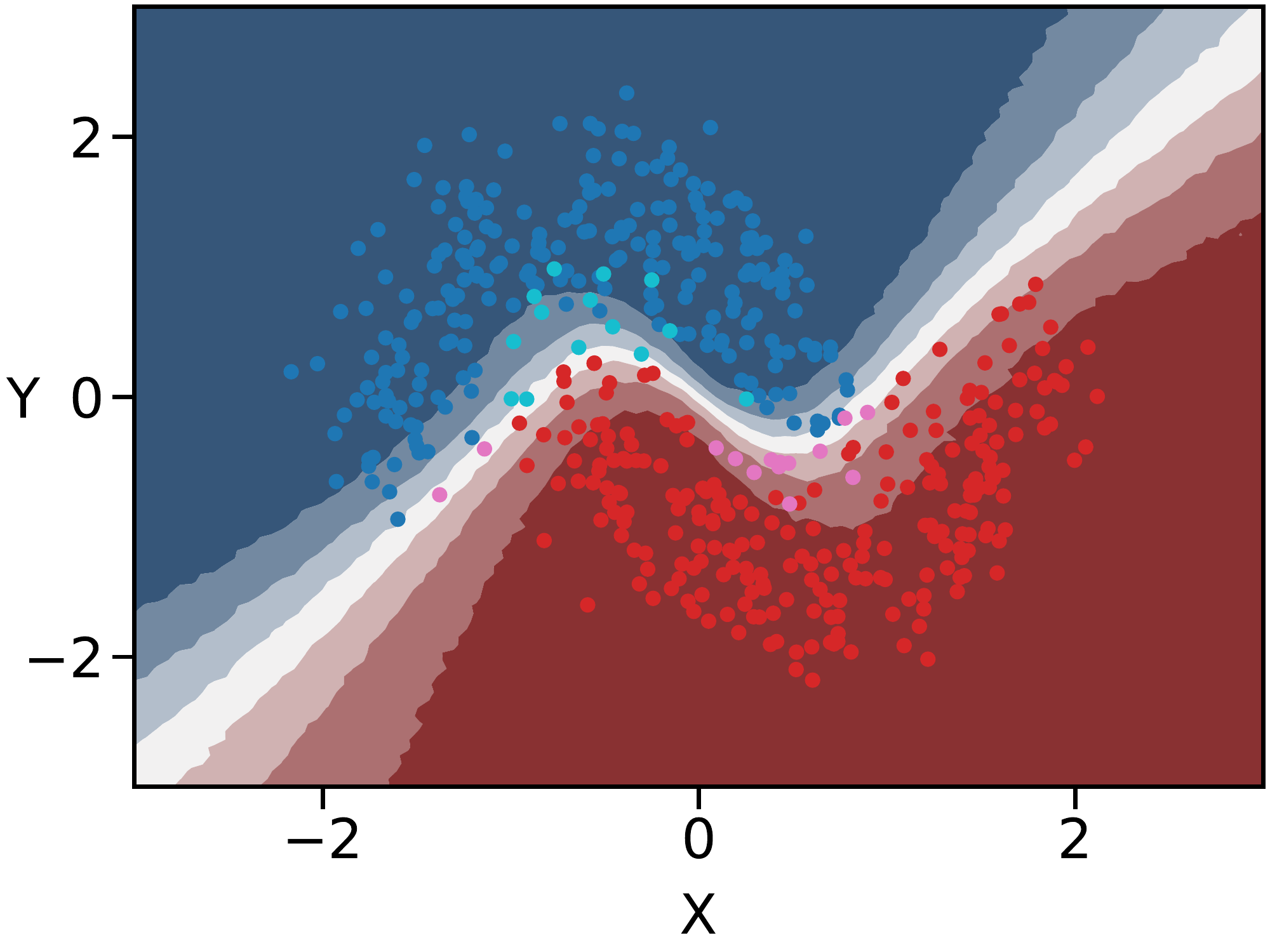}
        \caption{CIQP, $n=13$}
    \end{subfigure}
    \begin{subfigure}[b]{0.24\linewidth}
        \includegraphics[width=\textwidth]{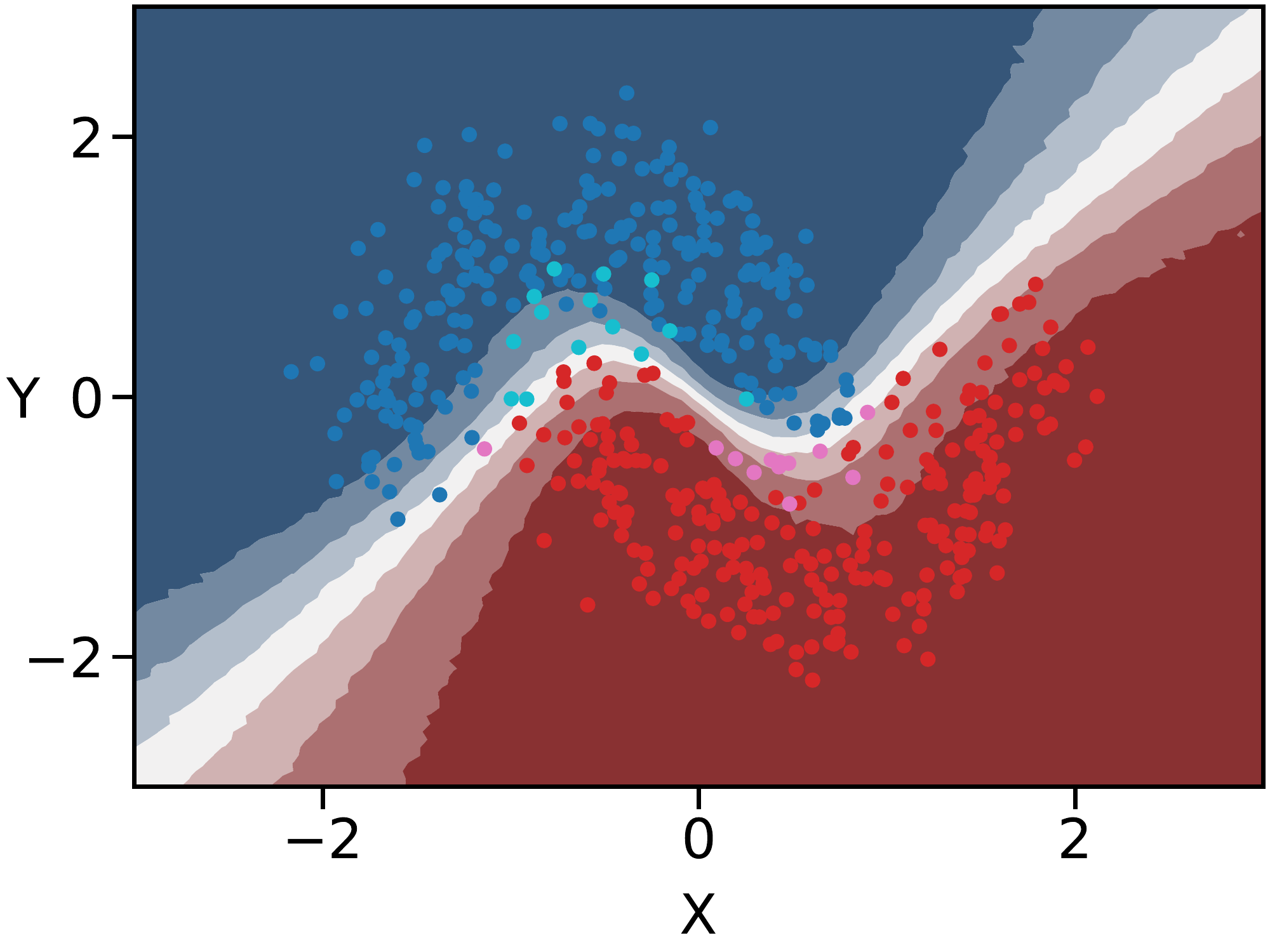}
        \caption{QICP, $n=13$}
    \end{subfigure}
    \begin{subfigure}[b]{0.24\linewidth}
        \includegraphics[width=\textwidth]{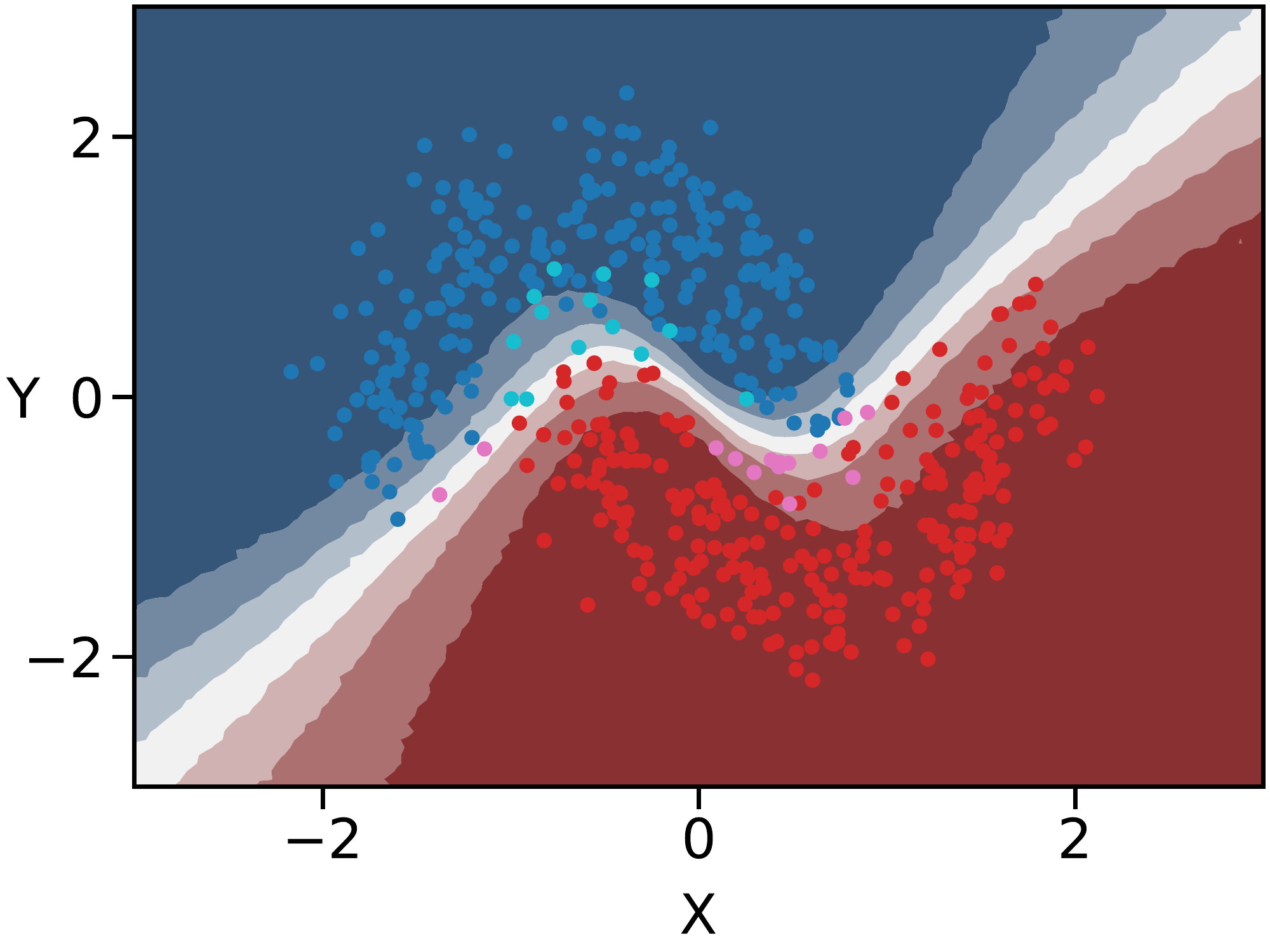}
        \caption{QIQP, $n=13$}
    \end{subfigure}
    
\caption{Additional Results for Binary Classification with \ac{bnn}: \emph{C} and \emph{Q} stand for \emph{Classical} and \emph{Quantum} respectively. \emph{I} and \emph{P} stand for \emph{Inference} and \emph{Prediction}. The Figure shows the expected increase in accuracy for higher qubit numbers $n$.}
\label{fig:bc_mean_add}
\end{center}
\vskip -0.2in
\end{figure}

\begin{figure}[t]
\vskip 0.2in
\begin{center}
\begin{subfigure}[b]{0.24\linewidth}
        \includegraphics[width=\textwidth]{figs/binary_classification/std/cicp_nh_5_seed_0_fr.pdf}
        \caption{CICP (Reference)}
    \end{subfigure}
    \begin{subfigure}[b]{0.24\linewidth}
        \includegraphics[width=\textwidth]{figs/binary_classification/std/ciqp_n_5_nh_5_seed_0_fr.pdf}
        \caption{CIQP, $n=5$}
    \end{subfigure}
    \begin{subfigure}[b]{0.24\linewidth}
        \includegraphics[width=\textwidth]{figs/binary_classification/std/qicp_n_5_nh_5_seed_0_fr.pdf}
        \caption{QICP, $n=5$}
    \end{subfigure}
    \begin{subfigure}[b]{0.24\linewidth}
        \includegraphics[width=\textwidth]{figs/binary_classification/std/qiqp_n_5_nh_5_seed_0_fr.pdf}
        \caption{QIQP, $n=5$}
    \end{subfigure}
    \vskip 1.0em
    \begin{subfigure}[b]{0.24\linewidth}
        \includegraphics[width=\textwidth]{figs/binary_classification/legend_std.pdf}
        \caption{Legend}
    \end{subfigure}
    \begin{subfigure}[b]{0.24\linewidth}
        \includegraphics[width=\textwidth]{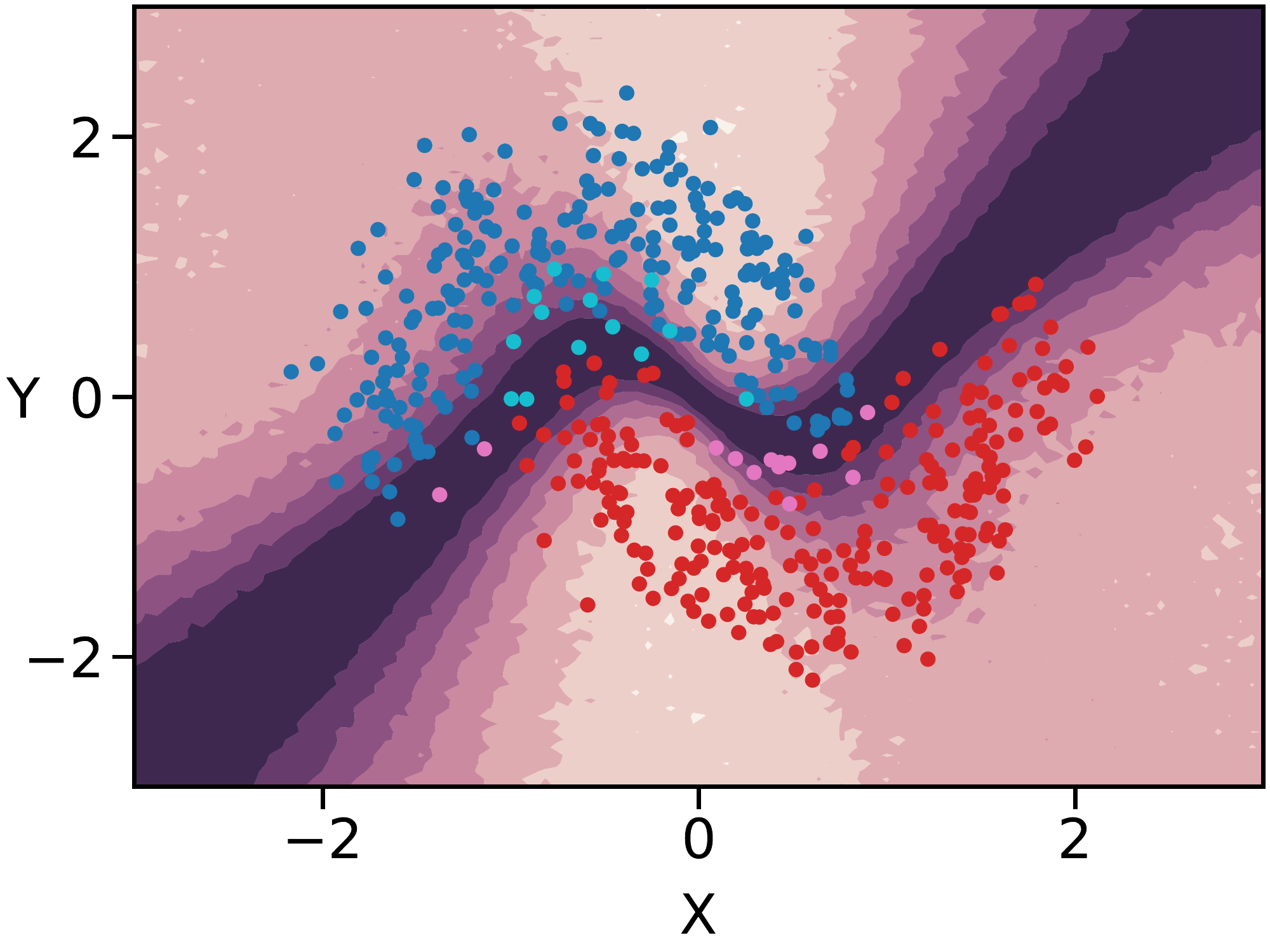}
        \caption{CIQP, $n=7$}
    \end{subfigure}
    \begin{subfigure}[b]{0.24\linewidth}
        \includegraphics[width=\textwidth]{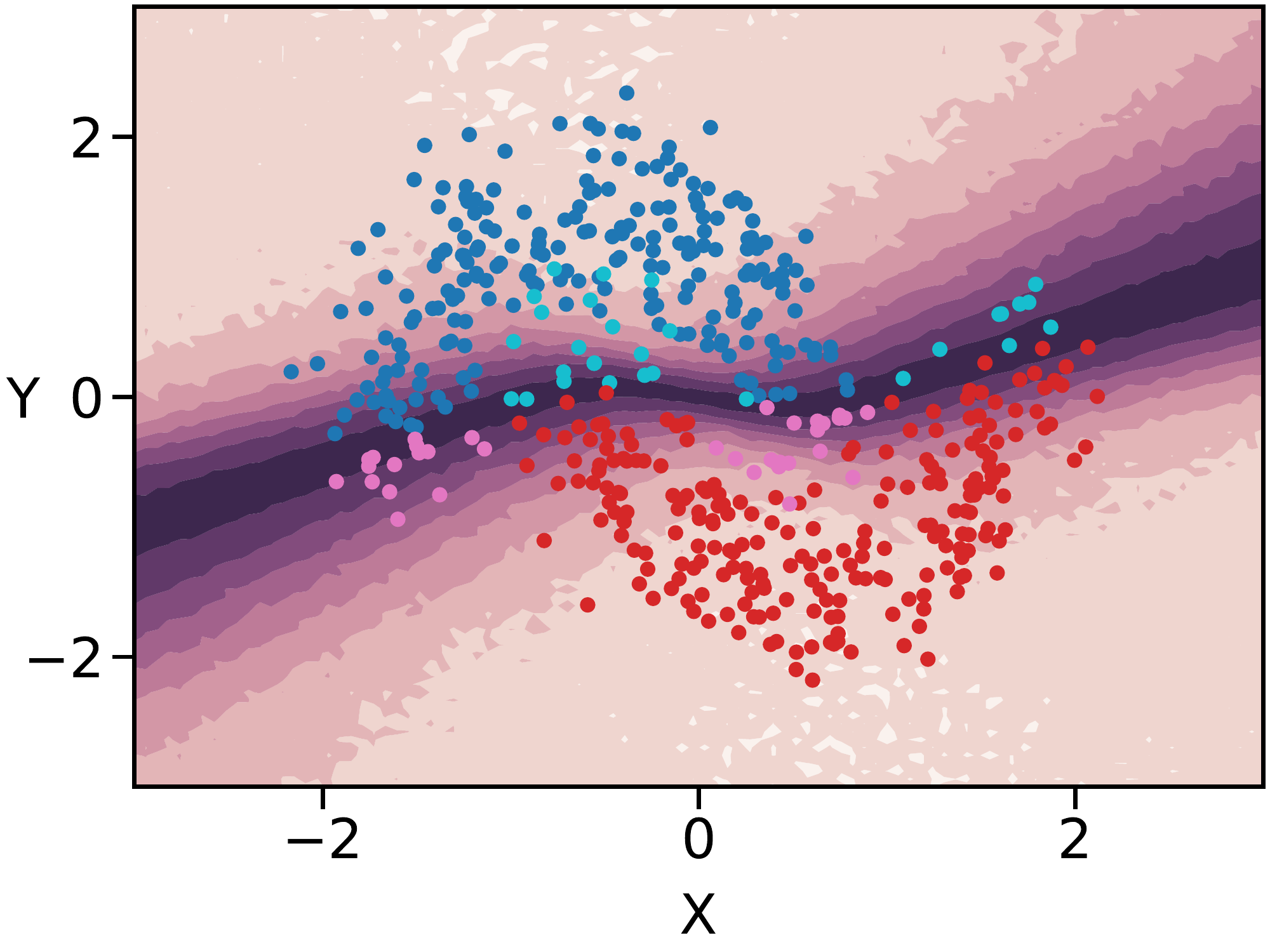}
        \caption{QICP, $n=7$}
    \end{subfigure}
    \begin{subfigure}[b]{0.24\linewidth}
        \includegraphics[width=\textwidth]{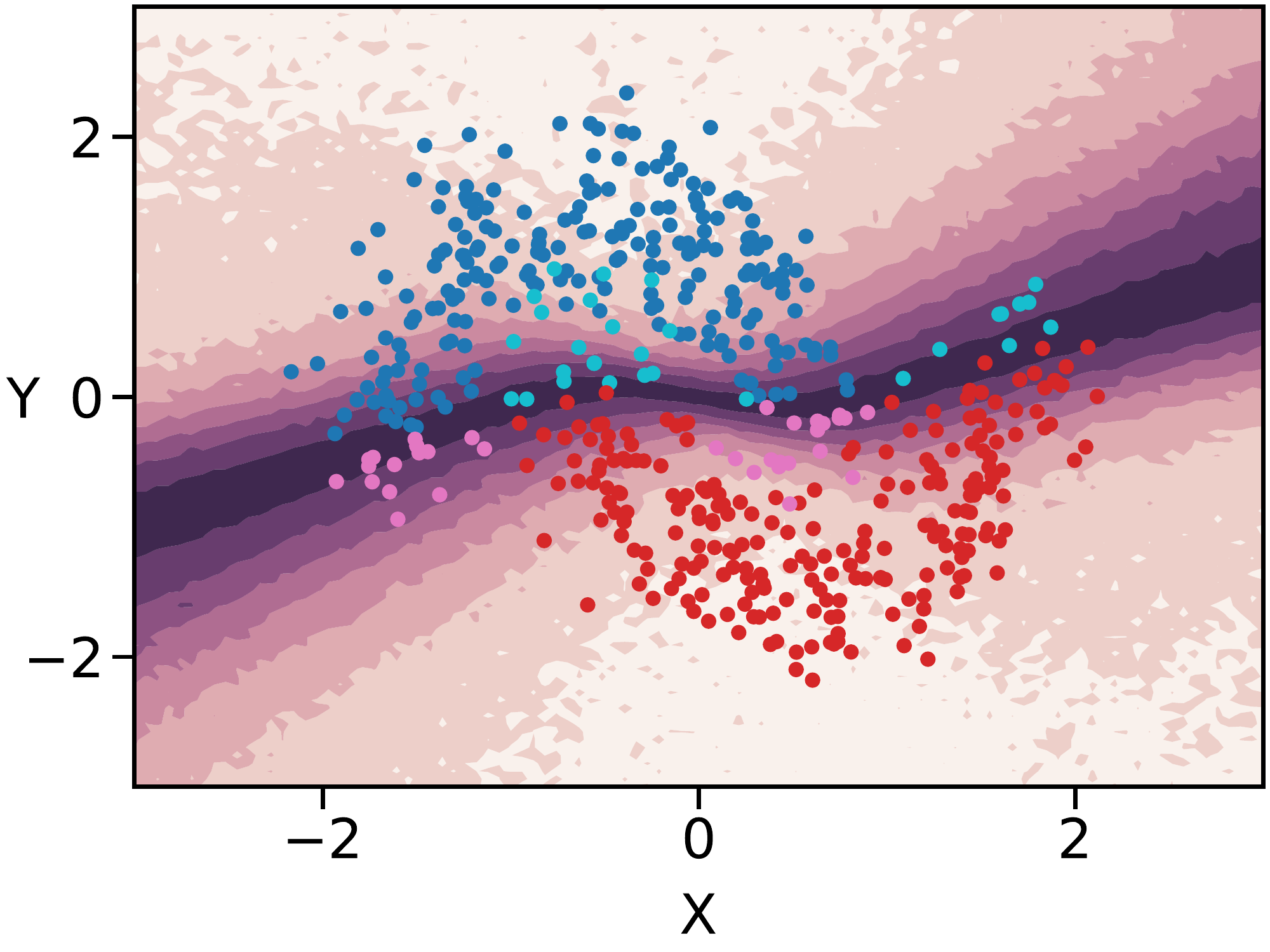}
        \caption{QIQP, $n=7$}
    \end{subfigure}
    \vskip 1.0em    
    \hspace{11.3em}
    \begin{subfigure}[b]{0.24\linewidth}
        \includegraphics[width=\textwidth]{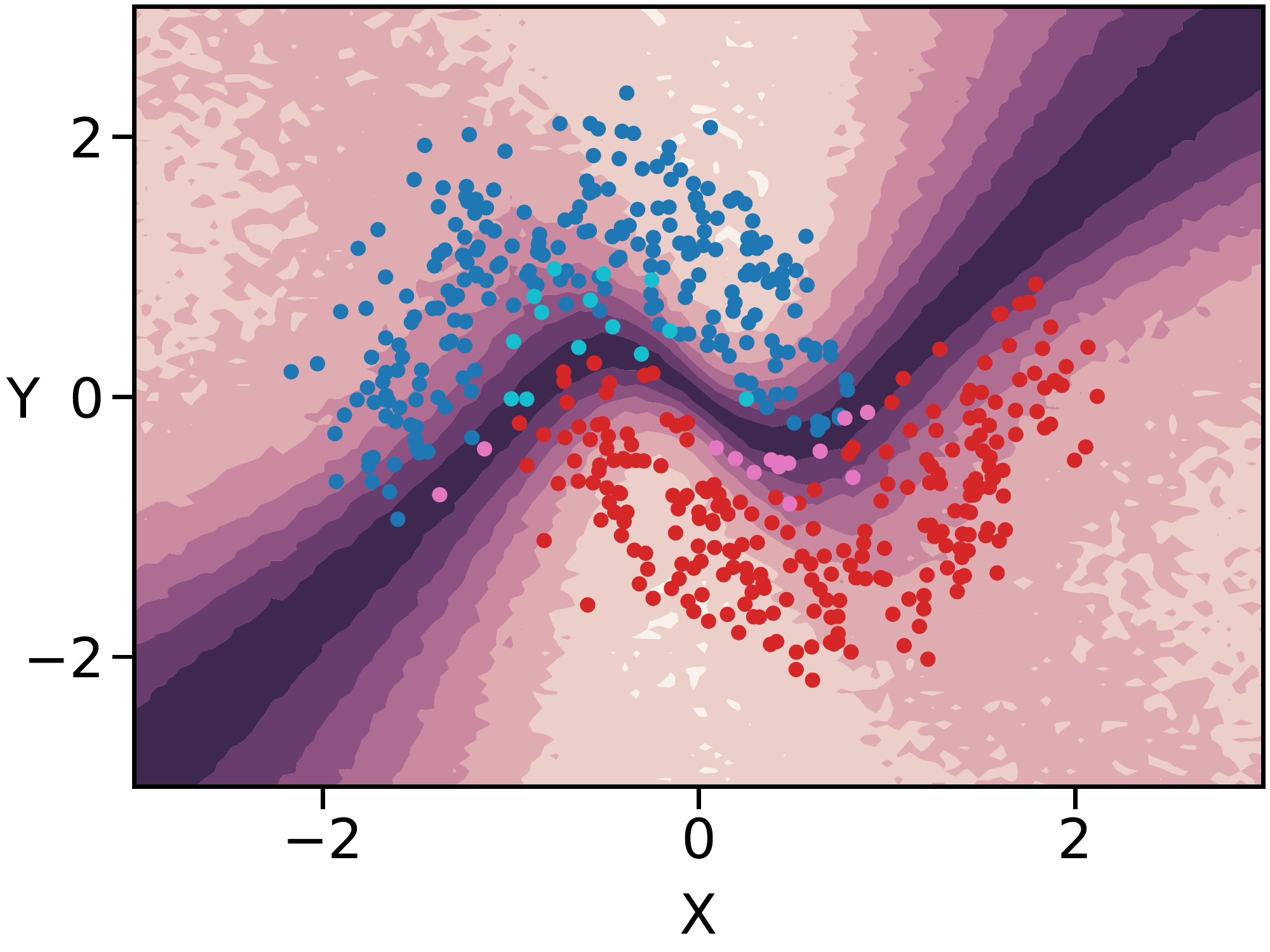}
        \caption{CIQP, $n=9$}
    \end{subfigure}
    \begin{subfigure}[b]{0.24\linewidth}
        \includegraphics[width=\textwidth]{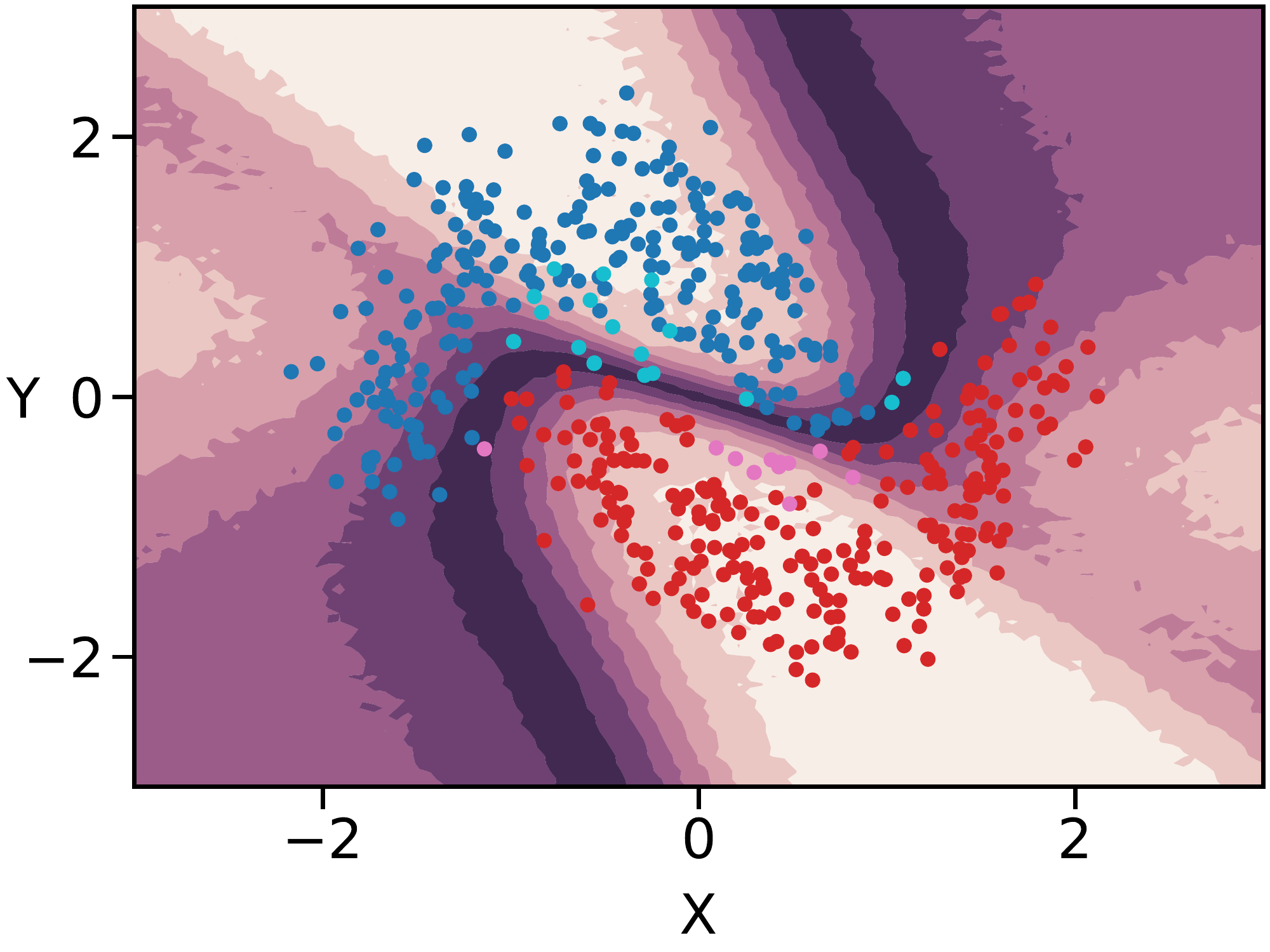}
        \caption{QICP, $n=9$}
    \end{subfigure}
    \begin{subfigure}[b]{0.24\linewidth}
        \includegraphics[width=\textwidth]{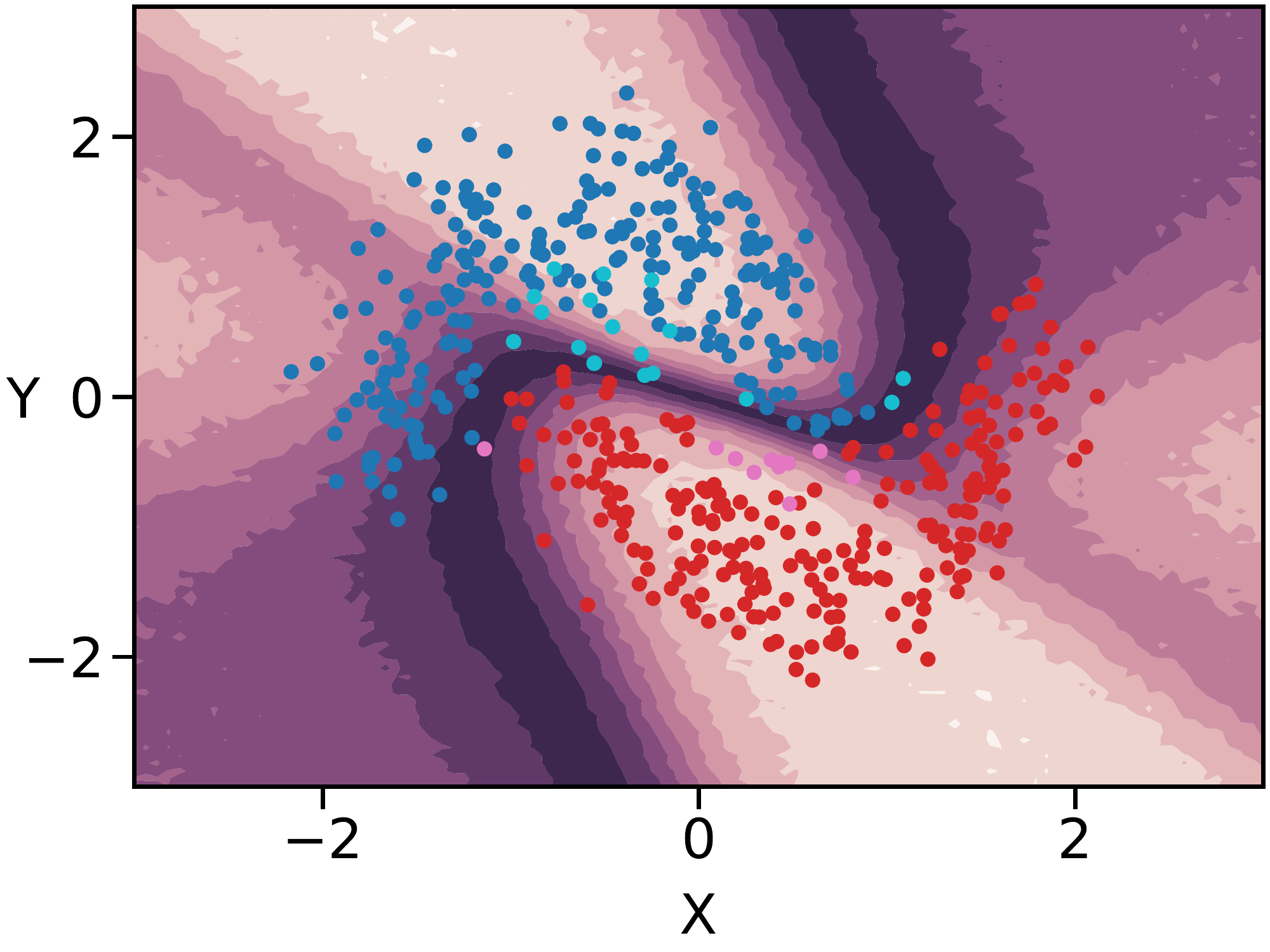}
        \caption{QIQP, $n=9$}
    \end{subfigure}
    \vskip 1.0em    
    \hspace{11.3em}
    \begin{subfigure}[b]{0.24\linewidth}
        \includegraphics[width=\textwidth]{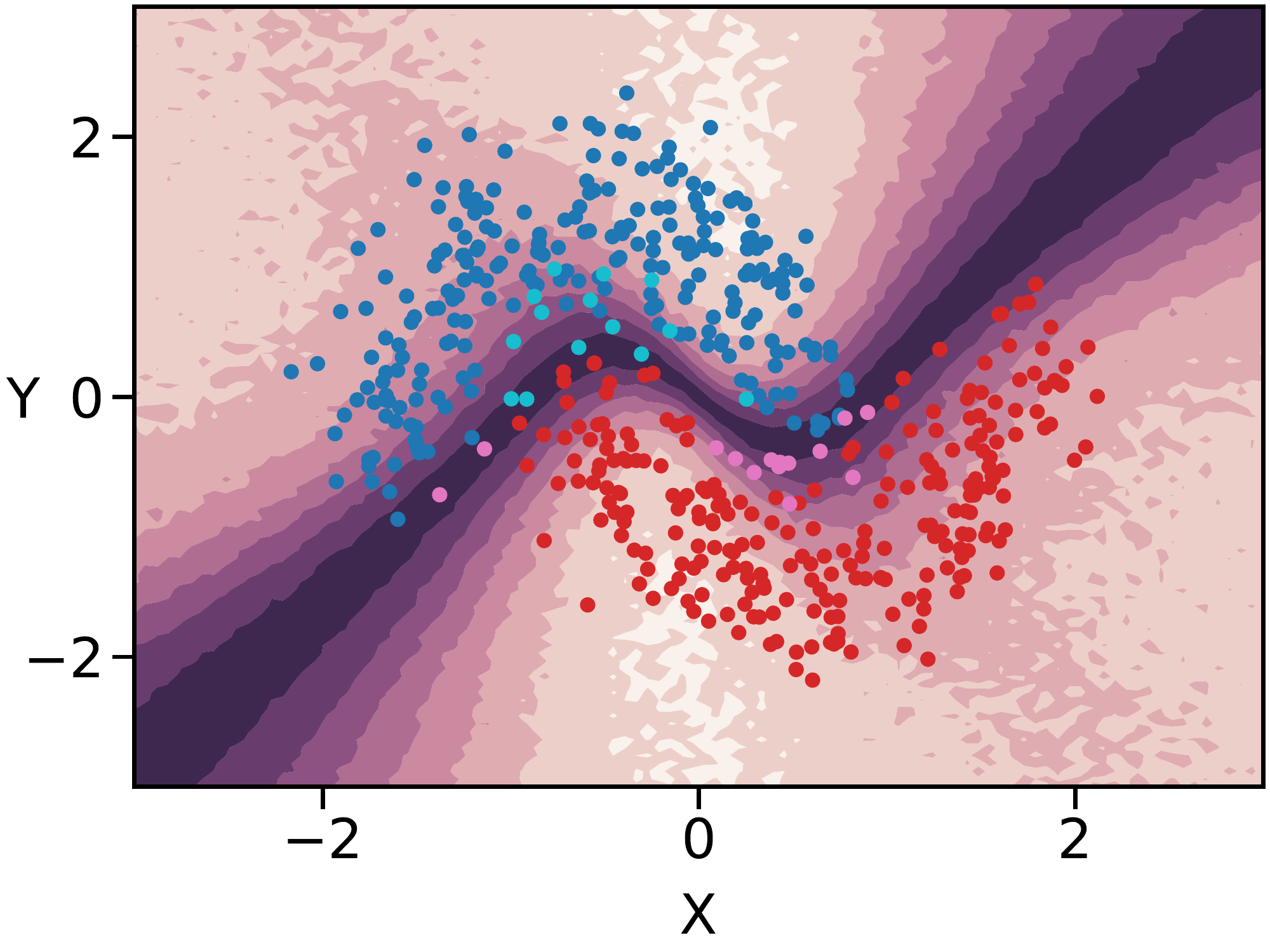}
        \caption{CIQP, $n=11$}
    \end{subfigure}
    \begin{subfigure}[b]{0.24\linewidth}
        \includegraphics[width=\textwidth]{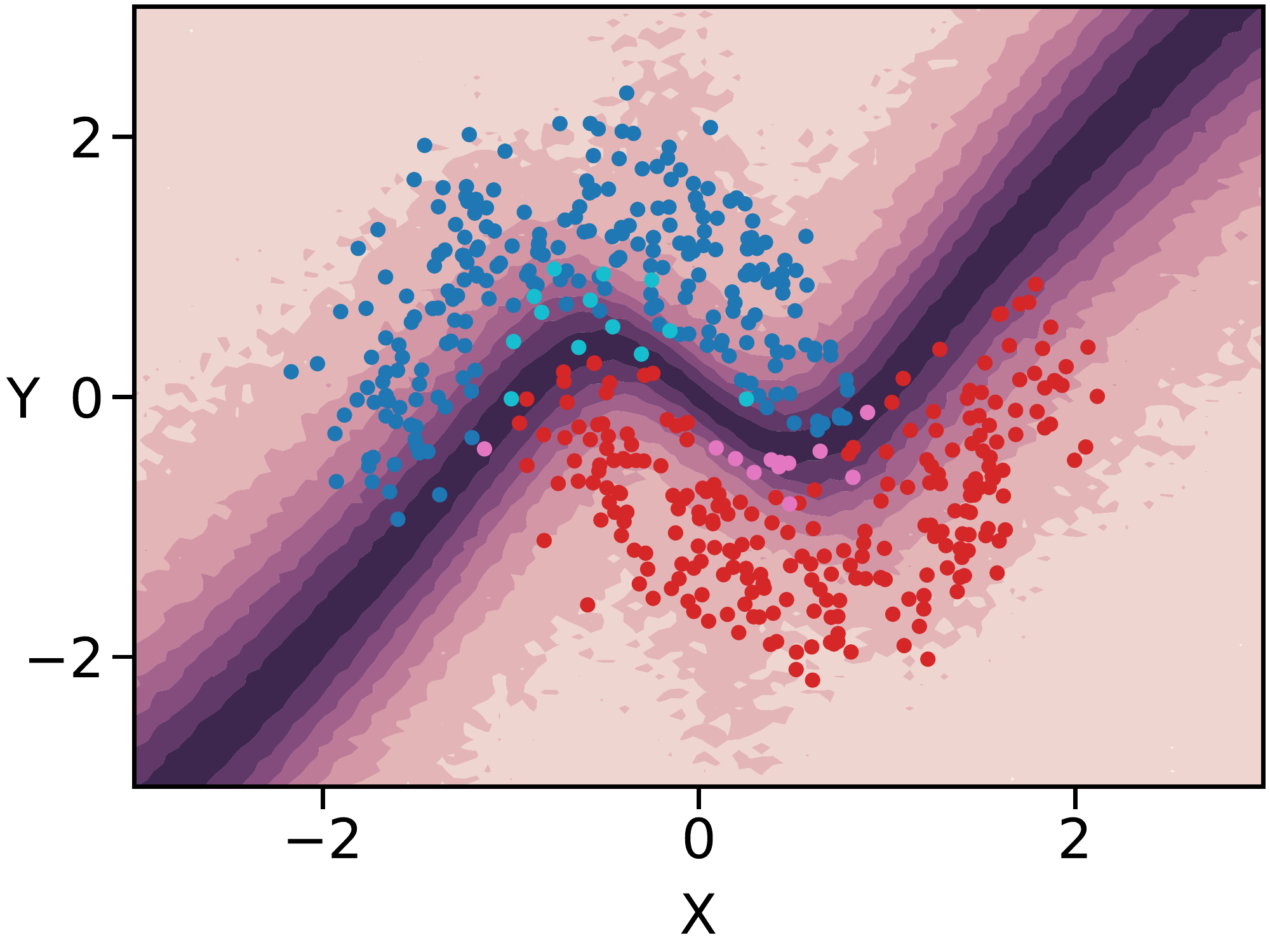}
        \caption{QICP, $n=11$}
    \end{subfigure}
    \begin{subfigure}[b]{0.24\linewidth}
        \includegraphics[width=\textwidth]{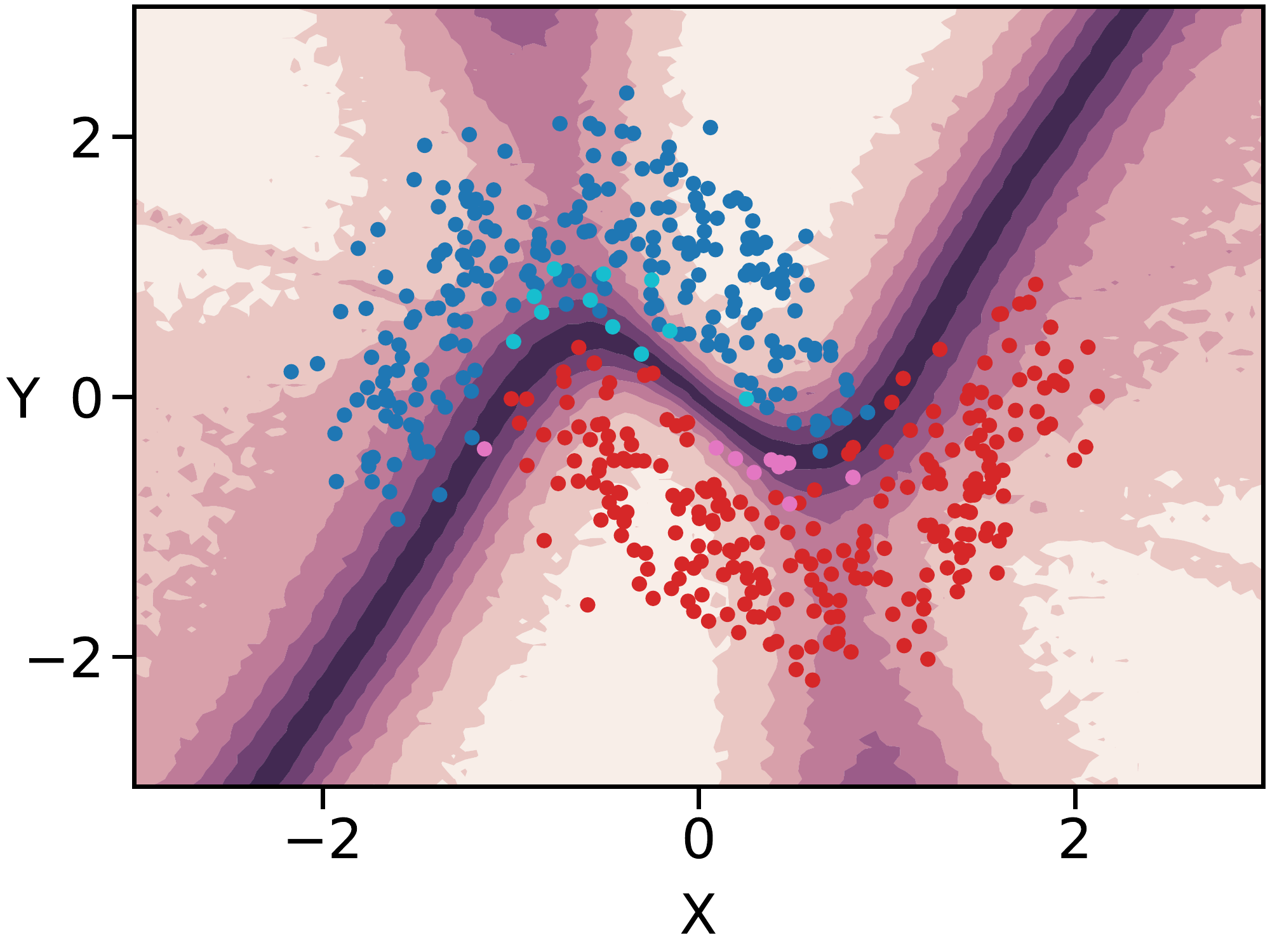}
        \caption{QIQP, $n=11$}
    \end{subfigure}
    \vskip 1.0em    
    \hspace{11.3em}
    \begin{subfigure}[b]{0.24\linewidth}
        \includegraphics[width=\textwidth]{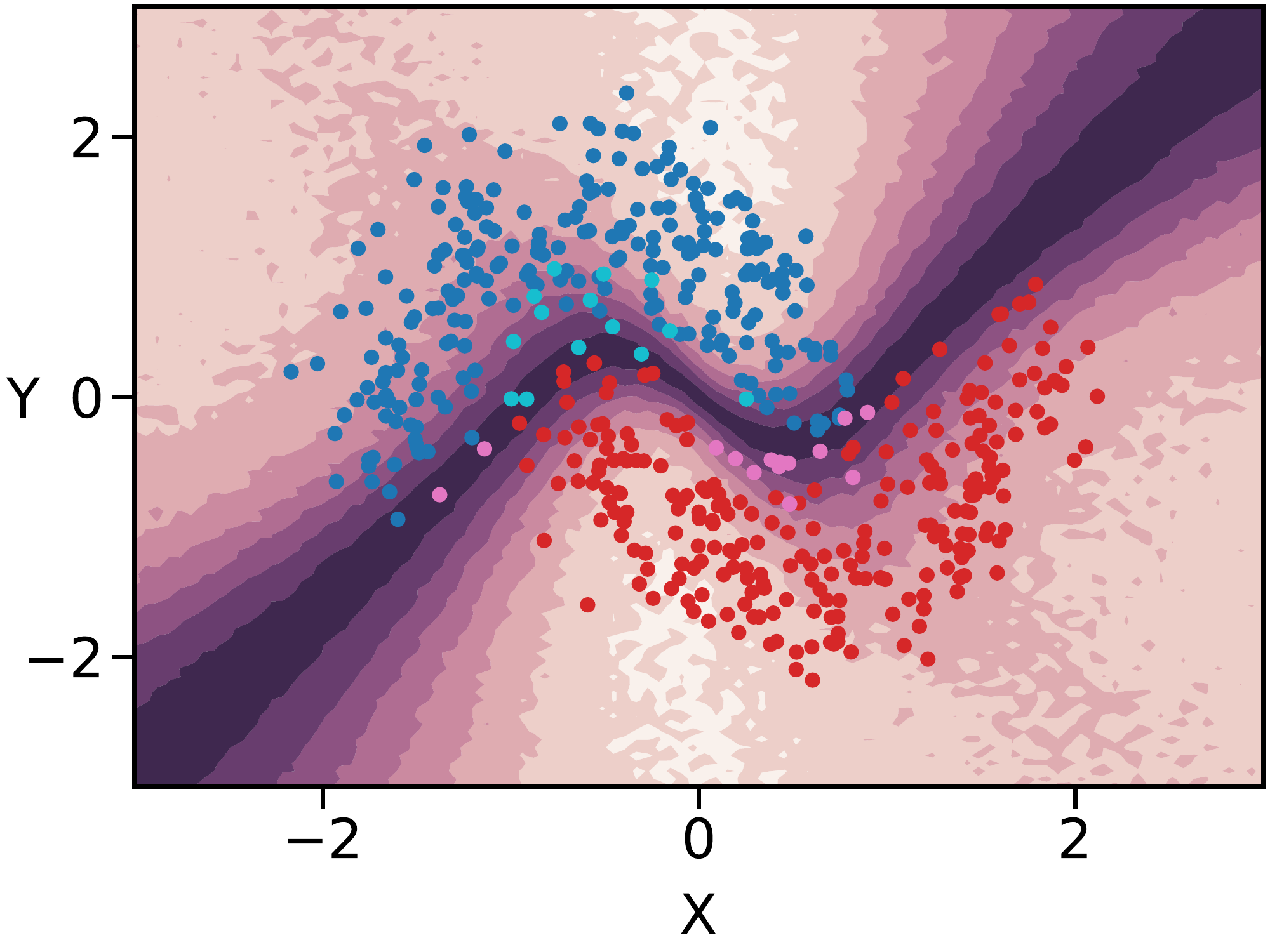}
        \caption{CIQP, $n=13$}
    \end{subfigure}
    \begin{subfigure}[b]{0.24\linewidth}
        \includegraphics[width=\textwidth]{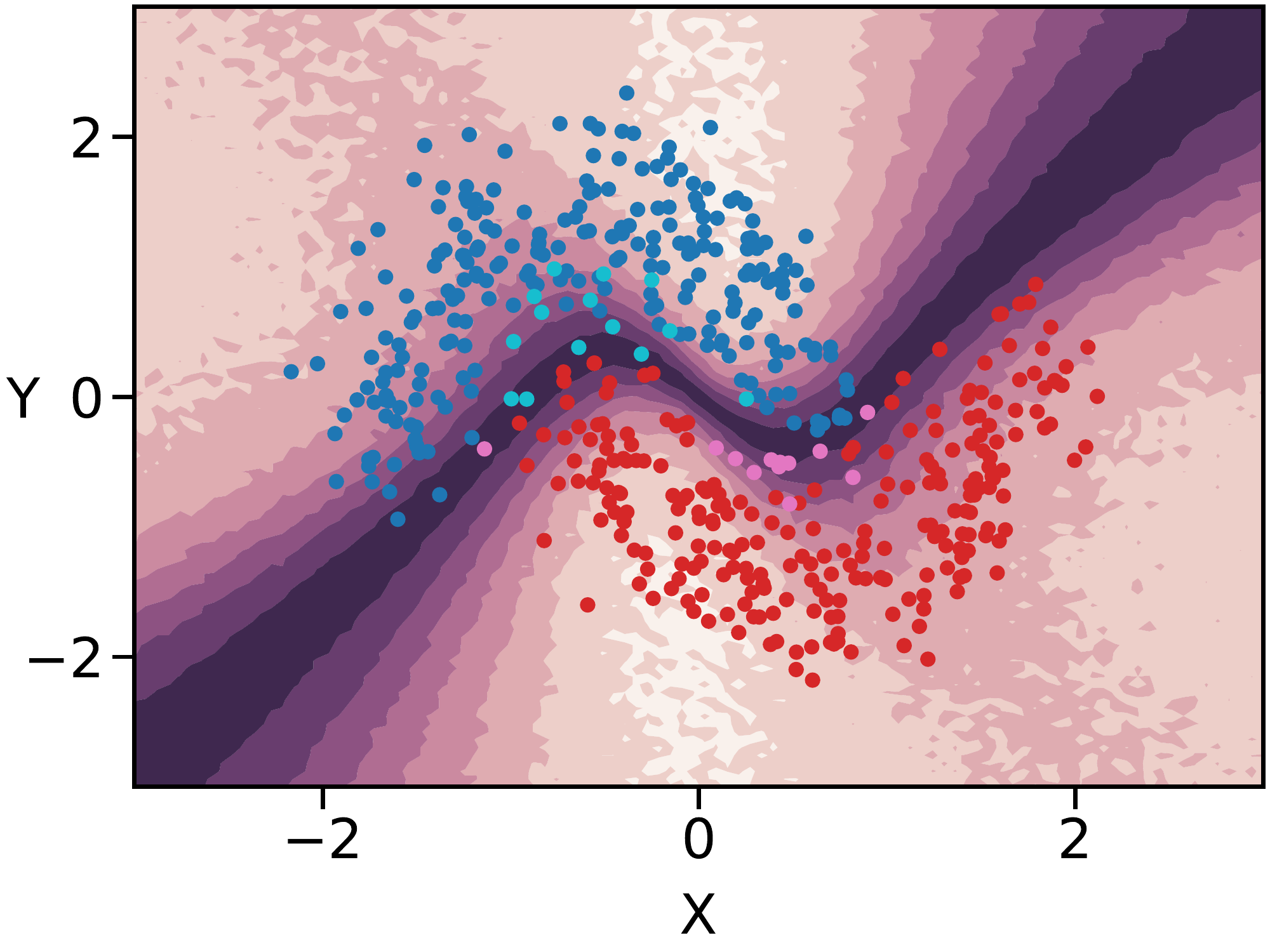}
        \caption{QICP, $n=13$}
    \end{subfigure}
    \begin{subfigure}[b]{0.24\linewidth}
        \includegraphics[width=\textwidth]{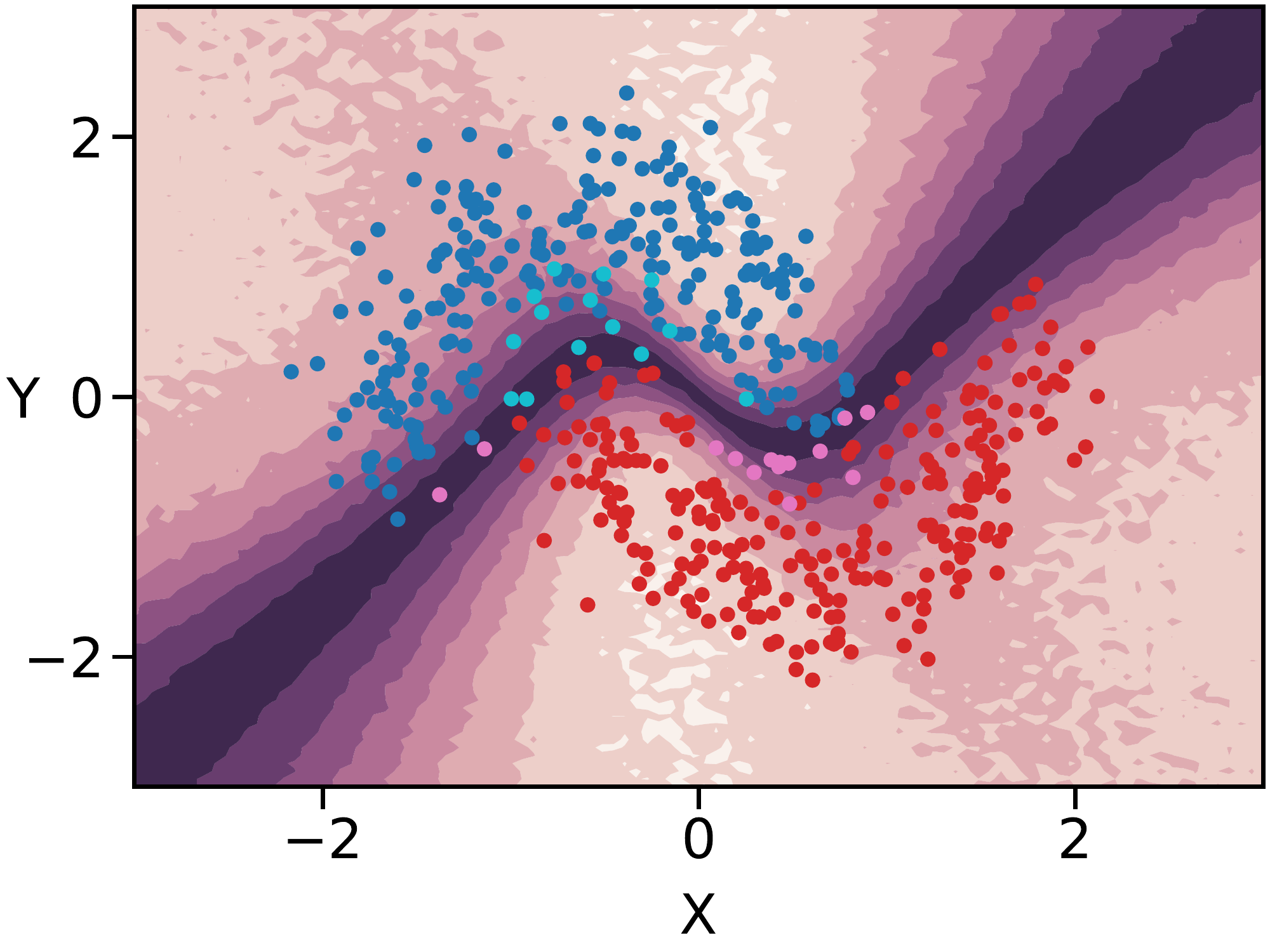}
        \caption{QIQP, $n=13$}
    \end{subfigure}
    
\caption{Additional Results for Posterior Predictive Standard Deviation of Binary Classification with \ac{bnn}: \emph{C} and \emph{Q} stand for \emph{Classical} and \emph{Quantum} respectively. \emph{I} and \emph{P} stand for \emph{Inference} and \emph{Prediction}. The Figure shows the expected increase in accuracy for higher qubit numbers $n$.}
\label{fig:bc_std_add}
\end{center}
\vskip -0.2in
\end{figure}

\cref{fig:lr_add} are further results for the linear regression task for a larger set of qubits. \cref{fig:bc_mean_add} and \cref{fig:bc_std_add} show the mean prediction and standard deviation of the binary classification task respectively, both for a larger set of qubits.

\end{document}